\documentclass[epsfig,12pt,onecolumn]{article}
\usepackage{amsfonts}
\usepackage{amssymb}
\usepackage{multicol}
\usepackage{graphicx}
\usepackage{float}
\usepackage{caption}
\usepackage{xcolor}
\usepackage[utf8]{inputenc}
\usepackage{amsmath}

\makeatletter \@addtoreset{equation}{section}
\renewcommand{\theequation}{\thesection.\arabic{equation}}
\def \be{\begin{equation}}
\def \ee{\end{equation}}
\def \bea{\begin{eqnarray}}
\def \eea{\end{eqnarray}}

\newcommand{\nc}{\newcommand}
\nc{\al}{\alpha} \nc{\bib}{\bibitem} \nc{\la}{\lambda}
\nc{\C}{\mbox{\hspace{1.24mm}\rule{0.2mm}{2.5mm}\hspace{-2.7mm} C}}
\nc{\R}{\mbox{\hspace{.04mm}\rule{0.2mm}{2.8mm}\hspace{-1.5mm} R}}
\input{tcilatex}
\begin{document}

\title{\rightline{\mbox{\small {LPHE-MS-March-23}} \vspace
{1 cm}} \textbf{Embedding Integrable Superspin Chain in String Theory}}
\author{Y. Boujakhrout, E.H Saidi, R. Ahl Laamara, L.B Drissi \\
{\small 1. LPHE-MS, Science Faculty}, {\small Mohammed V University in
Rabat, Morocco}\\
{\small 2. Centre of Physics and Mathematics, CPM- Morocco}}
\maketitle

\begin{abstract}
Using results on topological line defects of 4D Chern-Simons theory and the
algebra/cycle homology correspondence in complex surfaces $\mathcal{S}$ with
ADE singularities, we study the graded properties of the $sl(m|n)$ chain and
its embedding in string theory. Because of the $\mathbb{Z}_{2}$-grading of $%
sl(m|n)$, we show that the $\left( m+n\right) !/m!n!$ varieties of superspin
chains with underlying super geometries have different cycle homologies%
\textbf{. }We investigate the algebraic and homological \textrm{features} of
these integrable quantum chains and give a link between graded 2-cycles and
genus-g Rieman surfaces $\Sigma _{g}$. Moreover, using homology language, we
yield the brane realisation of the $sl(m|n)$ chain in type IIA string and
its uplift to M-theory. Other \textrm{aspects} like graded complex surfaces
with $sl(m|n)$ singularity as well as super magnons are also described.%
\newline
\textbf{Keywords:} Superspin chain, 4D CS theory with supergroups, Yangian
algebra and Magnons, Du Val singularities, Branes in type II strings and
M-theory.
\end{abstract}


\section{Introduction}

\begin{figure}[tbph]
\begin{center}
\includegraphics[width=14cm]{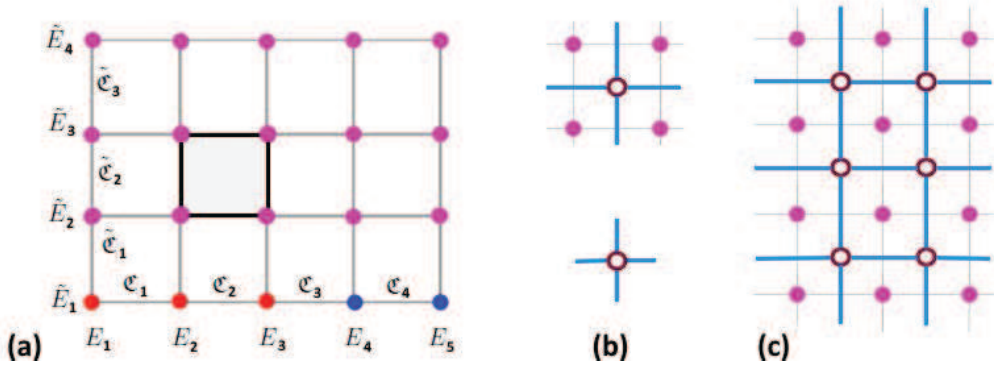}
\end{center}
\par
\vspace{-0.5cm}
\caption{(\textbf{a}) A lattice representation of the M-brane realisation of
the $sl(3|2)$ superspin chain with flavor symmetry $U\left( 4\right) _{f}$.
The vertices $E_{A}$ of the lattice represent gauge M5-branes and the $%
\tilde{E}_{l}$ describe flavor M5 branes. Line segments joining two vertices
are interpreted in terms of M2 branes stretching between the M5-branes. (%
\textbf{b}) a dual representation of a unit cell of the lattice given by a
point-like bold circle. (\textbf{c}) The dual representation of (\textbf{a})
with unit cell as in (\textbf{b}). The dual lattice (\textbf{c}) represents
intersecting M5 branes wrapping 4-cycles in ${\protect\large Y}_{4}$. }
\label{typeII}
\end{figure}

String theory represents a rich framework to investigate the quantum field
theories in diverse dimensions and quantum integrable models as well as the
connection between them \cite{1A}-\cite{2AA}. This link, commonly termed as
the Gauge/Integrability (G/I) correspondence \cite{3A,3Aa,3AA,3AB,3AC},
relates two areas of quantum physics and can be manifested in \textrm{the
framework} of string theory using brane realisations \textrm{and dualities} 
\cite{3AD}-\cite{nafiz}.\newline
First hints of the G/I\ correspondence linked Bethe Anzats soluble 2D
integrable systems to 2D $\mathcal{N}=2$ supersymmetric gauge theories \cite%
{3A,3Aa}. But more recently, these integrable models were shown to make a
natural appearance in the 4D bosonic Chern Simons (CS) theory equipped with
line and surface defects that generate lattice models \cite{wtn1}-\cite{6A}.
In this context, the XXX Heisenberg spin chain is realised using a set of
Wilson and 't Hooft lines such that their gauge configuration reproduces the
degrees of freedom of the spin chain and allows to recover solutions of this
integrable system such as the Yang-Baxter R-matrix and the Lax operator \cite%
{1B}-\cite{5BA}.\newline
In string theory language, the 4D CS describing integrable lattice models
can be realised via stacks of branes like D4s ending on NS5-brane in type
IIA string theory; while Yang-Baxter equation and the Yangian can be
obtained via the Hilbert space of a 6D gauge theory by applying T- and
S-duality \cite{2B}. Linking CS gauge theory to maximally supersymmetric and
topologically twisted Yang Mills theories; and applying dualities in the
string theory context, \textrm{allows for} the unification of integrability
in supersymmetric gauge theories \cite{3B}. For example, quantum integrable
systems such as the eight-vertex model and the XYZ spin chain \cite{6B} were
shown to appear in different field theory setups, all related by dualities
in string theory.\newline
All of \textrm{these }results represent building blocks for the study of the
integrable $sl(m|n)$ superspin chain that we are concerned about here. As a
generalisation of the $sl(m)$ bosonic \textrm{chain}, the $sl(m|n)$ \textrm{%
integrable} system should be endowed with an extended G/I correspondence
(super-G/I below). Indeed, in \cite{7B}, closed spin chains \textrm{%
characterized by} the super Yangian of $gl(m|n)$ were linked to a class of $%
\mathcal{N}=(2,2)$ supersymmetric gauge theories in 2D; such that the
supersymmetric vacua were identified with the Bethe states of the superspin
chain. On the other side, these integrable models are also identified with a
4D Chern Simons theory having a super gauge symmetry and equipped with super
line defects. \textrm{In} \cite{paper3}, the super-G/I allowed us to recover
the Lax operator of the $sl(m|n)$ spin chain \textrm{by studying} the
coupling of Wilson and 't Hooft super lines and properties of the gauge
field bundles in the CS theory with $SL(m|n)$ symmetry. Moreover, in the
recent inspiring work \cite{nafiz}, a brane realisation of the superspin
chain of $gl(m|n)$ Yangian was \textrm{employed} to \textrm{manifest the
equivalence of the} gauge theories appearing in the gauge side of the
super-G/I and \textrm{the dualities relating them} in string theory \cite%
{9B,10B}. Another result of \cite{nafiz} concerns the brane and
supersymmetric quiver interpretations of magnons for the \textrm{family} of $%
sl(m|n)$ superspin chains. These super magnons are generalisations of the\
standard $sl\left( m\right) $\ magnons which are quasi-particles of the
integrable quantum spin chains given by excitation modes of the ground state 
$\left\vert \Omega _{\mathbf{\Lambda }}\right\rangle $ labeled by a highest
weight vector $\Lambda $\ of the symmetry group. For the $SU\left( 2\right) $%
\ spin, the magnons are quantised spin waves that was extensively studied in
physical literature with regards to the magnetic properties of materials and
the entanglement \cite{1CA}-\cite{1C}. In particular, the superspin chain
models that we are interested in here are of \textrm{particular} utility due
to the special structure of super algebras and the rich \textrm{structure}
they hold. Thanks to super-G/I, these interesting algebraic features lead to
exotic phenomenons to discover on the string theory side and at the level of
gauge field configurations as well as in realisations using topological line
defects of the super CS theory. In fact, it was shown in \cite{2C}, that for
the $psu(2,2|4)$ integrable super spin chain underlying the AdS/CFT
correspondence, the different Dynkin diagrams and Cartan matrices of the
super Lie algebra are related via fermionic dualities. In \cite{3C}, the
authors encoded the possible sets of Bethe equations of the $SU(m|n)$ spin
chain arising from different Dynkin diagrams in the QQ-system, and described
the transformation properties of this system under fermionic and bosonic
dualities allowing to move between two different descriptions of the same
spin chain.\newline
In this paper, we make a \textrm{further} step towards the understanding of
the super-G/I correspondence with insertion in string theory. \textrm{We}
focuss on the specific aspects of the $sl\left( m|n\right) $ superspin chain
family arising from the $\mathbb{Z}_{2}$-grading of the super algebra and
the multiplicity of its Dynkin super diagrams; and \textrm{investigate}
connections with super line defects of the CS theory. We also study the
embedding of the super chain in M-theory and in type II strings with two
main objectives.\newline
First, we revisit the setup of the integrable super chain and use its
algebraic properties to $\left( \mathbf{i}\right) $ Link the quantum chain
states to topological line defects interpreted here in terms of the Wilson
and the 't Hooft super lines. These lines offer another way to realise the
real super chain \textrm{in 4D CS} by crossing lines as given by the Figure %
\ref{TH}. $\left( \mathbf{ii}\right) $ Motivate brane realisations of the
super chain in type II strings and M-theory as given by Tables \textbf{\ref%
{IIA}}, \textbf{\ref{NS}}, \textbf{\ref{M52}} as well as the Figure \ref{M25}%
. $\left( \mathbf{iii}\right) $ Relate the graded spin states to \textrm{%
remarkable} properties captured by Dynkin super diagrams of Lie
superalgebras like those given in Figure \textbf{\ref{10} }regarding the
representative $sl\left( 3|2\right) $ symmetry. \newline
\textrm{Secondly}, we take advantage of results on complex surfaces with ADE
singularities used in string compactifications \textrm{\cite{1D}-\cite{5D}}
to study their generalisation to complex manifolds with \emph{super
singularity} given by the $sl(m|n)$ gauge symmetry. We derive the Super
Algebra/Homology correspondence (Super-A/H) in complex surfaces with super
singularity extending the Du Val geometries of the Table \textbf{\ref{418}}.
We show that, unlike ADE singularities involving 2-cycles $\mathfrak{\tilde{C%
}}$ with self intersections $\mathfrak{\tilde{C}}^{2}=-2$, the super
geometries have three kinds of 2-cycles $\mathfrak{C}$ with self
intersections $\mathfrak{C}^{2}=0,\pm 2$ including the usual $\mathfrak{C}%
^{2}=-2$ in addition the two exotic $\mathfrak{C}^{2}=0,+2.$ Regarding the
embedding of the closed super chain $\mathbb{R}_{t}\times \mathbb{S}_{\theta
}^{1}$ in M-theory, we construct the structure of the 11D space time $%
\mathcal{M}_{1,10}$ and show that it has the fibration $\mathbb{R}_{t}\times 
\mathbb{S}_{\theta }^{1}\times \mathbb{S}_{M}^{1}\times {\large Y}_{4}$ with
complex 4D manifold as depicted by the graphs of Figure \textbf{\ref{typeII} 
}giving a lattice description of ${\large Y}_{4}.$%
The $\mathbb{S}_{M}^{1}$ is the usual circle of M-theory. In addition 
\textrm{to} the (1+1)D space time of the super chain, the algebraic geometry
properties of $\mathcal{M}_{1,10}$ are mainly captured by ${\large Y}_{4}$
which can be roughly imagined as given by the fibration $\mathbb{R}%
^{4}\times {\large C}_{4}$ with real 4-cycle ${\large C}_{4}$ itself fibered
like ${\large C}_{2}\times {\large \tilde{C}}_{2}$. This fibration is
generated by $\mathfrak{C}_{A}\times \mathfrak{\tilde{C}}_{l}$ as
schematically exhibited by the graph \textbf{\ref{typeII}}-\textbf{a}; the
irreducible 2-cycles $\mathfrak{C}_{A}$\textrm{\ }are the generators of $%
{\large C}_{2}$ while the $\mathfrak{\tilde{C}}_{l}$'s are the generators of 
${\large \tilde{C}}_{2}$. In the graphics \textbf{\ref{typeII}}, the
vertices of the lattice represent gauge M5$_{A}$ and flavor M5$_{l}$ branes
while the edges describe (M2)$_{A,A+1}$ and (M2)$_{l,l+1}$ branes stretching
between neighboring M5 branes. The two pictures (down and up) of the Figure 
\textbf{\ref{typeII}-b} describe a dual description of the unit cell\textrm{%
\textbf{\ }}$\mathfrak{C}_{A}\times \mathfrak{\tilde{C}}_{l}$ (4-cycle); it
is represented by a 4-vertex with a top given by the bold circle from which 
\textrm{come out} four legs for interactions. The graph \textbf{\ref{typeII}}%
-\textbf{c} gives a dual description of the graph (\textbf{a}) with bold
circle nodes describing M5 branes wrapping 4-cycles in ${\large Y}_{4}$ and
edges for intersections.\ 

\ \ \ \newline
The presentation is as follows. In section 2, we describe the degrees of
freedom of the $sl(m|n)$ chain and propose a short path for representing
this integrable chain in terms of line defects of 4D Chern-Simons theory and
branes in type II strings and in M-theory. In section 3, we \textrm{show}
that there are $\left( m+n\right) !/m!n!$ varieties of $sl(m|n)$ superspin
chains and give their properties. In section 4, we study the bridge between
the bosonic $sl(m+n)$ and the graded $sl(m|n)$ by using their Cartan
matrices $\mathcal{A}^{sl_{m+n}}$ and $\mathcal{K}^{sl_{m|n}}.$ We introduce
colored Dynkin super diagrams and use them to describe the 2-cycle homology
of complex manifolds with $sl(m|n)$ singularity. In section 5, we use
results obtained above to embed the $sl(m|n)$ chain in type II strings and
in M-theory. In section 6, we construct the geometry of the \textrm{M-theory}
manifold $\mathcal{M}_{1,10}$ \textrm{hosting} the M-brane system realising
the $SL(m|n)$ chain with flavor symmetry $U(N_{\mathrm{f}})$. In section 7,
we give a conclusion and \textrm{comments}. In the appendix, we describe the
super magnons by using the super Yangian algebraic method. The study
concerning \textrm{magnons was reported in the} appendix in order to \textrm{%
preserve} the sequence of ideas in the presentation of our work.

\section{Superspin chain: Degrees of freedom and defects}

In this section, we investigate the degrees of freedom of the $sl(m|n)$
superspin chain family and \textrm{employ them} to model this integrable
chain in terms of line defects of the 4D Chern-Simons (CS) theory. We use
the algebraic properties of this superspin chain to $\left( i\right) $
introduce the symmetries of the underlying topological field theory; in
particular the $SL(m|n)$ gauge invariance of the CS action and the $U\left(
N_{f}\right) $ flavor symmetry of the line defects. $\left( ii\right) $
motivate the brane realisations in type II strings and in M-theory.\newline
We will use convenient terminologies like for example\ super atoms and super 
\textrm{chains} for homogeneous chains with sites having the same superspin;
also the $sl(m|n)$ super Cartan matrix \textrm{associated to }Dynkin super
diagrams of $sl(m|n)$.

\subsection{$sl(m|n)$ superspin chain revisited}

\begin{figure}[tbph]
\begin{center}
\includegraphics[width=10cm]{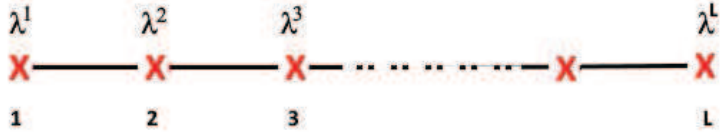}
\end{center}
\par
\vspace{-0.5cm}
\caption{Finite $sl\left( m|n\right) $ super chain with L-sites. At each
position, live quantum states characterised by the HW vector $\mathbf{%
\protect\lambda }$ of the $sl\left( m|n\right) $ fundamental representation. 
}
\label{HC}
\end{figure}

We begin by considering a finite integrable $sl(m|n)$ superspin chain with $%
L $ sites $\Upsilon _{l}$ to which we refer\ as the superspin "atoms" or
simply super atoms. We assume that the super chain has the following
specific properties:

\begin{description}
\item[$\left( \protect\alpha \right) $] It is closed with length $L$; that
is having $L$ super atoms $\Upsilon _{1},\Upsilon _{2},...,\Upsilon _{L}.$

\item[$\left( \protect\beta \right) $] The closure ($\Upsilon _{L+1}\equiv
\Upsilon _{1}$) is realised by twisted boundary conditions as described in 
\cite{nafiz}; we refer to the space dimension of this chain by the circle $%
\mathbb{S}_{\theta }^{1}$.

\item[$\left( \protect\gamma \right) $] It is an homogenous chain in the
sense that all of its\ super atoms $\Upsilon _{l}$ are identical; thus
having the same values for the superspin states.
\end{description}

\ \ \newline
Graphically, the super chain $\mathbb{S}_{\theta }^{1}$ is represented by
the Figure \textbf{\ref{HC} }with\textbf{\ }red cross points referring to
the $\Upsilon _{l}$'s. 
The superspin atoms of the chain have two kinds of degrees of freedom:
"extrinsic" degrees and "intrinsic" ones. \textrm{We} describe them below
with some details by first considering a representative super atom $\Upsilon
,$ say\ the first super atom\ $\Upsilon _{1}$, and then turning to the chain 
$\mathbb{S}_{\theta }^{1}=\left\{ \Upsilon _{l}\right\} _{1\leq l\leq L}.$

\ 

\textbf{A) Degrees of freedom of a super atom}\newline
Given a super atom $\Upsilon $ of the superspin chain, we have two classes
of degrees of freedom; namely \newline
$\left( \mathbf{1}\right) $ Its classical position $\xi $ in the real space $%
\mathbb{S}_{\theta }^{1}$. \newline
$\left( \mathbf{2}\right) $ Two kinds of intrinsic degrees of freedom
denoted below as $\mathbf{\lambda }_{\eta }$ and $z$ \textrm{which} play an
important role in our modeling: \newline
$\left( \mathbf{a}\right) $ The $\mathbf{\lambda }_{\eta }$'s are weight
vectors describing the superspin states for the atom $\Upsilon ;$ they form
a highest weight (HW) representation of $sl(m|n)$. For the interesting case
where the states of $\Upsilon $ sit in the fundamental representation, the
weight vectors are given by the $\left( m+n\right) $ unit weight vectors $%
\mathbf{\epsilon }_{A},...,\mathbf{\epsilon }_{m+n}.$ For this natural
choice, we have $\mathbf{\lambda }_{\eta }\equiv \mathbf{\lambda }_{A}$ with%
\begin{equation}
\mathbf{\lambda }_{A}=\mathbf{\epsilon }_{A}\qquad ,\qquad A=1,...,m+n
\label{fun}
\end{equation}%
However, for generic $sl(m|n)$ representations, the weight vectors $\mathbf{%
\lambda }_{\eta }$ are given by linear combinations like $\sum \mathbf{%
\lambda }_{\eta A}\mathbf{\epsilon }_{A}$. Below, we will mainly focuss on
the (m+n)- dimensional representation (\ref{fun}) to which we refer as $%
\mathcal{R}\left( \mathbf{\omega }_{1}\right) $ because the HW vector $%
\mathbf{\epsilon }_{1}$ is given by the fundamental coweight $\mathbf{\omega 
}_{1}$; i.e%
\begin{equation}
\mathbf{\epsilon }_{1}=\mathbf{\omega }_{1}  \label{22}
\end{equation}%
In this regard, notice that $sl(m|n)$ has $\left( m+n-1\right) $ fundamental
coweights $\mathbf{\omega }_{A};$ i.e:%
\begin{equation}
\mathbf{\omega }_{1},...,\mathbf{\omega }_{m+n-1}
\end{equation}%
these are duals of the simple roots $\mathbf{\alpha }_{A}$ to be described
in details later. We have $\left( m+n-1\right) $ simple roots%
\begin{equation}
\mathbf{\alpha }_{1},...,\mathbf{\alpha }_{m+n-1}
\end{equation}%
$\left( \mathbf{b}\right) $ The spectral parameter $z$ termed sometimes as
rapidity of the atom, with reference to the spectral parameter of scattering
theory \cite{3AG,6D,7D}. This complex $z$\ is also identified with the usual
complex parameter introduced in the\textrm{\ realisation }of representations
of the Yangian algebra $\mathcal{Y}\left[ sl(m|n)\right] $ in relation to
the study of quantum integrable spin chains. Recall that the Yangian
superalgebra is generated by the transfer matrix $T_{AB}\left( z\right) $
obeying the famous RTT equation \cite{8D}%
\begin{equation}
R\left( z_{1}-z_{2}\right) T\left( z_{1}\right) T\left( z_{2}\right)
=T\left( z_{2}\right) T\left( z_{1}\right) R\left( z_{1}-z_{2}\right)
\label{23}
\end{equation}%
with $R\left( z_{1}-z_{2}\right) $\ being the R-matrix and $T\left( z\right)
=E^{AB}\oplus T_{AB}\left( z\right) $. Notice that the spectral parameter $z$
is also used as a complex coordinate giving the position of a line defect $%
\mathcal{L}\left( z\right) $ in the holomorphic plane of 4D Chern-Simons
gauge theory as formulated in \cite{wtn1}.

\ \ 

\textbf{B) Degrees of freedom of the super chain}\newline
Before approaching the super chain, notice first that from the above
description, a given super atom $\Upsilon $ is characterised by the degrees
of freedom $\left( \xi ;\mathbf{\lambda }_{\mathbf{\omega }_{1}},z\right) $
that describe the position $\xi ,$ the HW vector $\mathbf{\lambda }_{\mathbf{%
\omega }_{1}}$ and the spectral parameter $z.$ For convenience, we rewrite
these degrees like 
\begin{equation}
\left( \xi ;\mathbf{\lambda }_{\mathbf{\omega }_{1}},z\right) =\left( \xi ;%
\mathbf{\lambda }_{A},z\right)   \label{ap}
\end{equation}%
where we exhibited the super label $A$ of (\ref{fun}). The $A=1,...,m+n$ is
the label of the fundamental coweight $\mathbf{\omega }_{1}.$ Eq(\ref{ap})
means that at the quantum level, we can think of the quantum state $%
\left\vert \Upsilon \right\rangle $ describing the ground state of the super
atom $\Upsilon $ as follows%
\begin{equation}
\left\vert \Psi _{A}\right\rangle =\left\vert \xi ;\mathbf{\lambda }%
_{A},z\right\rangle   \label{pa}
\end{equation}%
To fix the ideas and for later use, we give in Table \textbf{\ref{z}} 
\begin{table}[h]
\begin{center}
$%
\begin{tabular}{|c|l|l|}
\hline
super atom & real position & quantum state \\ \hline
$\Upsilon _{1}$ & \multicolumn{1}{|c|}{$\xi _{1}$} & \multicolumn{1}{|c|}{$%
|\Psi _{A}^{(1)}>=|\xi _{1};\lambda _{A}^{1},z_{1}>$} \\ \hline
$\Upsilon _{2}$ & \multicolumn{1}{|c|}{$\xi _{2}$} & \multicolumn{1}{|c|}{$%
|\Psi _{A}^{(2)}=|\xi _{2};\lambda _{A}^{2},z_{2}>$} \\ \hline
$\vdots $ & \multicolumn{1}{|c|}{$\vdots $} & \multicolumn{1}{|c|}{$\vdots $}
\\ \hline
$\Upsilon _{L}$ & \multicolumn{1}{|c|}{$\xi _{L}$} & \multicolumn{1}{|c|}{$%
|\Psi _{A}^{(L)}=|\xi _{L};\lambda _{A}^{L},z_{L}>$} \\ \hline
\end{tabular}%
$%
\end{center}
\par
\vspace{-0.5cm}
\caption{Quantum states of the L individual super atoms in the sl(m%
\TEXTsymbol{\vert}n) super chain.}
\label{z}
\end{table}
the classical positions $\xi _{1},\xi _{2},...,\xi _{L}$\ and the intrinsic
labeling of the quantum states of the $L$ super atoms\textrm{\ }$\Upsilon
_{1},\Upsilon _{2},...,\Upsilon _{L}$\ making the linear $sl(m|n)$\ chain. 
\newline
Returning to the super chain, notice that by using the atomic parameters (%
\ref{ap}), the degrees of freedom of the super chain are obtained by adding
an extra index to label the $L$ atoms as $\left\{ \Upsilon _{l}\right\} .$ 
\textrm{Thus,} the (extrinsic and intrinsic) degrees of freedom of the super
chain are given by%
\begin{equation}
\left( \xi _{l};\mathbf{\lambda }_{A}^{l},z_{l}\right) \qquad ,\qquad 1\leq
l\leq L  \label{dof}
\end{equation}%
with (\ref{pa}) promoted like%
\begin{equation}
\left\vert \Psi _{A}^{l}\right\rangle =\left\vert \xi ^{l};\mathbf{\lambda }%
_{A}^{l},z^{l}\right\rangle 
\end{equation}%
The ground state of the super chain minimizing the energy of the chain model
is given by a\ linear combination $\left\vert \Psi \left[ \varrho \right]
\right\rangle =\sum_{A,l}\varrho _{l}^{A}\left\vert \Psi
_{A}^{l}\right\rangle $ with some probability amplitude values $\varrho
_{l}^{A}.$ In what follows, we will refer to this $\left\vert \Psi \left[
\varrho \right] \right\rangle $ state as the pseudo-vacuum of the super
chain.

\ \ \ 

\begin{figure}[tbph]
\begin{center}
\includegraphics[width=6cm]{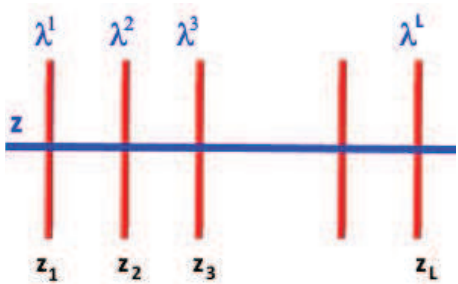}
\end{center}
\par
\vspace{-0.5cm}
\caption{Intrinsic degrees of freedom of the super chain interpreted in
terms of crossing line defects. The vertical lines are given by Wilson lines
with spectral parameters $z_{l}.$ The coupling between the Wilsons is
mediated by a horizontal 't Hooft line.}
\label{TH}
\end{figure}

\textbf{C) Symmetries and Super Magnons}\newline
To smooth the path towards the interpretation in terms of line defects and
the brane realisation of super chains in type II string theory and M-
theory, we need to replace the classical variable $\xi _{l}$ by a \textrm{one%
} suitable for this purpose. As one of th\textrm{e roles o}f the classical $%
\xi _{l}$s is to discriminate the $L$ super atoms of the super chain, we can
use instead a flavor symmetry group which turns out to be given by the
unitary%
\begin{equation}
U\left( N_{f}\right) =U\left( 1\right) \times SU\left( N_{f}\right)
\end{equation}%
where for convenience we have set $L=N_{f}.$ In this way of doing, the $%
N_{f} $ positions $\left\{ \xi _{l}\right\} $ are replaced by the weight
vectors $\left\{ \mathbf{e}_{l}\right\} $ of the fundamental representation
of $U\left( N_{f}\right) ;$ that is by substituting with%
\begin{equation}
\xi _{l}\rightarrow \mathbf{e}_{l}
\end{equation}%
in eq(\ref{dof}). Therefore, we obtain the following quantum description of
the pseudo-vacuum of the super chain $\sum_{A,l}\varrho _{l}^{A}\left\vert
\Psi _{A}^{l}\right\rangle $ where now 
\begin{equation}
\left\vert \Psi _{A}^{l}\right\rangle =\left\vert \mathbf{e}_{l};\mathbf{%
\lambda }_{A}^{l},z_{l}\right\rangle  \label{zd}
\end{equation}%
To deal with these super atomic states, we will think about them in terms of
the factorisation 
\begin{equation}
\begin{tabular}{lll}
$\left\vert \Psi _{A}^{l}\right\rangle $ & $=$ & $\left\vert \mathbf{e}%
_{l}\right\rangle \otimes \left\vert \mathbf{\lambda }_{A}^{l}\right\rangle
\otimes \left\vert z_{l}\right\rangle $ \\ 
& $=$ & $\left\vert \mathbf{e}_{l}\right\rangle \otimes \left\vert \mathbf{%
\Omega }_{A}^{l}\right\rangle $%
\end{tabular}
\label{fs}
\end{equation}%
with%
\begin{equation}
\left\vert \mathbf{\Omega }_{A}^{l}\right\rangle =\left\vert \mathbf{\lambda 
}_{A}^{l}\right\rangle \otimes \left\vert z_{l}\right\rangle
\end{equation}%
Notice the following three features: $\left( \mathbf{i}\right) $ the three
blocks $\left\vert \mathbf{e}_{l}\right\rangle ,$ $\left\vert \mathbf{%
\lambda }_{A}^{l}\right\rangle $ and $\left\vert z_{l}\right\rangle $ are
respectively related with the flavor $U\left( N_{f}\right) $, the gauge $%
SL\left( m|n\right) $ and the Yangian $\mathcal{Y}_{SL\left( m|n\right) }$. $%
\left( \mathbf{ii}\right) $ The properties of the composite bloc $\left\vert 
\mathbf{e}_{l};\mathbf{\lambda }_{A}^{l}\right\rangle =\left\vert \mathbf{e}%
_{l}\right\rangle \otimes \left\vert \mathbf{\lambda }_{A}^{l}\right\rangle $
are associated with the finite dimensional symmetry,%
\begin{equation}
U\left( N_{f}\right) \times SL\left( m|n\right)  \label{US}
\end{equation}%
while the $\left\vert \mathbf{e}_{l}\right\rangle \otimes \left\vert \mathbf{%
\Omega }_{A}^{l}\right\rangle $ is described\ by\ the infinite dimensional
invariance%
\begin{equation}
U\left( N_{f}\right) \times \mathcal{Y}_{SL\left( m|n\right) }  \label{UY}
\end{equation}%
We focus here on the sector $\left\vert \mathbf{e}_{l};\mathbf{\lambda }%
_{A}^{l}\right\rangle $ with finite symmetry (\ref{US}); and report to the
Appendix the analysis of the sector $\left\vert \mathbf{e}_{l}\right\rangle
\otimes \left\vert \mathbf{\Omega }_{A}^{l}\right\rangle $ with full
symmetry (\ref{UY}). The full sector (\ref{fs}) is \textrm{useful for the
study of} super magnon states (\ref{BE}-\ref{EB})\textrm{\ that w}e give
here below 
\begin{equation}
\prod\limits_{i=1}^{\nu }T_{A_{i}B_{i}}\left( z_{i}\right) \left\vert \Omega
_{\mathbf{\Lambda }}\right\rangle \qquad ,\qquad A_{i}>B_{i}
\end{equation}%
with%
\begin{equation}
\begin{tabular}{lll}
$T_{AA}\left( z\right) \left\vert \Omega _{\mathbf{\Lambda }}\right\rangle $
& $=$ & $b_{\lambda _{A}}\left( z\right) \left\vert \Omega _{\mathbf{\Lambda 
}}\right\rangle $ \\ 
$T_{AB}\left( z\right) \left\vert \Omega _{\mathbf{\Lambda }}\right\rangle $
& $=$ & $0,\qquad A<B$%
\end{tabular}%
\end{equation}%
Notice that to fix the ideas and as an anticipation of the sector $%
\left\vert z_{l}\right\rangle $; one can think about the spectral parameters 
$z_{l}$ in (\ref{dof}) as the zeros of the following holomorphic polynomial%
\begin{equation}
\mathcal{P}\left( z\right) =\prod\limits_{l=1}^{N_{f}}\mathcal{P}_{l}\left(
z\right) \qquad ,\qquad \mathcal{P}_{l}\left( z\right) =c_{l}\left(
z-z_{l}\right)  \label{pl}
\end{equation}%
to be further discussed later in details; see eqs(\ref{sf},\ref{WT}) and (%
\ref{s22}). The numbers $c_{l}$ are constants here. In the meantime, notice
that the $z_{l}$ zeros in (\ref{pl}) will give the positions of Wilson lines
(W$\left[ z_{l}\right] $ for short) while the $z$ is the position of a 't
Hooft line tH$\left[ z\right] $ as illustrated by Figure \textbf{\ref{TH}}
where the crossing of the two topological lines \textrm{takes} place at $%
z=z_{l}$.

\ \ \ 

\textbf{D) More on flavor symmetry, line defects and branes}\newline
We seen before that the $N_{f}$ super atoms of the chain,\ classically
described by their positions $\xi _{l}$, are\ quantum mechanically
discriminated by the $U\left( N_{f}\right) $ flavor symmetry. The
correspondence atom/flavor can be further presented as follows: To each
classical atom position $\xi _{l},$ we associate a global phase 
\begin{equation}
e^{i\theta ^{l}h_{l}}\qquad ,\qquad l=1,...,N_{f}
\end{equation}%
with abelian charge operator $h_{l}.$ As such, the $N_{f}$ phases of the
super chain can be combined into the following compact operator%
\begin{equation}
D=e^{i\mathbf{\theta .h}}\qquad ,\qquad \mathbf{\theta .h}%
=\sum_{l=1}^{N_{f}}\theta ^{l}h_{l}
\end{equation}%
with $N_{f}$ diagonal matrices $h_{l}$ as 
\begin{equation}
h_{l}=\left\vert l\right\rangle \left\langle l\right\vert \qquad ,\qquad 
\left[ h_{l},h_{l^{\prime }}\right] =0
\end{equation}%
These $h_{l}$ charge operators generate an abelian $U\left( 1\right) ^{N_{f}}
$ symmetry which is isomorphic to the Cartan subgroup of the non abelian $%
U\left( N_{f}\right) $ flavor symmetry with group elements as 
\begin{equation}
U=e^{i\mathbf{\Theta .\Sigma }}\qquad ,\qquad \mathbf{\Theta .\Sigma }%
=\sum_{l,l^{\prime }=1}^{N_{f}}\Theta ^{ll^{\prime }}\Sigma _{ll^{\prime }}
\end{equation}%
with 
\begin{equation}
\Sigma _{ll^{\prime }}=\left\vert l\right\rangle \left\langle l^{\prime
}\right\vert 
\end{equation}%
standing for the $N_{f}^{2}$ generators of $U\left( N_{f}\right) $.\newline
The use of the flavor symmetry to deal with the features of the superspin
chain has interesting consequences. In particular, the super chain has the
larger symmetry%
\begin{equation}
SL\left( m|n\right) \times U\left( N_{f}\right) 
\end{equation}%
Moreover, the presence of the complex $z_{l}$'s and the flavor invariance in
the formulation of degrees of freedom of the super chain is very suggestive.
The $z_{l}$'s are interpreted below as the positions of $N_{f}$ parallel
(vertical) topological line defects $\mathrm{\gamma }_{z_{l}}$ in the
holomorphic plane of the 4D Chern-Simons theory. We will think about these $%
N_{f}$ topological lines as describing electrically charged Wilson lines
that we denote as%
\begin{equation}
W_{l}\equiv W[\mathrm{\gamma }_{z_{l}},\mathbf{\lambda }_{l}]  \label{th}
\end{equation}%
such that the electric charges are given by highest weight vectors $\mathbf{%
\lambda }^{l}$ of the superspin representations $\mathcal{R}\left( \mathbf{%
\lambda }^{l}\right) $ of $sl(m|n)$. 
Regarding the brane realisation of\ these super chain states, the
topological lines are imagined as living at the intersection of two or more
branes \textrm{\cite{ABC}}. From the type IIA string view, the super atoms,
thought of as line defects W$_{l}$, are represented by $N_{f}$ D4 branes
intersecting NS5 and D2. \textrm{In }type IIB string, these W$_{l}$'s
involve D5 branes; and in M-theory, they involve M5 branes. These
representations in the brane language \textrm{are} summarised in the Table 
\textbf{\ref{IIA}}. 
\begin{table}[h]
$%
\begin{tabular}{c||c|c|c}
branes & type IIA & M theory & type IIB \\ \hline\hline
magnetic & $\left. 
\begin{array}{c}
\text{D4} \\ 
\text{NS5}%
\end{array}%
\right. $ & $\left. 
\begin{array}{c}
\text{M5} \\ 
\text{M5/S}_{\text{M}}^{1}%
\end{array}%
\right. $ & $\left. 
\begin{array}{c}
\text{D5/NS5} \\ 
\text{D3}%
\end{array}%
\right. $ \\ \hline
electric & $\left. 
\begin{array}{c}
\text{D2} \\ 
\text{F1}%
\end{array}%
\right. $ & $\left. 
\begin{array}{c}
\text{M2} \\ 
\text{M2/S}_{\text{M}}^{1}%
\end{array}%
\right. $ & $\left. 
\begin{array}{c}
\text{D1/F1} \\ 
\text{D3}%
\end{array}%
\right. $ \\ \hline\hline
\end{tabular}%
$\centering \renewcommand{\arraystretch}{1.2}
\caption{Type II strings and M-theory brane candidates for the description
of the Wilson and the't Hooft lines using brane intersections. As an
example, see Figure \textbf{\protect\ref{M25}.}}
\label{IIA}
\end{table}
In these regards, notice that graphically speaking, the interaction between
the super atoms of the chain is insured by a horizontal line \textrm{tH}$%
\left[ \mathrm{\gamma }_{z}\right] $ crossing all the \textrm{W}$_{l}$'s. 
\newline
As a \textrm{result} of this description, the superspin chain of the Figure 
\textbf{\ref{HC}} gets mapped to the line defect system given by the Figure 
\textbf{\ref{TH} }and the stringy realisation with intersecting brane as in
the Table \textbf{\ref{IIA}}. For the example of the M- brane picture, see
Figure \textbf{\ref{M25}}.

\subsection{Topological field description}

Here, we build the bridge between the super chain degrees of freedom
described above and the field modeling of the 4D Chern-Simons (CS) theory.
This bridge is based on the line defects \textrm{W}$_{\gamma }$\textrm{/tH}$%
_{l}$\textrm{\ }and the spectral parameter $z$ that appears in (\ref{pl}).
The complex $z$ is the coordinate variable of the complex line $\mathcal{C}$
in the CS theory. \newline
Recall that the action $\mathcal{S}\left[ \mathcal{A}\right] $ of the
topological CS gauge theory reads in absence of external sources as follows 
\cite{wtn1}, 
\begin{equation}
\mathcal{S}\left[ \mathcal{A}\right] =\int_{\mathbb{R}^{2}\times \mathcal{C}%
}dz\wedge str\left( \mathcal{A}\wedge d\mathcal{A}+\frac{2}{3}\mathcal{A}%
\wedge \mathcal{A}\wedge \mathcal{A}\right)  \label{ac}
\end{equation}%
In this gauge field action, $\mathcal{C}$ is the complex holomorphic line;
and the $\mathbb{R}^{2}\left( x,y\right) $ is the real topological plane
where live the external line defects such as the electrically charged Wilson
lines and the magnetically charged 't Hooft line. \newline
Moreover, the 1-form gauge potential $\mathcal{A}$ in (\ref{ac}) is a
function of the variables $\left( x,y;z\right) $ parameterising $\mathbb{R}%
^{2}\times \mathcal{C}$, it is a $\mathbb{Z}_{2}$- graded matrix. For the
case of 3D CS with supergroups, see \cite{8B}.%
\begin{equation}
\mathcal{A}=\left( 
\begin{array}{cc}
\mathfrak{A}_{m\times m} & \mathcal{\psi }_{m\times n} \\ 
\mathcal{\psi }_{n\times m} & \mathfrak{A}_{n\times n}%
\end{array}%
\right)
\end{equation}%
This potential also expands like $\mathcal{A}=t_{a}\mathcal{A}^{a}$ where $%
t_{a}$ generate the Lie superalgebra $sl\left( m|n\right) $ and $\mathcal{A}%
^{a}$ is a partial gauge connection given by 
\begin{equation}
\mathcal{A}^{a}=dx\mathcal{A}_{x}^{a}+dy\mathcal{A}_{y}^{a}+d\bar{z}\mathcal{%
A}_{\bar{z}}^{a}
\end{equation}%
In the absence of external line defects, the gauge action is as in eq(\ref%
{ac}), and the field equation of $\mathcal{A}$ following from the variation $%
\delta \mathcal{S}\left[ \mathcal{A}\right] =0$ reads as 
\begin{equation}
\mathcal{F}=d\mathcal{A}+\mathcal{A}\wedge \mathcal{A}=0
\end{equation}%
in agreement with the topological nature of the 4D Chern-Simons theory
without external sources. \newline
In the presence of 't Hooft lines as in Figure \textbf{\ref{HC}}, the gauge
curvature $F$\ is no longer trivial and looks like the curvature of a Dirac
monopole of 4D Yang Mills theory \cite{4B,1I,2I,3I}. In this situation, the
gauge connection $A$\ on the complement of the lines defines a topologically
non trivial bundle on the 2-spheres surrounding these lines.\emph{\newline
}We end this section by giving some useful algebraic tools regarding the $%
sl(m|n)$ chain with super atoms $\Upsilon ^{l}$ carrying generic
representation charges with HW vectors $\mathbf{\lambda }^{l}$. First,
recall that the HW vectors $\mathbf{\lambda }^{l}$ are expanded in terms of
the basis vectors $\epsilon _{A}$ (\ref{fun}) like,%
\begin{equation}
\mathbf{\lambda }^{l}=\sum_{A=1}^{m+n}\lambda _{A}^{l}\epsilon _{A}
\label{LL}
\end{equation}%
As the $\epsilon _{A}$'s can be also expressed in terms of the $\left(
m+n-1\right) $ fundamental coweights $\mathbf{\omega }_{A}$ of $sl(m|n),$ or
equivalently in terms of its simple roots $\mathbf{\alpha }_{A};$ we can
rewrite (\ref{LL}) either as a sum over the $\mathbf{\omega }_{A}$'s or as a
sum over the $\mathbf{\alpha }_{A}$'s. By using the $\mathbf{\omega }_{A}$%
's, we have 
\begin{equation}
\mathbf{\lambda }^{l}=\sum_{A=1}^{m+n-1}n_{A}^{l}\mathbf{\omega }_{A}
\end{equation}%
with the $\mathbf{\lambda }^{l}$'s defining HW representations $\mathcal{R}%
\left( \mathbf{\lambda }^{l}\right) $ of the Lie superalgebra $sl(m|n).$ As
the quantum state $\left\vert \Psi _{A}^{l}\right\rangle $ of the super
atoms given in (\ref{zd}) transforms in the representation $\mathcal{R}%
\left( \mathbf{\lambda }^{l}\right) \equiv \mathcal{R}^{l}$, the full state
spectrum of the super chain transforms in the tensor product representation 
\begin{equation}
\boldsymbol{R}\left( \mathbf{\Lambda }\right) =\prod\limits_{l=1}^{N_{f}}%
\mathcal{R}^{l}\qquad ,\qquad \mathbf{\Lambda }=\sum_{l=1}^{N_{f}}\mathbf{%
\lambda }^{l}
\end{equation}%
that is reducible as sum over irreducible representations of $sl(m|n)$. By
using the weight vector basis $\mathbf{\epsilon }_{A}$ of the superalgebra $%
sl(m|n)$ with inner product $\left\langle \mathbf{\epsilon }_{A},\mathbf{%
\epsilon }_{B}\right\rangle =\left( -\right) ^{\left\vert A\right\vert
}\delta _{AB}$ where $\left\vert A\right\vert $ refers to the two possible
degrees $\bar{0}$ and $\bar{1}$, we can expand the HW vectors $\mathbf{%
\lambda }^{l}$s as follows%
\begin{equation}
\mathbf{\lambda }^{l}=\sum_{A=1}^{m+n}\lambda _{A}^{l}\mathbf{\epsilon }_{A}
\end{equation}%
and%
\begin{equation}
\mathbf{\Lambda }=\sum_{A=1}^{m+n}\Lambda _{A}\mathbf{\epsilon }_{A}\qquad
,\qquad \Lambda _{A}=\sum_{l=1}^{L}\lambda _{A}^{l}  \label{LA}
\end{equation}%
To concretize our investigation, we think it interesting to \textrm{consider}
a particular Lie superalgebra and realize the objectives of this study on 
\textrm{that} example. In what follows, we will focus on the $sl(3|2)$
superspin model and comment on the extension of the results to the $sl(m|n)$
family with $m>n$.

\section{$sl(3|2)$ superspin models: algebraic set up}

In this section, we study the algebraic \textrm{setup} of the superspin
chain with length $N_{f}$ in order to $\left( i\right) $ classify the
varieties of such type of integrable super chains; and $\left( ii\right) $
give a front matter towards their embedding in type II strings that will be
investigated in sections 5 and 6.\newline
First, we show that there are 10 varieties of $sl(3|2)$ superspin chains and
explore their properties. In general, for the generic $sl(m|n)$ chain with $%
m>n$, there are%
\begin{equation}
\frac{\left( m+n\right) !}{m!n!}
\end{equation}%
varieties of $sl(m|n)$ graded chains. This diversity constitutes a special
feature of superspin chains due to the $\mathbb{Z}_{2}$-grading of Lie
superalgebras that does not occur in the bosonic $sl(m)$ chain. We also show
that among the ten $sl(3|2)$ superspin models, four of them are \textrm{%
somehow} redundant; thus leaving six basic $sl(3|2)$ super chains. One of
these super chains is very special; it is termed as the \emph{distinguished
chain},\ and \textrm{will} be the subject of a detailed study. Related ideas
regarding the study of the $gl(n|m)$ spin chain has been developed in 
\textrm{\cite{Volin}}.

\subsection{$sl(3|2)$ super chain models}

The $sl(3|2)$ superspin chain is a representative model of the $sl(m|n)$
family with $m>n$ or equivalently $m\neq n.$ So the result derived in this
section holds for the whole members of this family. Here, we use Lie super
algebraic properties of $sl(3|2)$ to show that there are 10 varieties of $%
sl(3|2)$ super chains and that only 6 of them are effectively different; the
additional 4 are related to their homologues by a mirror symmetry.

\subsubsection{Ten varieties of $sl(3|2)$ chains}

\begin{figure}[tbph]
\begin{center}
\includegraphics[width=14cm]{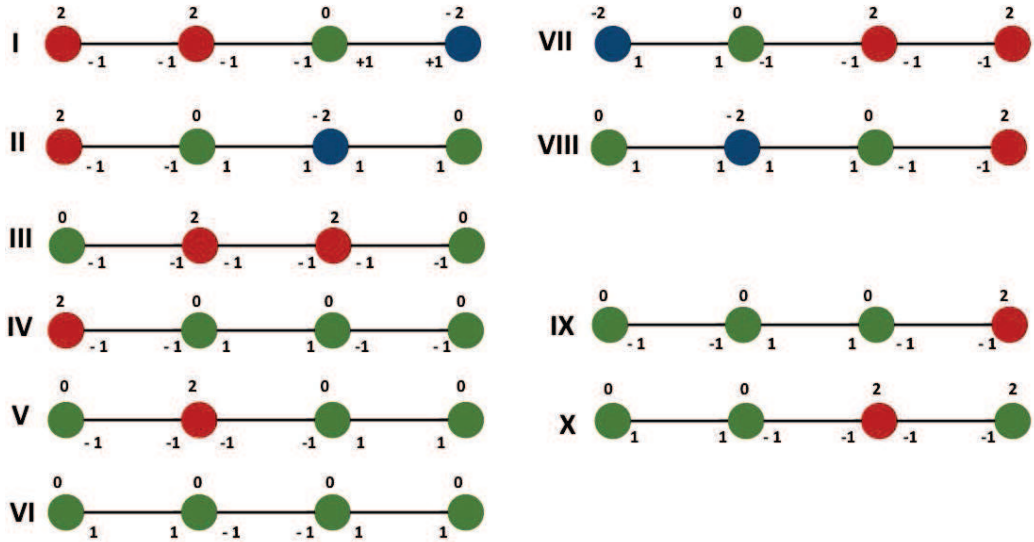}
\end{center}
\par
\vspace{-0.5cm}
\caption{The ten varieties of Dynkin super diagrams for the $sl(3|2)$ Lie
superalgebra. They can be organised like $\left( 4_{+}+1_{0}+1_{0}\right)
+4_{-}$ with regards to reflection acting as $N_{q}\rightarrow N_{-q}$ given
by eq(\protect\ref{ref}). So, the four $4_{-}$ of the 10 \emph{DSD}s are
redundant as they can be recovered by reflection of the four $4_{+}$. }
\label{10}
\end{figure}

We start by recalling that the $sl(3|2)$ chain given by the Figure \textbf{%
\ref{HC} }has $N_{f}$ super atoms with degrees of freedom as in (\ref{dof}-%
\ref{zd}). Each super atom $\Upsilon $ is specified by a set of quantum
charges; in particular the five $sl(3|2)$ ones indicating that $\left\vert
\Upsilon \right\rangle $ is generated by the following quantum states%
\begin{equation}
\left\vert \mathbf{\epsilon }_{1}\right\rangle ,\quad \left\vert \mathbf{%
\epsilon }_{2}\right\rangle ,\quad \left\vert \mathbf{\epsilon }%
_{3}\right\rangle ,\quad \left\vert \mathbf{\epsilon }_{4}\right\rangle
,\quad \left\vert \mathbf{\epsilon }_{5}\right\rangle  \label{e5}
\end{equation}%
These weight states $\left\vert \mathbf{\epsilon }_{A}\right\rangle $
describe the superspin representation of $sl(3|2)$ with highest weight
vector $\mathbf{\epsilon }_{1}.$ In representation theory language, this HW
vector corresponds to the fundamental coweight $\mathbf{\omega }_{1}$ of $%
sl(3|2)$; this means that $\mathbf{\epsilon }_{1}=\mathbf{\omega }_{1}$ is
as in (\ref{22}) while the other four in (\ref{e5}) descend from it. In this
regard, recall that, as far as the algebraic fundamentals of $sl(3|2)$ are
concerned, the $\mathbf{\omega }_{1}$ is the dual of the simple root $%
\mathbf{\alpha }_{1}$ of the Lie superalgebra $sl(3|2).$ Obviously, this is
not the unique fundamental coweight as $sl(3|2)$ has other fundamental
coweights $\mathbf{\omega }_{A}$ dual to the other simple roots $\mathbf{%
\alpha }_{A}$ in the sense that they obey $\mathbf{\omega }_{A}.\mathbf{%
\alpha }_{B}=\delta _{AB}$. This duality relation is remarkably solved by 
\begin{equation}
\mathbf{\alpha }_{A}=\mathcal{K}_{AB}^{sl_{3|2}}\mathbf{\omega }_{B}\qquad
,\qquad \mathbf{\omega }_{B}=(\mathcal{K}_{AB}^{sl_{3|2}})^{-1}\mathbf{%
\alpha }_{A}\qquad ,\qquad \det \mathcal{K}_{AB}^{sl_{3|2}}\neq 0  \label{K}
\end{equation}%
where the invertible $\mathcal{K}_{AB}^{sl_{3|2}}$ is the super Cartan
matrix of $sl(3|2)$. However, because of the $\mathbb{Z}_{2}$- grading
implying the decomposition,%
\begin{equation}
sl(3|2)=sl(3|2)_{\bar{0}}\oplus sl(3|2)_{\bar{1}}
\end{equation}%
the $\mathcal{K}_{AB}^{sl_{3|2}}$ is not uniquely defined. In fact, the $%
sl(3|2)$ superalgebra has $\left( i\right) $ 10 possible Dynkin-like super
diagrams (\emph{DSD}) as shown in Figure \textbf{\ref{10};} 
and then $\left( ii\right) $ 10 associated super Cartan matrices that we
denote like 
\begin{equation}
\begin{tabular}{lllllllll}
$\mathcal{K}_{sl_{3|2}}^{I}$ & $,\quad $ & $\mathcal{K}_{sl_{3|2}}^{II}$ & $%
,\quad $ & $\mathcal{K}_{sl_{3|2}}^{III}$ & $,\quad $ & $\mathcal{K}%
_{sl_{3|2}}^{IV}$ & $,\quad $ & $\mathcal{K}_{sl_{3|2}}^{V}$ \\ 
$\mathcal{K}_{sl_{3|2}}^{VI}$ & $,\quad $ & $\mathcal{K}_{sl_{3|2}}^{VII}$ & 
$,\quad $ & $\mathcal{K}_{sl_{3|2}}^{VIII}$ & $,\quad $ & $\mathcal{K}%
_{sl_{3|2}}^{IX}$ & $,\quad $ & $\mathcal{K}_{sl_{3|2}}^{X}$%
\end{tabular}
\label{SKM}
\end{equation}%
\textrm{Since} the roots, weights and super Cartan \textrm{matrice}s $%
\mathcal{K}_{sl_{3|2}}$ are highly involved in the study of the quantum
properties of\textrm{\ superspins}, we end up with\ ten classes of $sl(3|2)$
super chains. However, not all of them are different because of the mirror
symmetry exhibited by the Figure \textbf{\ref{10}} and corresponding to the
reflection%
\begin{equation}
\boldsymbol{r}:\mathbf{\alpha }_{A}\rightarrow -\mathbf{\alpha }_{5-A}\qquad
,\qquad \boldsymbol{r}^{2}=I_{id}  \label{ref}
\end{equation}%
To get more insight into this special feature of superspin chains compared
to the usual bosonic-like chains, notice that simple roots $\mathbf{\alpha }%
_{A}$ of $sl(3|2)$\ are realised in terms of graded unit weight vectors as
follows 
\begin{equation}
\mathbf{\alpha }_{A}=\epsilon _{A}-\epsilon _{A+1}  \label{alp}
\end{equation}%
just like for $sl\left( N\right) $ algebras. In particular, here $N=m+n=5$;
therefore this relation extends the usual realisation of simple roots for
the bosonic Lie algebra $sl\left( N\right) $ to the\textrm{\ }$sl(m|n)$%
\textrm{\ }superalgebra with the difference that now (\ref{alp}) depends on
the grading of the weight vectors $\epsilon _{A}$. In fact, depending on the
ordering of the $\epsilon _{A}$'s, we distinguish 
\begin{equation}
\frac{5!}{3!2!}=10
\end{equation}%
possibilities given by the permutations of the weight basis vectors (\ref{e5}%
). Notice that this basis is made of three even $\mathbb{Z}_{2}$- vectors $%
b_{A_{1}},b_{A_{2}},b_{A_{3}}$ termed as bosonic; and two odd $\mathbb{Z}%
_{2} $- vectors $f_{A_{4}},f_{A_{5}}.$ Notice also that a direct consequence
of the 10 possible realisations of $\mathbf{\alpha }_{A}$ is the existence
of 10 varieties \textrm{of} the super Cartan matrix defined as%
\begin{equation}
\mathcal{K}_{AB}^{sl_{3|2}}=\mathbf{\alpha }_{A}\mathbf{.\alpha }_{B}
\label{sc}
\end{equation}%
By using (\ref{K}), we deduce that as for the simple roots and $\mathcal{K}%
_{AB}^{sl_{3|2}}$; there are also 10 ways to realise the fundamental
coweights $\mathbf{\omega }_{A}$. \newline
In conclusion, the finite dimensional Lie superalgebra $sl(3|2)$ has
apparently 10 different \emph{DSD}s. These super diagrams are characterised,
amongst others, by the number $n_{f}$ of fermionic roots ($1\leq n_{f}\leq 4$%
). We give in Table \textbf{\ref{TAB1}} \textrm{the} interesting six \emph{%
DSD}s with fermionic nodes represented by the green color. We also give the
corresponding weight vector bases and the "lengths" of the simple roots%
\textbf{.} 
\begin{table}[h]
\centering \renewcommand{\arraystretch}{1.2} $%
\begin{tabular}{||c||c||c||c||c||c||}
\hline\hline
basis & $\left( e_{1},e_{2},e_{3},e_{4},e_{5}\right) $ & $\alpha _{\text{a}%
}^{2}=2$ & $\alpha _{\text{a}}^{2}=-2$ & $\alpha _{\text{a}}^{2}=0$ & Dynkin
diagram \\ \hline\hline
\ \ I & $\left. \left( b_{1},b_{2},b_{3},f_{1},f_{2}\right) \right. $ & 2 & 1
& 1 & \includegraphics[width=3cm]{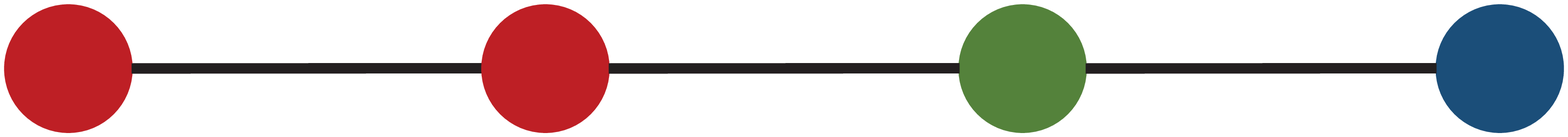} \\ \hline\hline
\ II & $\left( b_{1},b_{2},f_{1},f_{2},b_{3}\right) $ & 1 & 1 & 2 & %
\includegraphics[width=3cm]{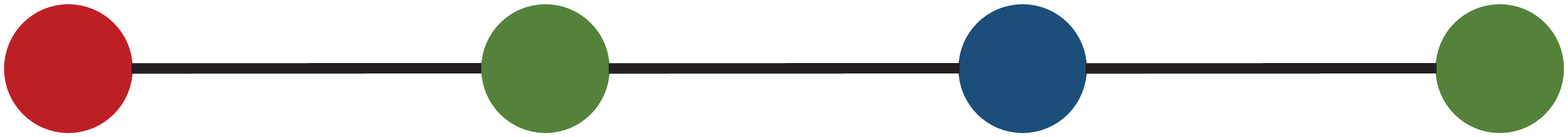} \\ \hline\hline
\ III & $\left( f_{1},b_{1},b_{2},b_{3},f_{2}\right) $ & 2 & 0 & 2 & %
\includegraphics[width=3cm]{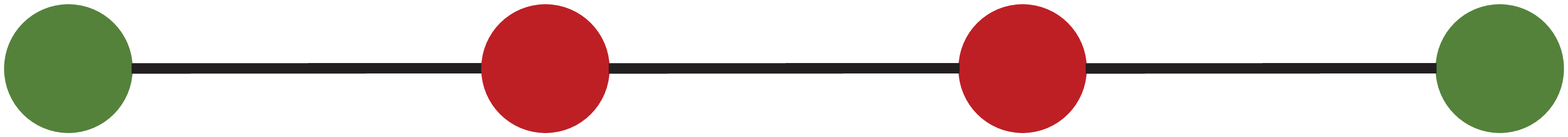} \\ \hline\hline
\ IV & $\left( b_{1},b_{2},f_{1},b_{3},f_{2}\right) $ & 1 & 0 & 3 & %
\includegraphics[width=3cm]{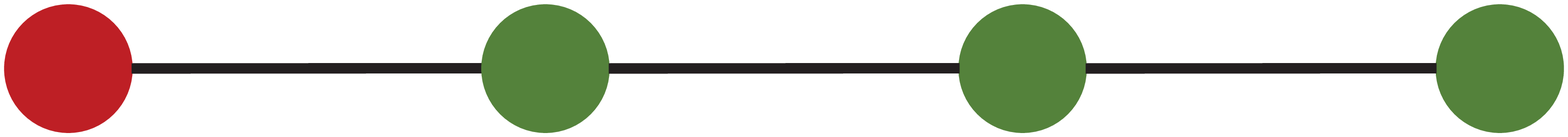} \\ \hline\hline
V & $\left( f_{1},b_{1},b_{2},f_{2},b_{3}\right) $ & 1 & 0 & 3 & %
\includegraphics[width=3cm]{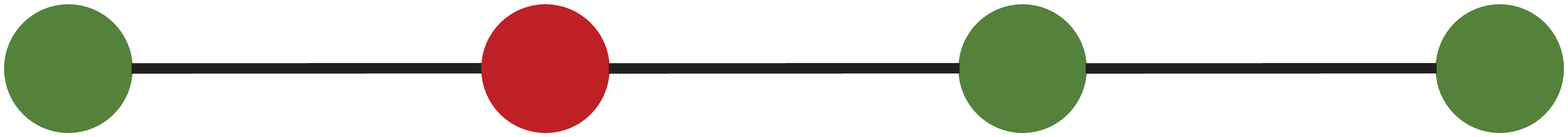} \\ \hline\hline
VI & $\left( b_{1},f_{1},b_{2},f_{2},b_{3}\right) $ & 0 & 0 & 4 & %
\includegraphics[width=3cm]{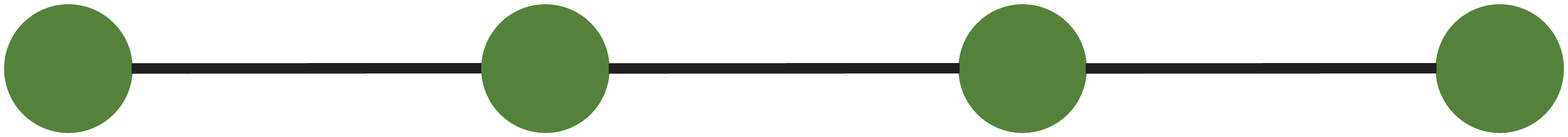} \\ \hline\hline
\end{tabular}%
$%
\caption{ Six of the ten Dynkin super diagrams of the Lie superalgebra $%
sl(3|2)$. This list is ordered according to the number of green nodes.}
\label{TAB1}
\end{table}
The first \emph{DSD} in the Table \textbf{\ref{TAB1}} has three bosonic
nodes (two reds and one blue); and one fermionic green node. The other \emph{%
DSD}s have more than one fermionic (green) node.

\subsubsection{Building the super Cartan matrices (\protect\ref{SKM})}

Here, we construct the explicit expressions of the super Cartan matrices $%
\mathcal{K}_{sl_{3|2}}^{I}$-$\mathcal{K}_{sl_{3|2}}^{VI}$ associated with
the six different \emph{DSD}s listed in the Table \textbf{\ref{TAB1}}. We
also give the expression of their inverses as they are important for the
determination of the coweights $\mathbf{\omega }_{A}$ in terms of the roots $%
\mathbf{\alpha }_{A}$.

\begin{itemize}
\item \textbf{Super Cartan matrix} $\mathcal{K}_{sl_{3|2}}^{I}$\newline
In this case, the simple roots are realised as%
\begin{equation}
\mathbf{\alpha }_{1}=\mathbf{\varepsilon }_{1}-\mathbf{\varepsilon }%
_{2},\quad \mathbf{\alpha }_{2}=\mathbf{\varepsilon }_{2}-\mathbf{%
\varepsilon }_{3},\quad \mathbf{\alpha }_{3}=\mathbf{\varepsilon }_{3}-%
\mathbf{\delta }_{1},\quad \mathbf{\alpha }_{4}=\mathbf{\delta }_{1}-\mathbf{%
\delta }_{2}
\end{equation}%
The\ corresponding super Cartan matrix $\mathcal{K}_{sl_{3|2}}^{I}$
describing the first \emph{DSD}$_{I}$ in the table \textbf{\ref{TAB1}} reads
as follows 
\begin{equation}
\mathcal{K}_{sl_{3|2}}^{I}=\left( 
\begin{array}{cccc}
2 & -1 & 0 & 0 \\ 
-1 & 2 & -1 & 0 \\ 
0 & -1 & 0 & +1 \\ 
0 & 0 & +1 & -2%
\end{array}%
\right)   \label{I}
\end{equation}%
It has $\det \mathcal{K}_{sl_{3|2}}^{I}=1$. The associated \emph{DSD}$_{I}$
has three bosonic simple roots $\mathbf{\alpha }_{1},\mathbf{\alpha }_{2},%
\mathbf{\alpha }_{4}$ and one fermionic $\mathbf{\alpha }_{3}$; they are as
follows: \newline
$\left( \mathbf{a}\right) $ two simple roots \textrm{with} length $\mathbf{%
\alpha }_{1}^{2}=\mathbf{\alpha }_{2}^{2}=2,$ they correspond to the simple
roots of $sl\left( 3\right) $ and are given by the two red nodes in the
first row of Table \textbf{\ref{TAB1}}. \newline
$\left( \mathbf{b}\right) $ one bosonic simple root given by the blue node,
it corresponds to $sl\left( 2\right) $ but with $\mathbf{\alpha }_{4}^{2}=-2$
making its geometrical interpretation very interesting; a proposal using the
Euler characteristic $\mathrm{\chi }$ will be given later on. \newline
These three bosonic simple roots indicate that $\mathcal{K}_{sl_{3|2}}^{I}$
concerns the bosonic sub-symmetry 
\begin{equation}
sl\left( 3|0\right) \oplus sl\left( 0|2\right) \simeq sl\left( 3\right)
\oplus sl\left( 2\right) 
\end{equation}%
$\left( \mathbf{c}\right) $ One fermionic-like simple root $\mathbf{\alpha }%
_{3}$ given by the green node having the remarkable property $\mathbf{\alpha 
}_{3}^{2}=0;$ it is the unique odd simple root in the root system of $sl(3|2)
$ with super Cartan given by $\mathcal{K}_{sl_{3|2}}^{I}$. This root can be
interpreted in terms of the following Lie sub-superalgebra of $sl(3|2),$%
\begin{equation}
sl\left( 1|1\right) 
\end{equation}%
This is the fermionic homologue of the bosonic $sl\left( 2\right) $. A
geometric interpretation of this root in terms of 2-cycles will be given
later. \newline
The inverse of (\ref{I}) is given by%
\begin{equation}
(\mathcal{K}_{sl_{3|2}}^{I})^{-1}=\left( 
\begin{array}{cccc}
0 & -1 & -2 & -1 \\ 
-1 & -2 & -4 & -2 \\ 
-2 & -4 & -6 & -3 \\ 
-1 & -2 & -3 & -2%
\end{array}%
\right)   \label{car}
\end{equation}%
it has negative entries.

\item \textbf{Super Cartan matrix} $\mathcal{K}_{sl_{3|2}}^{II}$\newline
The simple roots are realised as%
\begin{equation}
\mathbf{\alpha }_{1}=\mathbf{\varepsilon }_{1}-\mathbf{\varepsilon }%
_{2},\quad \mathbf{\alpha }_{2}=\mathbf{\varepsilon }_{2}-\mathbf{\delta }%
_{1},\quad \mathbf{\alpha }_{3}=\mathbf{\delta }_{1}-\mathbf{\delta }%
_{2},\quad \mathbf{\alpha }_{4}=\mathbf{\delta }_{2}-\mathbf{\varepsilon }%
_{3}
\end{equation}%
The Cartan matrix $\mathcal{K}_{sl_{3|2}}^{II}$ corresponding to the second 
\emph{DSD}$_{II}$ and its inverse read as follows, 
\begin{equation}
\mathcal{K}_{sl_{3|2}}^{II}=\left( 
\begin{array}{cccc}
2 & -1 & 0 & 0 \\ 
-1 & 0 & +1 & 0 \\ 
0 & +1 & -2 & +1 \\ 
0 & 0 & +1 & 0%
\end{array}%
\right) ,\quad (\mathcal{K}_{sl_{3|2}}^{II})^{-1}=\left( 
\begin{array}{cccc}
0 & -1 & 0 & 1 \\ 
-1 & -2 & 0 & 2 \\ 
0 & 0 & 0 & 1 \\ 
1 & 2 & 1 & 0%
\end{array}%
\right)
\end{equation}%
We have $\det \mathcal{K}_{sl_{3|2}}^{II}=1$. \newline
The \emph{DSD}$_{II}$ has four nodes: two bosonic and two fermionic. The two
bosonic roots are given by $\mathbf{\alpha }_{1}$ and $\mathbf{\alpha }_{3}$%
; the first obeys $\mathbf{\alpha }_{1}^{2}=2$ while the second has $\mathbf{%
\alpha }_{3}^{2}=-2$. These roots indicate \textrm{that} the underlying
bosonic symmetry of $sl(3|2)$ with super Cartan $\mathcal{K}_{sl_{3|2}}^{II}$
contains 
\begin{equation}
sl\left( 2|0\right) \oplus sl\left( 0|2\right) \simeq sl\left( 2\right)
\oplus sl\left( 2\right)
\end{equation}%
The two fermionic nodes are given by the simple roots $\mathbf{\alpha }_{2}$
and $\mathbf{\alpha }_{4};$ they have vanishing lengths $\mathbf{\alpha }%
_{2}^{2}=\mathbf{\alpha }_{4}^{2}=0$ \textrm{and} correspond to%
\begin{equation}
sl\left( 1|1\right) \oplus sl\left( 1|1\right)
\end{equation}

\item \textbf{Super Cartan matrix} $\mathcal{K}_{sl_{3|2}}^{III}$\newline
The simple roots are realised as%
\begin{equation}
\mathbf{\alpha }_{1}=\mathbf{\delta }_{1}-\mathbf{\varepsilon }_{1},\quad 
\mathbf{\alpha }_{2}=\mathbf{\varepsilon }_{1}-\mathbf{\varepsilon }%
_{2},\quad \mathbf{\alpha }_{3}=\mathbf{\varepsilon }_{2}-\mathbf{%
\varepsilon }_{3},\quad \mathbf{\alpha }_{4}=\mathbf{\varepsilon }_{3}-%
\mathbf{\delta }_{2}
\end{equation}%
The Cartan matrix $\mathcal{K}_{sl_{3|2}}^{III}$ describing the third \emph{%
DSD}$_{III}$ and its inverse are given by, 
\begin{equation}
\mathcal{K}_{sl_{3|2}}^{III}=\left( 
\begin{array}{cccc}
0 & -1 & 0 & 0 \\ 
-1 & 2 & -1 & 0 \\ 
0 & -1 & 2 & -1 \\ 
0 & 0 & -1 & 0%
\end{array}%
\right) ,\quad (\mathcal{K}_{sl_{3|2}}^{III})^{-1}=\left( 
\begin{array}{cccc}
-2 & -1 & 0 & 1 \\ 
-1 & 0 & 0 & 0 \\ 
0 & 0 & 0 & -1 \\ 
1 & 0 & -1 & -2%
\end{array}%
\right)
\end{equation}%
with $\det \mathcal{K}_{sl_{3|2}}^{III}=1.$ \newline
This \emph{DSD}$_{III}$ has two bosonic simple roots with $\mathbf{\alpha }%
_{2}^{2}=\mathbf{\alpha }_{3}^{2}=2$ describing the bosonic symmetry 
\begin{equation}
sl\left( 3|0\right) \simeq sl\left( 3\right)
\end{equation}%
The \emph{DSD}$_{III}$ has also two fermionic simple roots with $\mathbf{%
\alpha }_{1}^{2}=\mathbf{\alpha }_{4}^{2}=0$ corresponding to%
\begin{equation}
sl\left( 1|1\right) \oplus sl\left( 1|1\right)
\end{equation}

\item \textbf{Super Cartan matrix} $\mathcal{K}_{sl_{3|2}}^{IV}$\newline
The simple roots are realised by%
\begin{equation}
\mathbf{\alpha }_{1}=\mathbf{\varepsilon }_{1}-\mathbf{\varepsilon }%
_{2},\quad \mathbf{\alpha }_{2}=\mathbf{\varepsilon }_{2}-\mathbf{\delta }%
_{1},\quad \mathbf{\alpha }_{3}=\mathbf{\delta }_{1}-\mathbf{\varepsilon }%
_{3},\quad \mathbf{\alpha }_{4}=\mathbf{\varepsilon }_{3}-\mathbf{\delta }%
_{2}
\end{equation}%
The Cartan matrix $\mathcal{K}_{sl_{3|2}}^{IV}$ for the fourth \emph{DSD}$%
_{IV}$ and its inverse read as follows, 
\begin{equation}
\mathcal{K}_{sl_{3|2}}^{IV}=\left( 
\begin{array}{cccc}
2 & -1 & 0 & 0 \\ 
-1 & 0 & +1 & 0 \\ 
0 & +1 & 0 & -1 \\ 
0 & 0 & -1 & 0%
\end{array}%
\right) ,\quad (\mathcal{K}_{sl_{3|2}}^{IV})^{-1}=\left( 
\begin{array}{cccc}
0 & -1 & 0 & -1 \\ 
-1 & -2 & 0 & -2 \\ 
0 & 0 & 0 & -1 \\ 
-1 & -2 & -1 & -2%
\end{array}%
\right)
\end{equation}%
with $\det \mathcal{K}_{sl_{3|2}}^{IV}=1.$ \newline
The \emph{DSD}$_{IV}$ has one bosonic-like simple root with $\mathbf{\alpha }%
_{1}^{2}=2$ underlying the bosonic subsymmetry 
\begin{equation}
sl\left( 2|0\right) \simeq sl\left( 2\right)
\end{equation}%
The \emph{DSD}$_{IV}$ also has three fermionic-like simple roots with $%
\mathbf{\alpha }_{2}^{2}=\mathbf{\alpha }_{3}^{2}=\mathbf{\alpha }%
_{4}^{2}=0. $

\item \textbf{Super Cartan matrix} $\mathcal{K}_{sl_{3|2}}^{V}$\newline
The simple roots are now realised by%
\begin{equation}
\mathbf{\alpha }_{1}=\mathbf{\delta }_{1}-\mathbf{\varepsilon }_{1},\quad 
\mathbf{\alpha }_{2}=\mathbf{\varepsilon }_{1}-\mathbf{\varepsilon }%
_{2},\quad \mathbf{\alpha }_{3}=\mathbf{\varepsilon }_{2}-\mathbf{\delta }%
_{2},\quad \mathbf{\alpha }_{4}=\mathbf{\delta }_{2}-\mathbf{\varepsilon }%
_{3}
\end{equation}%
The Cartan matrix $\mathcal{K}_{sl_{3|2}}^{V}$ for the fifth \emph{DSD}$_{V}$
and its inverse are given by,%
\begin{equation}
\mathcal{K}_{sl_{3|2}}^{V}=\left( 
\begin{array}{cccc}
0 & -1 & 0 & 0 \\ 
-1 & 2 & -1 & 0 \\ 
0 & -1 & 0 & +1 \\ 
0 & 0 & +1 & 0%
\end{array}%
\right) ,\quad (\mathcal{K}_{sl_{3|2}}^{V})^{-1}=\left( 
\begin{array}{cccc}
-2 & -1 & 0 & -1 \\ 
-1 & 0 & 0 & 0 \\ 
0 & 0 & 0 & 1 \\ 
-1 & 0 & 1 & 0%
\end{array}%
\right)
\end{equation}%
with $\det \mathcal{K}_{sl_{3|2}}^{IV}=1.$ \newline
The \emph{DSD}$_{V}$ has one bosonic-like simple root with $\mathbf{\alpha }%
_{2}^{2}=2$ and three fermionic-like simple roots with $\mathbf{\alpha }%
_{1}^{2}=\mathbf{\alpha }_{3}^{2}=\mathbf{\alpha }_{4}^{2}=0.$

\item \textbf{Super Cartan matrix} $\mathcal{K}_{sl_{3|2}}^{VI}$\newline
The simple roots read in this case like%
\begin{equation}
\mathbf{\alpha }_{1}=\mathbf{\varepsilon }_{1}-\mathbf{\delta }_{1},\quad 
\mathbf{\alpha }_{2}=\mathbf{\delta }_{1}-\mathbf{\varepsilon }_{2},\quad 
\mathbf{\alpha }_{3}=\mathbf{\varepsilon }_{2}-\mathbf{\delta }_{2},\quad 
\mathbf{\alpha }_{4}=\mathbf{\delta }_{2}-\mathbf{\varepsilon }_{3}
\end{equation}%
This super Cartan matrix is very special as it is associated to the purely
fermionic super Dynkin diagram \textrm{with all nodes odd}. \newline
The super $\mathcal{K}_{sl_{3|2}}^{VI}$\ and its inverse are given by, 
\begin{equation}
\mathcal{K}_{sl_{3|2}}^{VI}=\left( 
\begin{array}{cccc}
0 & +1 & 0 & 0 \\ 
+1 & 0 & -1 & 0 \\ 
0 & -1 & 0 & +1 \\ 
0 & 0 & +1 & 0%
\end{array}%
\right) ,\quad (\mathcal{K}_{sl_{3|2}}^{VI})^{-1}=\left( 
\begin{array}{cccc}
0 & 1 & 0 & 1 \\ 
1 & 0 & 0 & 0 \\ 
0 & 0 & 0 & 1 \\ 
1 & 0 & 1 & 0%
\end{array}%
\right)
\end{equation}%
with $\det \mathcal{K}_{sl_{3|2}}^{VI}=1.$
\end{itemize}

\ \ \ \ \newline
In what follows, we will focuss on the distinguished $sl(3|2)$\ chain model
with \emph{DSD}$_{I}$ as an example.\ This super chain is also imagined as a
representative of the distinguished class of the $sl(m|n)$\ family $(m>n)$.
The corresponding Distinguished Dynkin Super-Diagram (for short \emph{DDSD})
has one fermionic simple root given by the green node in \emph{DSD}$_{I}$\
of eq(\ref{TAB1}). The \emph{DDSD} is described by the generalised Cartan
matrix\textrm{\ }%
\begin{equation}
\mathcal{K}_{sl_{3|2}}^{{\small (I)}}=\mathbf{\alpha }_{A}^{{\small (I)}}%
\mathbf{.\alpha }_{B}^{{\small (I)}}
\end{equation}%
with graded simple roots realised as in eq(\ref{alp}), namely $\alpha
_{A}=\epsilon _{A}-\epsilon _{A+1}$\ with 
\begin{equation}
\epsilon _{A}=\left( \varepsilon _{a},\delta _{i}\right) ,\qquad \varepsilon
_{a}=\left( \varepsilon _{1},\varepsilon _{2},\varepsilon _{3}\right)
,\qquad \epsilon _{4}=\delta _{1},\quad \epsilon _{5}=\delta _{2}
\end{equation}%
In matrix representation, the $K_{AB}^{sl_{3|2}}$\ reads like in eq(\ref{I})
with inverse as in (\ref{car}).

\subsection{Distinguished superspin chain}

Here, we investigate the quantum properties of the distinguished $sl(3|2)$
super chain by comparing it with the bosonic $sl(5)$ chain. The
distinguished $sl(3|2)$ and the $sl(5)$ share some basic features that we
will use to build the algebraic geometry interpretation. An example of the
shared properties, besides rank and dimension, are given by the bosonic $%
sl(3)\oplus sl(2)$ which appears as \textrm{the} maximal Lie subalgebra.

\subsubsection{Bosonic $sl(5)$ spin chain}

We begin by recalling that for the bosonic- like $sl(5)$ spin chain, the
quantum states of the atoms are generated by five weight vectors $\mathbf{%
\tilde{\varepsilon}}_{\bar{a}}$ labeled by $\bar{a}=1,2,3,4,5$. This is the
same number as for the five $\mathbf{\epsilon }_{A}$'s \ regarding the super
atoms of the distinguished $sl(3|2)$ superspin chain. The $sl(5)$ atomic
weight charges generate a 5D HW representation of $sl(5)$ with basis vectors
as 
\begin{equation}
\left\vert \mathbf{\tilde{\varepsilon}}_{1}\right\rangle ,\quad \left\vert 
\mathbf{\tilde{\varepsilon}}_{2}\right\rangle ,\quad \left\vert \mathbf{%
\tilde{\varepsilon}}_{3}\right\rangle ,\quad \left\vert \mathbf{\tilde{%
\varepsilon}}_{4}\right\rangle ,\quad \left\vert \mathbf{\tilde{\varepsilon}}%
_{5}\right\rangle  \label{ee}
\end{equation}%
By using the four fundamental coweights $\mathbf{\tilde{\omega}}_{\bar{a}}$
of the bosonic $sl(5),$ the five weight vectors $\mathbf{\tilde{\varepsilon}}%
_{\bar{a}}$ are realised as follows%
\begin{equation}
\begin{tabular}{lll}
$\mathbf{\tilde{\varepsilon}}_{1}$ & $=$ & $\mathbf{\tilde{\omega}}_{1}$ \\ 
$\mathbf{\tilde{\varepsilon}}_{2}$ & $=$ & $\mathbf{\tilde{\omega}}_{1}-%
\mathbf{\tilde{\alpha}}_{1}$ \\ 
$\mathbf{\tilde{\varepsilon}}_{3}$ & $=$ & $\mathbf{\tilde{\omega}}_{1}-%
\mathbf{\tilde{\alpha}}_{1}-\mathbf{\tilde{\alpha}}_{2}$ \\ 
$\mathbf{\tilde{\varepsilon}}_{4}$ & $=$ & $\mathbf{\tilde{\omega}}_{1}-%
\mathbf{\tilde{\alpha}}_{1}-\mathbf{\tilde{\alpha}}_{2}-\mathbf{\tilde{\alpha%
}}_{3}$ \\ 
$\mathbf{\tilde{\varepsilon}}_{5}$ & $=$ & $\mathbf{\tilde{\omega}}_{1}-%
\mathbf{\tilde{\alpha}}_{1}-\mathbf{\tilde{\alpha}}_{2}-\mathbf{\tilde{\alpha%
}}_{3}-\mathbf{\tilde{\alpha}}_{4}$%
\end{tabular}%
\end{equation}%
obeying\ the $sl\left( 5\right) $ traceless condition%
\begin{equation}
\mathbf{\tilde{\varepsilon}}_{1}+\mathbf{\tilde{\varepsilon}}_{2}+\mathbf{%
\tilde{\varepsilon}}_{3}+\mathbf{\tilde{\varepsilon}}_{4}+\mathbf{\tilde{%
\varepsilon}}_{5}=0
\end{equation}%
requiring in turns the following relationship with the simple roots $\mathbf{%
\tilde{\alpha}}_{\bar{a}}$ of $sl(5)$%
\begin{equation}
5\mathbf{\tilde{\omega}}_{1}=4\mathbf{\tilde{\alpha}}_{1}+3\mathbf{\tilde{%
\alpha}}_{2}+2\mathbf{\tilde{\alpha}}_{3}+\mathbf{\tilde{\alpha}}_{4}
\end{equation}%
Notice that the highest weight vector $\mathbf{\tilde{\varepsilon}}_{1}$ is
precisely the fundamental coweight $\mathbf{\tilde{\omega}}_{1}$ which is
dual to the simple root $\mathbf{\tilde{\alpha}}_{1}=\mathbf{\tilde{%
\varepsilon}}_{1}-\mathbf{\tilde{\varepsilon}}_{2}$; that is%
\begin{equation}
\mathbf{\tilde{\varepsilon}}_{1}=\mathbf{\tilde{\omega}}_{1}\qquad ,\qquad 
\mathbf{\tilde{\omega}}_{1}.\mathbf{\tilde{\alpha}}_{1}=1  \label{ef}
\end{equation}%
In this context, recall that the Lie algebra $sl(5)$ has four simple roots $%
\mathbf{\tilde{\alpha}}_{\bar{a}}=\mathbf{\tilde{\varepsilon}}_{\bar{a}}-%
\mathbf{\tilde{\varepsilon}}_{\bar{a}+1}$ with intersection matrix $\mathcal{%
A}_{\bar{a}\bar{b}}^{sl_{5}}=\mathbf{\tilde{\alpha}}_{\bar{a}}.\mathbf{%
\tilde{\alpha}}_{\bar{b}}$ given by%
\begin{equation}
\mathcal{A}^{sl_{5}}=\left( 
\begin{array}{cccc}
2 & -1 & 0 & 0 \\ 
-1 & 2 & -1 & 0 \\ 
0 & -1 & 2 & -1 \\ 
0 & 0 & -1 & 2%
\end{array}%
\right)  \label{a5}
\end{equation}%
Its determinant is equal to $5$ and its inverse is given by 
\begin{equation}
\mathcal{A}_{sl_{5}}^{-1}=\frac{1}{5}\left( 
\begin{array}{cccc}
4 & 3 & 2 & 1 \\ 
3 & 6 & 4 & 2 \\ 
2 & 4 & 6 & 3 \\ 
1 & 2 & 3 & 4%
\end{array}%
\right)
\end{equation}%
In what follows, we will use properties of the $sl(5)$ spin chain captured 
\textrm{by} (\ref{ee}) to \textrm{unveil characteristics of} the
distinguished $sl(3|2)$ super chain. Notice that $sl(5)$ and the\
distinguished $sl(3|2)$ have the same rank 4 and the same dimension 24.
Moreover, they\ both belong to the finite dimensional special linear
\textquotedblleft $sl$\textquotedblright\ family manifested by a linear
Dynkin diagram.

\subsubsection{Super atoms in the distinguished $sl(3|2)$ chain}

Using the algebraic properties of the bosonic $sl\left( 5\right) $ described
above, we construct below those homologous properties in the distinguished $%
sl(3|2)$ which are satisfied by the super weights $\mathbf{\epsilon }_{A}$.
This approach gives a short way towards the interpretation of exotic
properties regarding the super atoms and the distinguished $sl(3|2)$
superspin chain. \textrm{By the word exotic, we mean the special values of
the elements of the Cartan matrices} $\mathcal{A}^{sl_{5}}$ and $\mathcal{K}%
^{sl_{3|2}}$%
\begin{equation}
\mathcal{A}^{sl_{5}}=\left( 
\begin{array}{cccc}
2 & -1 & 0 & 0 \\ 
-1 & 2 & -1 & 0 \\ 
0 & -1 & 2 & -1 \\ 
0 & 0 & -1 & 2%
\end{array}%
\right) ,\qquad \mathcal{K}^{sl_{3|2}}=\left( 
\begin{array}{cccc}
2 & -1 & 0 & 0 \\ 
-1 & 2 & -1 & 0 \\ 
0 & -1 & 0 & +1 \\ 
0 & 0 & +1 & -2%
\end{array}%
\right)
\end{equation}%
Now, we think about the invertible $\mathcal{A}^{sl_{5}}$ and the invertible 
$\mathcal{K}^{sl_{3|2}}$ as two cousin matrices that offer a bridge between
the $sl(5)$ spin chain and the $sl(3|2)$ super chain. This link implies that
the quantum states (\ref{e5}) namely $\left\vert \mathbf{\epsilon }%
_{1}\right\rangle ,\left\vert \mathbf{\epsilon }_{2}\right\rangle
,\left\vert \mathbf{\epsilon }_{3}\right\rangle ,\left\vert \mathbf{\epsilon 
}_{4}\right\rangle ,\left\vert \mathbf{\epsilon }_{5}\right\rangle $ can be
put in correspondence with the $\left\vert \mathbf{\tilde{\varepsilon}}%
_{1}\right\rangle ,\left\vert \mathbf{\tilde{\varepsilon}}_{2}\right\rangle
,\left\vert \mathbf{\tilde{\varepsilon}}_{3}\right\rangle ,\left\vert 
\mathbf{\tilde{\varepsilon}}_{4}\right\rangle ,\left\vert \mathbf{\tilde{%
\varepsilon}}_{5}\right\rangle $ satisfying (\ref{ee}-\ref{ef}); i.e:%
\begin{equation}
\mathbf{\tilde{\varepsilon}}_{\bar{a}}\quad \leftrightarrow \quad \mathbf{%
\epsilon }_{A}
\end{equation}%
As the $\mathbf{\tilde{\varepsilon}}_{\bar{a}}$ of the $sl(5)$ are nicely
related to the simple roots $\mathbf{\tilde{\alpha}}_{\bar{a}}$ and the
simple coweights $\mathbf{\tilde{\omega}}_{\bar{a}}$, we investigate below
the extension of this feature to the $\mathbf{\epsilon }_{A}$'s of $sl(3|2)$.

\ \ \ 

\textbf{A) root/weight duality in} $sl(3|2)$\newline
As for $sl(5)$, the four fundamental coweights $\mathbf{\omega }_{1},$ $%
\mathbf{\omega }_{2},$ $\mathbf{\omega }_{3},$ $\mathbf{\omega }_{4}$ of the
distinguished $sl(3|2)$ are dual to the four simple roots $\mathbf{\alpha }%
_{1},$ $\mathbf{\alpha }_{2},$ $\mathbf{\alpha }_{3},$ $\mathbf{\alpha }_{4}$%
. These two basic quantities are related by the super Cartan matrix like $%
\mathbf{\alpha }_{A}=\mathcal{K}_{AB}^{sl_{3|2}}\mathbf{\omega }_{B}.$ By
substituting, we have 
\begin{equation}
\begin{tabular}{lll}
$\mathbf{\alpha }_{1}$ & $=$ & $2\mathbf{\omega }_{1}-\mathbf{\omega }_{2}$
\\ 
$\mathbf{\alpha }_{2}$ & $=$ & $2\mathbf{\omega }_{2}-\mathbf{\omega }_{2}-%
\mathbf{\omega }_{3}$ \\ 
$\mathbf{\alpha }_{3}$ & $=$ & $\mathbf{\omega }_{4}-\mathbf{\omega }_{2}$
\\ 
$\mathbf{\alpha }_{4}$ & $=$ & $\mathbf{\omega }_{3}-2\mathbf{\omega }_{4}$%
\end{tabular}%
\end{equation}%
Moreover, using (\ref{K}) with the inverse $\mathcal{K}_{sl_{3|2}}^{-1}$ as
in (\ref{car}), we also have 
\begin{equation}
\begin{tabular}{lll}
$\mathbf{\omega }_{1}$ & $=$ & $-\mathbf{\alpha }_{2}-2\mathbf{\alpha }_{3}-%
\mathbf{\alpha }_{4}$ \\ 
$\mathbf{\omega }_{2}$ & $=$ & $-\mathbf{\alpha }_{1}-2\mathbf{\alpha }_{2}-4%
\mathbf{\alpha }_{3}-2\mathbf{\alpha }_{4}$ \\ 
$\mathbf{\omega }_{3}$ & $=$ & $-2\mathbf{\alpha }_{1}-4\mathbf{\alpha }%
_{2}-6\mathbf{\alpha }_{3}-3\mathbf{\alpha }_{4}$ \\ 
$\mathbf{\omega }_{4}$ & $=$ & $-\mathbf{\alpha }_{1}-2\mathbf{\alpha }_{2}-3%
\mathbf{\alpha }_{3}-2\mathbf{\alpha }_{4}$%
\end{tabular}
\label{321}
\end{equation}

\ \ \ 

\textbf{B) simple roots and coweights in terms of}\emph{\ }$\mathbf{\epsilon 
}_{A}$\emph{'s}\newline
The expressions of the simple roots $\mathbf{\alpha }_{A}$ in terms of the
basic weights $\mathbf{\epsilon }_{A}$ are given by $\mathbf{\alpha }_{A}=%
\mathbf{\epsilon }_{A}-\mathbf{\epsilon }_{A+1}.$ Putting these relations
into (\ref{321}), and using the super traceless condition, 
\begin{equation}
\mathbf{\epsilon }_{1}+\mathbf{\epsilon }_{2}+\mathbf{\epsilon }_{3}-\mathbf{%
\epsilon }_{4}-\mathbf{\epsilon }_{5}=0
\end{equation}%
we end up with the following expressions%
\begin{equation}
\begin{tabular}{lll}
$\mathbf{\omega }_{1}$ & $=$ & $\mathbf{\epsilon }_{1}$ \\ 
$\mathbf{\omega }_{2}$ & $=$ & $\mathbf{\epsilon }_{1}+\mathbf{\epsilon }%
_{2} $%
\end{tabular}%
,\qquad 
\begin{tabular}{lll}
$\mathbf{\omega }_{3}$ & $=$ & $\mathbf{\epsilon }_{4}+\mathbf{\epsilon }%
_{5} $ \\ 
$\mathbf{\omega }_{4}$ & $=$ & $\mathbf{\epsilon }_{5}$%
\end{tabular}
\label{eps}
\end{equation}%
Below, we take advantage of this formal similarity between $\mathcal{K}%
^{sl_{3|2}}$ and $\mathcal{A}^{sl_{5}}$\ to pave the way for the
interpretation of the intersection $\mathbf{\alpha }_{A}\mathbf{.\alpha }%
_{B}=\mathcal{K}_{AB}^{sl_{3|2}}$ in terms of intersecting 2-cycles as a
shortcut\ \textrm{to} brane realisation using algebraic geometry and methods
of singularity theory \cite{1D,2D,3D,4D,5D}.

\section{Distinguished DSD and super geometry}

In this section, we develop the study of the bridge between $sl(5)$ and $%
sl(3|2)$ by using their Cartan matrices $\mathcal{A}^{sl_{5}}$ and $\mathcal{%
K}^{sl_{3|2}}.$ Then, we use this bridging to construct the 2-cycle homology
associated with the distinguished $sl(3|2)$ singularity and propose a
complex super geometry to use later for embedding the superspin chain into
type II strings and M- theory compactified on {\large Y}$_{4}$, as
represented by the Figure \textbf{\ref{typeII}}.

\subsection{Colored Dynkin diagrams}

\begin{figure}[tbph]
\begin{center}
\includegraphics[width=12cm]{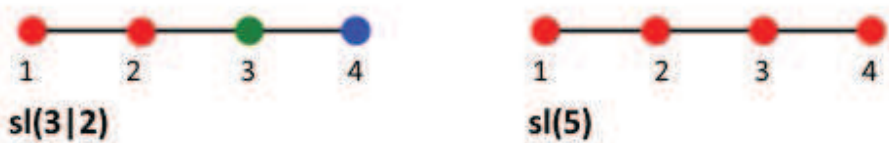}
\end{center}
\par
\vspace{-0.5cm}
\caption{On the left, the Dynkin super diagram of the distinguished sl(3%
\TEXTsymbol{\vert}2). On the right, the Dynkin diagram of sl(5).}
\label{DSD}
\end{figure}

The Dynkin diagrams associated to the Cartan matrix $\mathcal{A}^{sl_{5}}$
and its super homologue $\mathcal{K}^{sl_{3|2}}$ look very close to each
other; a property that deserves to be examined more closely. The $\mathcal{K}%
^{sl_{3|2}}$ and $\mathcal{A}^{sl_{5}}$ are graphically given by the
pictures \textrm{of }Figure \textbf{\ref{DSD}}. 
They have four nodes with the same \emph{linear} shape; but with different
colors. The bosonic $\mathcal{A}^{sl_{5}}$ has a unified color, say the red
color as in the right picture of the Figure \textbf{\ref{DSD}}. The $%
\mathcal{K}^{sl_{3|2}}$ involves three different colors as in Table \textbf{%
\ref{TAB1}} and the left picture of Figure \textbf{\ref{DSD}}. The purpose
for using these colors is to depict their differences while emphasizing
their similarities.

\subsubsection{Bridging $\mathcal{A}^{sl_{5}}$ towards $\mathcal{K}%
^{sl_{3|2}}$}

Compared to the usual Cartan matrix of the bosonic Lie algebra $sl\left(
5\right) $ given by%
\begin{equation}
\mathcal{A}^{sl_{5}}=\mathbf{\tilde{\alpha}}_{a}\mathbf{.\tilde{\alpha}}_{b}
\end{equation}%
with $\tilde{\alpha}_{a}^{2}=+2$ and $\tilde{\alpha}_{a}.\tilde{\alpha}%
_{a+1}<0$, the super $\mathcal{K}^{sl_{3|2}}$ has exotic values; in
particular the three following 
\begin{equation}
\left( i\right) :\mathbf{\alpha }_{3}^{2}=0\qquad ,\qquad \left( ii\right) :%
\mathbf{\alpha }_{4}^{2}=-2\qquad ,\qquad \left( iii\right) :\mathbf{\alpha }%
_{3}\mathbf{.\alpha }_{4}=+1
\end{equation}%
This feature makes the geometric engineering method of the super chain in
type II strings on local manifolds \textrm{somehow} special. In this regard,
it is interesting to recall that in the singularity theory\textrm{\footnote{%
\ To fix the ideas; see the list of ADE surfaces collected in the Table of
eq(\ref{418}).}} of the complex ALE surfaces $\mathcal{\tilde{S}}$ with $%
sl\left( 5\right) $ geometry, bosonic Cartan matrices like $\mathcal{A}%
_{ab}^{sl_{5}}$ of (\ref{car}) have an interpretation in terms of
intersecting 2-cycles $\mathfrak{\tilde{C}}_{a}$ with intersection matrix 
\cite{1D}-\cite{5D} 
\begin{equation}
\mathcal{I}_{ab}^{sl_{5}}=\mathfrak{\tilde{C}}_{a}\mathfrak{.\tilde{C}}_{b}
\label{322}
\end{equation}%
precisely given by $\mathcal{I}_{ab}^{sl_{5}}=-\mathcal{A}_{ab}^{sl_{5}}$;
that is 
\begin{equation}
\mathcal{I}_{ab}^{sl_{5}}=\left( 
\begin{array}{cccc}
-2 & 1 & 0 & 0 \\ 
1 & -2 & 1 & 0 \\ 
0 & 1 & -2 & 1 \\ 
0 & 0 & 1 & -2%
\end{array}%
\right)  \label{sl5}
\end{equation}%
From this intersection matrix of 2-cycles, we learn the following features%
\begin{equation}
\mathfrak{\tilde{C}}_{a}^{2}=-2\qquad ,\qquad \mathfrak{\tilde{C}}_{a}.%
\mathfrak{\tilde{C}}_{a+1}=1  \label{2c}
\end{equation}%
indicating that the $\mathfrak{\tilde{C}}_{a}$'s are complex projective
lines $\mathbb{CP}_{a}^{1}$ (2-spheres) intersecting transversally according
to the Dynkin diagram of $sl(5)$ \cite{5A}. So the four simple roots $%
\mathbf{\tilde{\alpha}}_{a}=\mathbf{\tilde{\varepsilon}}_{a}-\mathbf{\tilde{%
\varepsilon}}_{a+1}$ ($a=1,2,3,4$) of the Lie algebra $sl\left( 5\right) $
are put in correspondence with four 2-cycles $\mathfrak{\tilde{C}}_{a}$ in
the resolved ALE surface $\mathcal{\tilde{S}}^{sl_{5}}$. Later on, we will
use the Algebra/Homology (A/H) correspondence given by 
\begin{equation}
\begin{tabular}{c|c}
roots \ \ \  & \ 2-cycles \\ \hline
$\mathbf{\tilde{\alpha}}_{a}$ & $\mathfrak{\tilde{C}}_{a}$ \\ \hline
$\mathbf{\tilde{\varepsilon}}_{a}$ & $\tilde{E}_{a}$ \\ \hline
\end{tabular}
\label{cr}
\end{equation}%
and%
\begin{equation}
\begin{tabular}{c|c}
{\small Lie algebra }${\small sl(5)}$ & {\small homology of ALE surfaces} $%
\mathcal{\tilde{S}}^{sl_{5}}$ \\ \hline
$\mathbf{\tilde{\alpha}}_{a}^{2}=2$ & $\mathfrak{\tilde{C}}_{a}^{2}=-2$ \\ 
\hline
$\mathbf{\tilde{\varepsilon}}_{a}^{2}=1$ & $\tilde{E}_{a}^{2}=-1$ \\ \hline
\end{tabular}
\label{rc}
\end{equation}%
\begin{equation*}
\end{equation*}%
to deal with the geometrisation of the symmetries and\textrm{\ therefore }%
build brane realisations of the superspin chain. As our super chain has $%
U\left( N_{f}\right) \times SL\left( 3|2\right) $ invariance, we also apply
the correspondence (\ref{cr}-\ref{rc}) to the two symmetry factors while
taking into account some specificities as described below.

\ \ \ 

\textbf{A)} \textbf{flavor symmetry} $U\left( N_{f}\right) =U\left( 1\right)
\times SU\left( N_{f}\right) $ \newline
As $su\left( N_{f}\right) $ is a bosonic Lie algebra contained into $%
sl\left( N_{f}\right) $, the correspondence (\ref{cr}-\ref{rc}) given for $%
sl\left( 5\right) $\textrm{\ }also applies to $su\left( N_{f}\right) $. In
this case, we have:

\begin{itemize}
\item $N_{f}-1$ intersecting 2-spheres $\mathfrak{\tilde{C}}_{a},$ labeled
by $a=1,...,N_{f}-1$ with self intersection of these \textrm{cycles} 
\begin{equation}
\mathfrak{\tilde{C}}_{a}^{2}=-2
\end{equation}

\item Being associated with a flavor symmetry and not a gauge invariance,
these 2-cycles $\mathfrak{\tilde{C}}_{a}$ must have large volumes. The
solution of this constraint will be given later.
\end{itemize}

\ \ \ 

\textbf{B)}\emph{\ }\textbf{graded}\emph{\ }$SL(3|2)$\emph{\ }\textbf{%
invariance} \newline
As the singular surfaces with $SL(m|n)$ gauge symmetries have not been
explored in the stringy literature, we use below the relationship between
the Dynkin diagrams of $SL(5)$ and $SL(3|2)$ to propose the following Super
Algebra/Homology (Super A/H) correspondence%
\begin{equation}
\begin{tabular}{c|c}
{\small Lie superalgebra }${\small sl(3|2)}$ & {\small graded surfaces }$%
\mathcal{S}^{sl_{{\small 3|2}}}$ \\ \hline
{\small graded roots} $\mathbf{\alpha }_{A}$ & {\small graded cycles} $%
\mathfrak{C}_{A}$ \\ \hline
{\small graded weights} $\mathbf{\epsilon }_{A}$ & {\small graded divisors} $%
E_{A}$ \\ \hline
\end{tabular}
\label{pcr}
\end{equation}%
extending the A/H bosonic-like one 
\begin{equation}
\begin{tabular}{c|c}
{\small Lie algebra} ${\small sl(5)}$ & {\small homology of surfaces} $%
\mathcal{\tilde{S}}^{sl_{5}}$ \\ \hline
{\small roots} $\mathbf{\tilde{\alpha}}_{\bar{a}}$ & {\small cycles} $%
\mathfrak{\tilde{C}}_{\bar{a}}$ \\ \hline
{\small weights} $\mathbf{\tilde{\varepsilon}}_{\bar{a}}$ & {\small divisors}
$\tilde{E}_{\bar{a}}$ \\ \hline
\end{tabular}
\label{crp}
\end{equation}

\subsubsection{More on roots and weights of $sl(3|2)$}

The distinguished $sl(3|2)$ Lie superalgebra is a $\mathbb{Z}_{2}$-graded
algebra decomposing like $sl(3|2)_{\bar{0}}\oplus sl(3|2)_{\bar{1}}$ with
two sectors:

\begin{enumerate}
\item an even sector given by the Lie algebra 
\begin{equation}
sl(3|2)_{\bar{0}}=sl(3)\oplus sl\left( 1\right) \oplus sl(2)
\end{equation}

\item an odd sector given by the vector space $sl(3|2)_{\bar{1}}$; it is an $%
sl(3|2)_{\bar{0}}$- module with 12 dimensions splitting as 
\begin{equation}
\left( \mathbf{3,\bar{2}}\right) \oplus \left( \mathbf{\bar{3},2}\right)
\end{equation}
with the labels $\mathbf{3}$ and $\mathbf{2}$ referring to the fundamental
representations of $sl(3)$ and $sl(2).$
\end{enumerate}

Notice that $sl(3|2)$ is the super-traceless Lie sub-superalgebra of $%
gl(3|2).$ It is characterised by the following algebraic properties 
\begin{equation}
\begin{tabular}{|c|c|c|c|c|}
\hline
$sl(3|2)$ & rank & dim & \# roots & simple roots \\ \hline
$sl(3|2)_{\bar{0}}$ & 4 & 12 & 8 & $\mathbf{\alpha }_{1}\mathbf{,\alpha }_{2}%
\mathbf{,\alpha }_{4}$ \\ \hline
$sl(3|2)_{\bar{1}}$ & 0 & 12 & 12 & $\mathbf{\alpha }_{3}$ \\ \hline
\end{tabular}
\label{32}
\end{equation}%
\begin{equation*}
\end{equation*}%
The simple roots $\mathbf{\alpha }_{A}$ generating the distinguished root
system $\Phi _{sl_{3|2}}$ play an important role in our construction. They
are realised as $\mathbf{\alpha }_{A}=\mathbf{\epsilon }_{A}\mathbf{%
-\epsilon }_{A+1}$ where the five unit weight vectors $\left( \mathbf{%
\epsilon }_{1}\mathbf{,\epsilon }_{2}\mathbf{,\epsilon }_{3}\mathbf{%
,\epsilon }_{4}\mathbf{,\epsilon }_{5}\right) $ are\ distinguishably ordered
like 
\begin{equation}
\mathbf{\epsilon }_{A}=\left( \mathbf{\varepsilon }_{1}\mathbf{,\varepsilon }%
_{2}\mathbf{,\varepsilon }_{3}\mathbf{,\delta }_{1}\mathbf{,\delta }%
_{2}\right)
\end{equation}%
with bosonic $\mathbf{\varepsilon }_{a}\mathbf{.\varepsilon }_{b}=\delta
_{ab}$ and fermionic $\mathbf{\delta }_{i}\mathbf{.\delta }_{j}=-\delta
_{ij} $ normalisation. These weights are graphically represented as depicted
by the picture of Table \textbf{\ref{T2}.} 
\begin{table}[h]
\begin{center}
$%
\begin{tabular}{ccccccccccc}
&  & $\varepsilon _{1}$ & \ \  & $\varepsilon _{2}$ & \ \ \  & $\varepsilon
_{3}$ & \ \ \  & $\delta _{1}$ & \ \ \  & $\delta _{2}$ \\ 
$\left( a\right) $ &  & $\ \textcolor{red}{|}$ &  & $\textcolor{red}{|}$ & 
& $\textcolor{red}{|}$ &  & $\textcolor{blue}{|}$ &  & $\textcolor{blue}{|}$
\\ 
&  &  &  &  &  &  &  &  &  &  \\ 
&  &  & $\alpha _{\text{1}}$ &  & $\alpha _{\text{2}}$ &  & $\alpha _{\text{3%
}}$ &  & $\alpha _{\text{4}}$ &  \\ 
$\left( b\right) $ &  & $\ \textcolor{red}{|}$ & $\textcolor{red}{\bigcirc}$
& $\textcolor{red}{|}$ & $\textcolor{red}{\bigcirc}$ & $\textcolor{red}{|}$
& $\textcolor{green}{\bigcirc}$ & $\textcolor{blue}{|}$ & $%
\textcolor{blue}{\bigcirc}$ & $\textcolor{blue}{|}$%
\end{tabular}%
$%
\end{center}
\caption{Top: the graphical representation of the weight vector basis of $%
sl\left( 3|2\right) $. The 5 unit weights are depiccted by vertical lines;
three even in red color and two odd in blue. Bottom: the four simple roots $%
\protect\alpha _{A}=\protect\epsilon _{A}-\protect\epsilon _{A+1}$ with the
fermionic one in blue.}
\label{T2}
\end{table}
To make the correspondence with the bosonic $sl(5)$ more transparent, we
give below the homologue of eq(\ref{32}). It reads as follows 
\begin{equation}
\begin{tabular}{c||c|c|c|c}
{\small Lie algebra} & rank & dim & \# roots & simple roots \\ \hline
$sl(5)$ & 4 & 24 & 20 & $\mathbf{\tilde{\alpha}}_{1}\mathbf{,\tilde{\alpha}}%
_{2}\mathbf{,\tilde{\alpha}}_{3}\mathbf{,\tilde{\alpha}}_{4}$%
\end{tabular}%
\end{equation}%
with $\mathbf{\tilde{\alpha}}_{a}=\mathbf{\tilde{\varepsilon}}_{a}\mathbf{-%
\tilde{\varepsilon}}_{a+1}$ and $\mathbf{\tilde{\varepsilon}}_{a}\mathbf{.%
\tilde{\varepsilon}}_{b}=\delta _{ab}$. By comparing the two weight bases $%
\left\{ \mathbf{\epsilon }_{A}\right\} $ and $\left\{ \mathbf{\tilde{%
\varepsilon}}_{a}\right\} ,$ we learn that the bridging from the $sl(5)$ to
the distinguished $sl(3|2)$ is obtained either $\left( i\right) $ by
promoting the Euclidian metric $\mathbf{\tilde{\varepsilon}}_{a}\mathbf{.%
\tilde{\varepsilon}}_{b}=\delta _{ab}$ to the pseudo-Euclidian $\mathbf{%
\epsilon }_{A}.\mathbf{\epsilon }_{B}=g_{AB}$ given by $\left( -\right)
^{A}\delta _{AB}$ and reading like 
\begin{equation}
g_{AB}=\left( 
\begin{array}{ccccc}
1 &  &  &  &  \\ 
& 1 &  &  &  \\ 
&  & 1 &  &  \\ 
&  &  & -1 &  \\ 
&  &  &  & -1%
\end{array}%
\right)  \label{4g}
\end{equation}%
or $\left( ii\right) $ by replacing the real\ weight vectors $\mathbf{\tilde{%
\varepsilon}}_{3+i}$ by pure imaginary weights as follows%
\begin{equation}
\mathbf{\varepsilon }_{a}=\left. \tilde{\varepsilon}_{a}\right\vert
_{a=1,2,3}\qquad ,\qquad \mathbf{\delta }_{1}=i\mathbf{\tilde{\varepsilon}}%
_{4}\qquad ,\qquad \mathbf{\delta }_{2}=i\mathbf{\tilde{\varepsilon}}_{5}
\end{equation}%
Below, we use the metric (\ref{4g}).

\subsection{From $\mathcal{K}_{AB}^{sl_{3|2}}$ to homology of $SL(3|2)$
singularity}

First, we give the typical structure of the complex surfaces with isolated
singularities classified by Dynkin diagrams of finite dimensional ADE Lie
algebras. Then, we present a proposal for the extension of $sl(n)$
geometries to $sl\left( m|n\right) $ while focussing on $sl\left( 3|2\right) 
$.

\subsubsection{Du Val singularities}

We begin by recalling the list of Du Val singularities regarding complex
surfaces $V/\Gamma $ with orbifold symmetry given by finite groups $\Gamma $
contained in $GL\left( 2\right) .$ The complex surfaces with Du Val
singularities can be defined by local equations \cite{1D,2D,5A,DV}, 
\begin{equation}
f\left( x,y,z\right) =0
\end{equation}%
embedded in $\mathbb{C}^{3}$ with local coordinates $\left( x,y,z\right) $.
The resolution of the Du Val singularities are classified by the Dynkin
diagrams of the simply laced ADE Lie algebras as listed in Table \textbf{\ref%
{418}}. 
\begin{table}[h]
\centering \renewcommand{\arraystretch}{1.2} $%
\begin{tabular}{||c||c||c||c||c||}
\hline\hline
Geometry & $f\left( x,y,z\right) $ & $\Gamma $ & Cartan & Resolution graph
and 2-cycles \\ \hline\hline
$A_{n-1}$ & $x^{2}+y^{2}+z^{n}$ & $Z_{n}$ & $A[A_{n-1}]$ & %
\includegraphics[width=3cm]{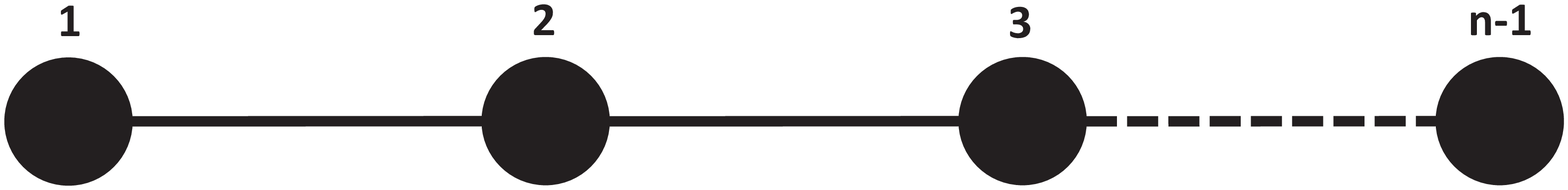} \\ \hline\hline
$D_{n}$ & $x^{2}+y^{2}z+z^{n-1}$ & BD$_{4n-8}$ & $A[D_{n}]$ & %
\includegraphics[width=3cm]{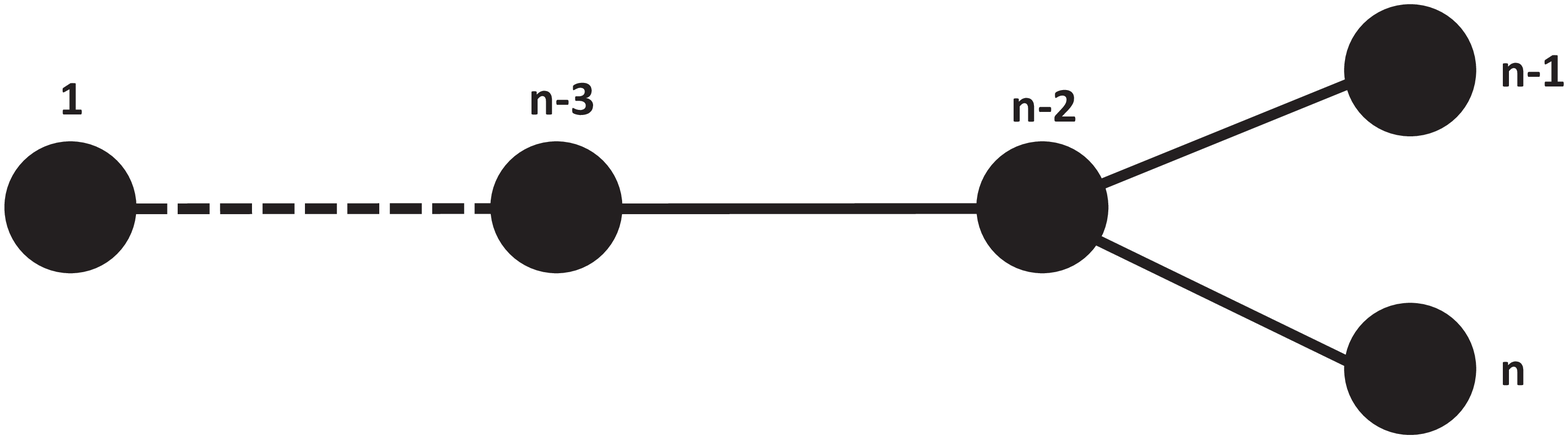} \\ \hline\hline
$E_{6}$ & $x^{2}+y^{3}+z^{4}$ & BT$_{24}$ & $A[E_{6}]$ & %
\includegraphics[width=3cm]{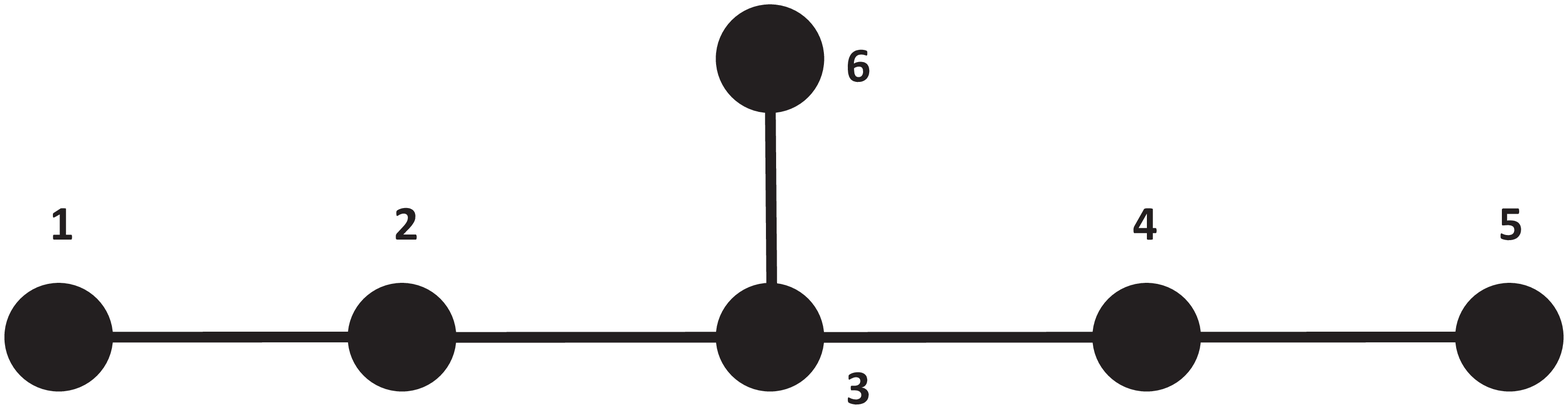} \\ \hline\hline
$E_{7}$ & $x^{2}+y^{3}+yz^{3}$ & BO & $A[E_{7}]$ & %
\includegraphics[width=3cm]{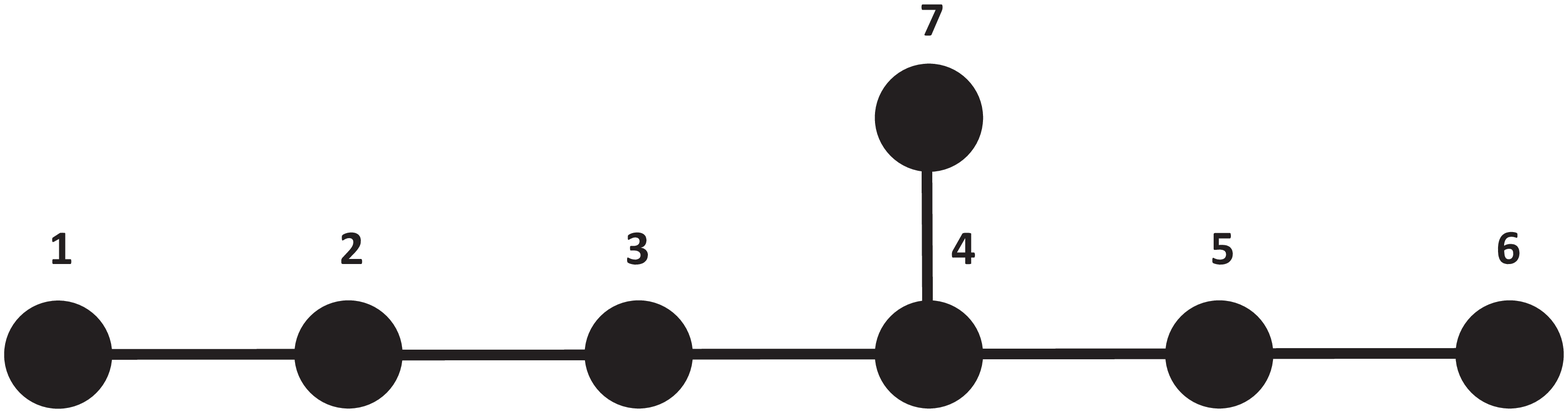} \\ \hline\hline
$E_{8}$ & $x^{2}+y^{3}+z^{5}$ & BIcosa & $A[E_{8}]$ & %
\includegraphics[width=3cm]{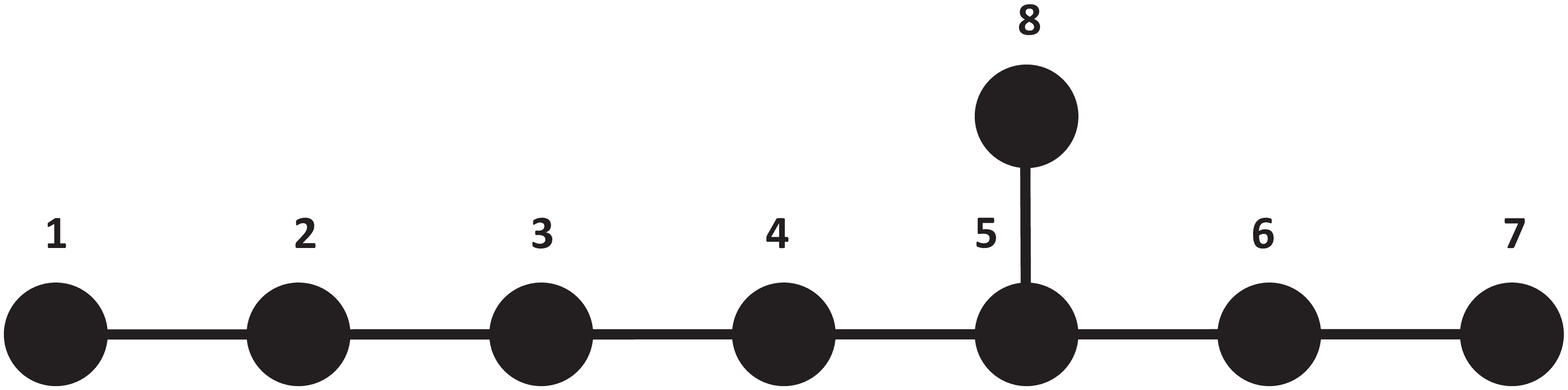} \\ \hline\hline
\end{tabular}%
$%
\caption{The defining equations of the complex surfaces with ADE geometries.
The 2-cycle homology of the surface is given by the ADE Dynkin diagrams with
roots describing the 2-cycles.}
\label{418}
\end{table}
\begin{equation*}
\end{equation*}%
In the Du Val list, the BD$_{4n-8}$ and the BT$_{24}$ are respectively the
Binary- Dihedral and Binary-Tetrahedral groups \textrm{\cite{HEPA,HEPB}}.
The BO is the \textrm{Binary-Octahedral} group and BIcosa is the Binary
Icosahedral. For the $A_{1}$ example, we have 
\begin{equation}
x^{2}+y^{2}+z^{2}=0
\end{equation}%
describing a vanishing sphere at $\left( x,y,z\right) =\left( 0,0,0\right) .$
By setting $u=i\left( x+iy\right) ,v=i\left( x-iy\right) $, we can express
the above $SU\left( 2\right) $ singularity like $uv=z^{2}$ corresponding to
the leading element of the $SU\left( n\right) $ family%
\begin{equation}
uv=z^{n}
\end{equation}%
to be used later on. The homology of compact 2-cycles of the resolved
singularities in (\ref{418}) involves intersecting 2-spheres according to
the ADE Dynkin diagrams.\newline
In the language of weights $\mathbf{\tilde{\varepsilon}}_{a}$ and roots $%
\mathbf{\tilde{\alpha}}_{a}=\mathbf{\tilde{\varepsilon}}_{a}-\mathbf{\tilde{%
\varepsilon}}_{a+1}$, the resolution of the $A_{N-1}$ singularity is
captured by the relation $\mathbf{\tilde{\varepsilon}}_{1}-\mathbf{\tilde{%
\varepsilon}}_{N}=\sum \mathbf{\tilde{\alpha}}_{a}.$ For the example of $%
sl\left( 5\right) ,$ this reads as follows%
\begin{equation}
\mathbf{\tilde{\varepsilon}}_{1}-\mathbf{\tilde{\varepsilon}}_{5}=\mathbf{%
\tilde{\alpha}}_{1}+\mathbf{\tilde{\alpha}}_{2}+\mathbf{\tilde{\alpha}}_{3}+%
\mathbf{\tilde{\alpha}}_{4}  \label{res}
\end{equation}%
By using A/H correspondence (\ref{crp}), we also have the homological
equation \textrm{\cite{5A}},%
\begin{equation}
\tilde{E}_{1}-\tilde{E}_{5}=\mathfrak{\tilde{C}}_{1}+\mathfrak{\tilde{C}}%
_{2}+\mathfrak{\tilde{C}}_{3}+\mathfrak{\tilde{C}}_{4}
\end{equation}

\subsubsection{Surface with $sl\left( 3|2\right) $ geometry}

To our knowledge, the extension of the classification (Table \textbf{\ref%
{418})} to graded simply laced superalgebras like $SL(3|2)$ have not been
studied in the stringy literature and is still an open problem in geometry
with a super singularity. Here, we develop a proposal for approaching
surfaces with $sl\left( m|n\right) $ geometries. This proposal is based on
the super A/H correspondence (\ref{pcr}) \textrm{allowing} to define the
resolution of the $SL(3|2)$ singularity as follows:

\begin{itemize}
\item Algebraically, in terms of super weights $\mathbf{\epsilon }_{A}$ and
super roots $\mathbf{\alpha }_{A}$ like%
\begin{equation}
\mathbf{\epsilon }_{1}-\mathbf{\epsilon }_{5}=\mathbf{\alpha }_{1}+\mathbf{%
\alpha }_{2}+\mathbf{\alpha }_{3}+\mathbf{\alpha }_{4}
\end{equation}%
where we have inserted $\mathbf{\epsilon }_{2},\mathbf{\epsilon }_{3},%
\mathbf{\epsilon }_{4}$ as 
\begin{equation*}
\mathbf{\epsilon }_{1}-\mathbf{\epsilon }_{5}=\left( \mathbf{\epsilon }_{1}-%
\mathbf{\epsilon }_{2}\right) +\left( \mathbf{\epsilon }_{2}-\mathbf{%
\epsilon }_{3}\right) +\left( \mathbf{\epsilon }_{3}-\mathbf{\epsilon }%
_{4}\right) +\left( \mathbf{\epsilon }_{4}-\mathbf{\epsilon }_{5}\right)
\end{equation*}%
interpreted in terms of blowing up the singularity.

\item Homologically in terms of the graded divisors $E_{A}$ and graded
cycles $\mathfrak{C}_{A}$ as follows%
\begin{equation}
E_{1}-E_{5}=\mathfrak{C}_{1}+\mathfrak{C}_{2}+\mathfrak{C}_{3}+\mathfrak{C}%
_{4}
\end{equation}
\end{itemize}

\begin{figure}[tbph]
\begin{center}
\includegraphics[width=14cm]{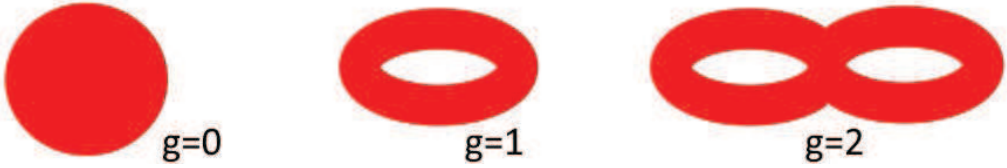}
\end{center}
\par
\vspace{-0.5cm}
\caption{Rieman surfaces with genus $g$ and Euler characteristics $2-2g$.
The $g=0$ corresponds to the 2-sphere. The $g=1$ gives the 2-torus. The $g=2$
describes the genus 2 Rieman surface.}
\label{g}
\end{figure}

The intersection of these graded cycles within the graded surface $\mathcal{S%
}^{sl_{{\small 3|2}}}$ having an $SL(3|2)$ geometry is given by%
\begin{equation}
\mathcal{I}_{AB}^{sl_{3|2}}=\mathfrak{C}_{A}.\mathfrak{C}_{B}
\end{equation}%
with 
\begin{equation}
\mathfrak{C}_{A}=E_{A}-E_{A+1}
\end{equation}%
By using 
\begin{equation}
E_{A}.E_{B}=-g_{AB}
\end{equation}%
where the metric $g_{AB}$ is as in (\ref{4g}), we end up with 
\begin{equation}
\mathcal{I}_{AB}^{sl_{3|2}}=\left( 
\begin{array}{cccc}
-2 & 1 & 0 & 0 \\ 
1 & -2 & 1 & 0 \\ 
0 & 1 & 0 & -1 \\ 
0 & 0 & -1 & 2%
\end{array}%
\right)  \label{sl32}
\end{equation}%
Comparing the intersection matrix (\ref{sl32}) to the bosonic intersection
matrix (\ref{sl5}), we learn that the self intersections of the 2-cycles of
the graded $\mathcal{S}^{sl_{{\small 3|2}}}$ have different values and
different signs namely 
\begin{equation}
\mathfrak{C}_{1}^{2}=\mathfrak{C}_{2}^{2}=-2\qquad ,\qquad \mathfrak{C}%
_{3}^{2}=0\qquad ,\qquad \mathfrak{C}_{4}^{2}=2  \label{ccc}
\end{equation}%
A way to think about these self intersections is by mimicking the 2-cycle
homology associated with Du Val singularities involving four 2-spheres
intersecting transversally according to the ADE Dynkin diagrams. By
extending this property to the graded surface $\mathcal{S}^{sl_{{\small 3|2}%
}}$, we imagine the graded 2-cycles in (\ref{ccc}) in terms of closed Rieman
surfaces $\Sigma _{g}$ without boundary. The topology of the $\Sigma _{g}$'s
is given by Euler characteristics reading as \cite{DW,DW1}, 
\begin{equation}
\mathrm{\chi }_{g}=2g-2
\end{equation}%
where the positive integer $g$ refers to the genus of $\Sigma _{g}$. Notice
that the topological $\mathrm{\chi }_{g}$ has an indefinite sign in perfect
agreement with what we are looking for. Indeed, for $g=0$ we have the
2-sphere with $\mathrm{\chi }_{0}=2$ while for $g=1$ we have a 2-torus.
However, for $g=2$ we have the double 2-torus with $\mathrm{\chi }_{2}=-2$
indicating that graded surfaces with $sl(m|n)$ singularity have higher genus
Rieman surfaces $\Sigma _{g}$ versus 2-spheres for Du Val singularities. The
three kinds of 2-cycles involved in the graded surface $\mathcal{S}^{sl_{%
{\small 3|2}}}$ are as depicted in the Figure \textbf{\ref{g}}. 

\section{Brane realisation of the super chain}

In this section, we use results obtained above to embed the distinguished $%
sl(3|2)$ superspin chain in type II strings and in M-theory. First, we study
the embedding of the super chain in type IIA string. Then, we give the uplift%
\textrm{\ to the }M-theory realisation.

\subsection{Embedding the super chain in type IIA string}

We start this investigation by giving three useful properties (A, B and C)
regarding the quantum states of the super chain. Then, we use these
properties to construct the type IIA brane realisation of the super chain.
The brane construction given below follows the approach of \cite{nafiz}
motivated by the Bethe/gauge correspondence\textrm{. }For other stringy
realizations and dualities as well as for useful tools concerning
supersymmetric twists and $\Omega $-\ deformation, \textrm{see} \cite{3B}
and \cite{SH1,SH2,SH3,SH4}.

\subsubsection{Ground state of the super chain}

We begin by recalling that the distinguished super chain of Figure \textbf{%
\ref{HC}} has an $SL(3|2)\times U\left( N_{f}\right) $ symmetry. The $%
SL(3|2) $ is the superspin group appearing as a gauge symmetry in the 4D
Chern-Simons gauge theory while the $U\left( N_{f}\right) $ is the flavor
symmetry distinguishing the N$_{f}$ atoms of the chain.

\ \ \ \ 

\textbf{A) HW states and vacuum}\newline
Algebraically speaking, the atomic states $\left \vert \Psi
_{A}^{l}\right
\rangle $ of the super chain have the tensor structure 
\begin{equation}
\left \vert \Psi _{A}^{l}\right \rangle =\left \vert \mathbf{\epsilon }%
_{A}\right \rangle _{sl_{3|2}}\otimes \left \vert \mathbf{e}^{l}\right
\rangle _{u_{N_{f}}}  \label{la}
\end{equation}%
They carry $5N_{f}$ degrees of freedom given by the $\left( \mathbf{5},%
\mathbf{N}_{f}\right) $ bi-fundamental of $SL(3|2)\times U\left(
N_{f}\right) .$ The quantum $\left \vert \mathbf{\epsilon }%
_{A}\right
\rangle $ with $A=1,2,3,4,5$ transform in the 5D representation
of $sl(3|2)$. The states $\left \vert \mathbf{e}_{l}\right \rangle $ with
label $l=1,...,N_{f}$ transform in the fundamental representation of $%
u\left( N_{f}\right) .$ The HW of the $\left( \mathbf{5},\mathbf{N}%
_{f}\right) $ bi-fundamental is given by%
\begin{equation}
\left \vert \Psi _{1}^{1}\right \rangle =\left \vert \mathbf{\epsilon }%
_{1}\right \rangle _{sl_{3|2}}\otimes \left \vert \mathbf{e}^{1}\right
\rangle _{u_{N_{f}}}
\end{equation}%
For the purpose of the brane construction we are interested in here, we
think it would be useful to give some technical details regarding the $%
\mathbf{\epsilon }_{A}$'s and the $\mathbf{e}_{l}$'s. To avoid confusion
with the previous section, we will use the following notation%
\begin{equation}
\begin{tabular}{c||c|c|c}
{\small algebra} & {\small simple roots} & {\small fundamental coweights} & 
{\small weight basis vectors} \\ \hline
$sl(3|2)$ & $\mathbf{\alpha }_{1},...,\mathbf{\alpha }_{4}$ & $\mathbf{%
\omega }_{1},...,\mathbf{\omega }_{4}$ & $\mathbf{\epsilon }_{1},...,\mathbf{%
\epsilon }_{5}$ \\ \hline
$su(N_{f})$ & $\mathbf{\tilde{\alpha}}_{1},...,\mathbf{\tilde{\alpha}}%
_{N_{f}-1}$ & $\mathbf{\tilde{\omega}}_{1},...,\mathbf{\tilde{\omega}}%
_{N_{f}-1}$ & $\mathbf{e}_{1},...,\mathbf{e}_{N_{f}}$%
\end{tabular}
\label{53}
\end{equation}

$\bullet $ \emph{the states} $\left\vert \mathbf{\epsilon }_{A}\right\rangle 
$\newline
Recall that the quantum states $\left\vert \mathbf{\epsilon }%
_{A}\right\rangle $ form the vector basis of the 5D representation $\mathcal{%
R}\left( \mathbf{\omega }_{1}\right) $ of $sl(3|2)$ \textrm{are} as in eq(%
\ref{eps}). The HW vector $\mathbf{\epsilon }_{1}$ is given by the
fundamental coweight $\mathbf{\omega }_{1},$ and the others \textrm{by}%
\begin{equation}
\begin{tabular}{lll}
$\mathbf{\epsilon }_{2}$ & $=$ & $\mathbf{\omega }_{2}-\mathbf{\omega }_{1}$
\\ 
$\mathbf{\epsilon }_{3}$ & $=$ & $\mathbf{\omega }_{3}-\mathbf{\omega }_{2}$%
\end{tabular}%
\qquad ,\qquad 
\begin{tabular}{lll}
$\mathbf{\epsilon }_{4}$ & $=$ & $\mathbf{\omega }_{3}-\mathbf{\omega }_{4}$
\\ 
$\mathbf{\epsilon }_{5}$ & $=$ & $\mathbf{\omega }_{4}$%
\end{tabular}
\label{re5}
\end{equation}%
These five $\mathbf{\epsilon }_{A}$'s obey the $sl(3|2)$ super-traceless
condition 
\begin{equation}
\sum_{A=1}^{5}\left( -\right) ^{\left\vert A\right\vert }\mathbf{\epsilon }%
_{A}=0\qquad ,\qquad \left\vert \mathbf{\epsilon }_{A}\right\vert \equiv
\left\vert A\right\vert  \label{str}
\end{equation}%
The relationship between $\mathbf{\epsilon }_{A}$ and the simple roots $%
\mathbf{\alpha }_{A}$ is obtained by\textrm{\ }using (\ref{321}).

\ \ \ \ 

$\bullet $ \emph{the states} $\left \vert \mathbf{e}_{l}\right \rangle $%
\newline
Similar relations to eqs(\ref{re5}-\ref{str}) can be written down for the
flavor symmetry. In this case, the $\mathbf{e}^{1}$ is the HW vector of $%
U\left( N_{f}\right) =U\left( 1\right) _{f}\times SU\left( N_{f}\right) .$
The traceless condition for $SU\left( N_{f}\right) $ is given by 
\begin{equation}
\sum_{l=1}^{N_{f}}\mathbf{e}^{l}=0
\end{equation}%
and is solved as%
\begin{equation}
\begin{tabular}{lll}
$\mathbf{e}_{1}$ & $=$ & $\mathbf{\tilde{\omega}}_{1}$ \\ 
$\mathbf{e}_{2}$ & $=$ & $\mathbf{\tilde{\omega}}_{2}-\mathbf{\tilde{\omega}}%
_{1}$ \\ 
& $\vdots $ &  \\ 
$\mathbf{e}_{N_{f}-1}$ & $=$ & $\mathbf{\tilde{\omega}}_{N_{f}-1}-\mathbf{%
\tilde{\omega}}_{N_{f}-2}$ \\ 
$\mathbf{e}_{N_{f}}$ & $=$ & $-\mathbf{\tilde{\omega}}_{N_{f}-1}$%
\end{tabular}%
\end{equation}%
\begin{equation*}
\end{equation*}

\textbf{B) Transitions}\emph{\ }$\left \vert \mathbf{\epsilon }%
_{A}\right
\rangle \rightarrow \left \vert \mathbf{\epsilon }%
_{B}\right
\rangle $ and $\left \vert \mathbf{e}_{l}\right \rangle
\rightarrow \left
\vert \mathbf{e}_{l^{\prime }}\right \rangle $\newline
As far as $sl(3|2)$ is concerned\footnote{%
\ \ In these kinds of transitions, we have ignored the effect of the
spectral parameter z of line defects. A rigourous description with magnons
requires the implementation of this parameter. This issue needs considering
the super Yangian $\mathcal{Y}_{sl(3|2)}$; it is developed in the appendix.}%
, the transition from the quantum $\left \vert \mathbf{\epsilon }%
_{A}\right
\rangle $ to its neighbour $\left \vert \mathbf{\epsilon }%
_{A+1}\right
\rangle $ is generated by the step operator $E_{-\alpha _{A}}$
of $sl(3|2)$ as follows, 
\begin{equation}
\left \vert \mathbf{\epsilon }_{A+1}\right \rangle =E_{-\mathbf{\alpha }%
_{A}}\left \vert \mathbf{\epsilon }_{A}\right \rangle  \label{ea1}
\end{equation}%
By using standard notations of Lie algebra representations, we can also
present $\left \vert \mathbf{\epsilon }_{A+1}\right \rangle $\ like $%
\left
\vert \mathbf{\epsilon }_{A}-\mathbf{\alpha }_{A}\right \rangle .$
For convenience, we use the relation $E_{-\mathbf{\alpha }_{A}}=\left \vert 
\mathbf{\epsilon }_{A+1}\right \rangle \left \langle \mathbf{\epsilon }%
_{A}\right \vert $ and the formal identification $E_{-\mathbf{\alpha }%
_{A}}\sim \underline{\mathbf{\alpha }}_{A}$ to think about the root $\mathbf{%
\alpha }_{A}$ as follows%
\begin{equation}
\underline{\mathbf{\alpha }}_{A}\sim \left \vert \mathbf{\epsilon }%
_{A+1}\right \rangle \left \langle \mathbf{\epsilon }_{A}\right \vert
\label{ea2}
\end{equation}%
For the generic transition $\left \vert \mathbf{\epsilon }_{A}\right \rangle
\rightarrow \left \vert \mathbf{\epsilon }_{B}\right \rangle ,$ the involved
root is given by $\mathbf{\alpha }_{AB}=\mathbf{\epsilon }_{A}-\mathbf{%
\epsilon }_{B}$. \newline
A similar description holds for the flavor symmetry sector. The homologue of
(\ref{ea1}-\ref{ea2}) reads as follows%
\begin{equation}
\begin{tabular}{ccc}
$\left \vert \mathbf{e}_{l+1}\right \rangle $ & $=$ & $\tilde{E}_{-\mathbf{%
\tilde{\alpha}}_{l}}\left \vert \mathbf{e}_{l}\right \rangle $ \\ 
$\underline{\mathbf{\tilde{\alpha}}}_{l}$ & $\sim $ & $\left \vert \mathbf{e}%
_{l+1}\right \rangle \left \langle \mathbf{e}_{l}\right \vert $%
\end{tabular}
\label{ea3}
\end{equation}

\ \ \ 

\textbf{C) More on the factorisation}\emph{\ }eq(\ref{la})\newline
The factorisation $\left\vert \mathbf{\epsilon }_{A}\right\rangle \otimes
\left\vert \mathbf{e}_{l}\right\rangle $ of the wave function (\ref{la}) 
\textrm{is justified by the fact that} $SL(3|2)$ and $U\left( N_{f}\right) $
commute, 
\begin{equation}
\left[ \tilde{g},g\right] =0,\qquad \tilde{g}\in u\left( N_{f}\right)
,\qquad g\in sl(3|2)
\end{equation}%
The transition from $\left\vert \mathbf{\epsilon }_{A}\right\rangle \otimes
\left\vert \mathbf{e}_{l}\right\rangle $ to the state $\left\vert \mathbf{%
\epsilon }_{B}\right\rangle \otimes \left\vert \mathbf{e}_{l^{\prime
}}\right\rangle $ is given by the tensor product $E_{-\mathbf{\alpha }%
_{AB}}\otimes \tilde{E}_{-\mathbf{\tilde{\alpha}}_{ll^{\prime }}}.$ As an
example, the jumping from $\left\vert \mathbf{\epsilon }_{A}\right\rangle
\otimes \left\vert \mathbf{e}_{l}\right\rangle $ to $\left\vert \mathbf{%
\epsilon }_{A+1}\right\rangle \otimes \left\vert \mathbf{e}%
_{l+1}\right\rangle $ is insured by%
\begin{equation}
E_{-\mathbf{\alpha }_{A}}\otimes \tilde{E}_{-\mathbf{\tilde{\alpha}}%
_{l}}\sim \underline{\mathbf{\alpha }}_{A}\otimes \underline{\mathbf{\tilde{%
\alpha}}}_{l}
\end{equation}

\subsubsection{Type IIA brane realisation}

In type II superstrings, the atomic states $\left\vert \mathbf{\epsilon }%
_{A},\mathbf{e}^{l}\right\rangle $ of the super chain are represented by a
system of branes whose directions expand in the 10D string spacetime
dimensions denoted as $\mathcal{M}_{1,9}$. Recall that in 10D type IIA
superstring, we have three pairs of p-branes namely: $\left( \mathbf{1}%
\right) $ the F1 string and the associated NS5. $\left( \mathbf{2}\right) $
the electric D0 and the magnetic D6. $\left( \mathbf{3}\right) $ the
electric D2 and the D4 dual.\newline
The brane system realising the states $\left\vert \Psi _{A}^{l}\right\rangle 
$ of the super chain can be derived using the dictionary of Table \textbf{%
\ref{NS}} 
\begin{table}[h]
\centering\renewcommand{\arraystretch}{1.2} $%
\begin{tabular}{c|ccc}
super chain & \multicolumn{3}{|c}{type IIA brane} \\ \hline\hline
$\left\vert \epsilon _{A}\right\rangle $ & NS5$_{\epsilon _{A}}$ & $\equiv $
& NS5$_{A}$ \\ 
\underline{$\alpha $}$_{A}$ & D2$_{\alpha _{A}}$ & $\equiv $ & D2$_{AA+1}$
\\ 
$\left\vert e^{l}\right\rangle $ & D6$_{e^{l}}$ & $\equiv $ & D6$^{l}$ \\ 
\underline{$\tilde{\alpha}$}$^{l}$ & D2$_{\tilde{\alpha}^{l}}$ & $\equiv $ & 
D2$^{ll+1}$ \\ 
$\left\vert \epsilon _{A},e^{l}\right\rangle $ & D4$_{\epsilon _{A},e^{l}}$
& $\equiv $ & D4$_{A}^{l}$ \\ \hline\hline
\end{tabular}%
$%
\caption{Lie superalgebra/Type IIA brane correspondence: Basic weight
vectors are associated with NS5 and \ D4 branes. Simple roots are associated
with D2 branes.}
\label{NS}
\end{table}
giving the algorithm for the embedding of the $sl\left( 3|2\right) $ chain
in type IIA string. The brane intersections are motivated by the algebraic
intersections 
\begin{equation}
\begin{tabular}{lll}
$\mathbf{\alpha }_{A}.\mathbf{\epsilon }_{A}$ & $=$ & $\left( -\right)
^{\left\vert A\right\vert }$ \\ 
$\mathbf{\alpha }_{A}\mathbf{.\epsilon }_{A+1}$ & $=$ & $-\left( -\right)
^{\left\vert A+1\right\vert }$ \\ 
$\mathbf{\alpha }_{A}\mathbf{.\alpha }_{B}$ & $=$ & $\mathcal{K}_{AB}$%
\end{tabular}
\label{int}
\end{equation}%
as well as similar relations for the $sl\left( N_{f}\right) $ flavor sector.
In these relations $\mathcal{K}_{AB}$ is the distinguished $sl\left(
3|2\right) $ Cartan matrix.

\ \ \ \ \ 

\textbf{A) Brane system and intersection}\newline
From the correspondence of Table \textbf{\ref{NS}} and the transitions
between the quantum states $\left\vert \mathbf{\epsilon }_{A}\right\rangle
\rightarrow \left\vert \mathbf{\epsilon }_{A+1}\right\rangle $ and $%
\left\vert \mathbf{e}_{l}\right\rangle \rightarrow \left\vert \mathbf{e}%
_{l+1}\right\rangle $ given by (\ref{ea1}-\ref{ea3}), we have the following
brane system:

\begin{description}
\item[$\left( \mathbf{i}\right) $] The transition between two neighboring NS5%
$_{\mathbf{A}}$ and NS5$_{\mathbf{A+1}}$ is mediated by the brane D2$_{%
\mathbf{AA+1}}$ stretching between them. The transition between D6$^{l}$ and
D6$^{l+1}$ is insured by the flavored D2$^{l,l+1}$ brane. For the generic
transition between NS5$_{\mathbf{A}}$ and NS5$_{\mathbf{B}}$, the natural
mediator is given by D2$_{\mathbf{AB}}$; and the transition between D6$^{l}$
and D6$^{l^{\prime }}$ is realised by the flavored D2$^{ll^{\prime }}$.

\item[$\left( \mathbf{ii}\right) $] Analogously, the transitions between
neighboring D4$_{\mathbf{A}}^{l}$ and D4$_{\mathbf{A+1}}^{l}$ is insured by
D2$_{\mathbf{AA+1}}$ while the transition between D4$_{\mathbf{A}}^{l}$ and
D4$_{\mathbf{A}}^{l+1}$\ is given by D2$^{ll+1}.$ This indicates that the
transition from D4$_{\mathbf{A}}^{l}$ and D4$_{\mathbf{A+1}}^{l+1}$ is
reached in two steps as D2$_{\mathbf{AA+1}}$D2$^{ll+1}$. This similarity
between NS5$_{\mathbf{A}}$ and D4$_{\mathbf{A}}^{l}$ becomes transparent in
M-theory as both NS5 and D4 are mapped to an M5 brane.

\item[$\left( \mathbf{iii}\right) $] The transition between the gauge NS5$_{%
\mathbf{A}}$ and the flavored D6$^{l}$ is given by the D4$_{\mathbf{A}}^{l}$
brane. General pictures involving multi-steps may be also drawn, for example
through two types of branes like D2$_{\mathbf{AB}}$D4$_{\mathbf{B}}^{l}$.
\end{description}

\ \ \ \newline
Notice that besides the transitions described by eqs(\ref{ea1}-\ref{ea3}),
there are other types of quantum excitations of the super chain which are
given by super magnons. The rigorous description of these quasi-particles
goes beyond the $sl\left( 3|2\right) $ gauge symmetry since they are given
by Verma modules of the $\mathcal{Y}_{sl_{3|2}}$ Yangian superalgebra (\ref%
{23}). The building of these quantum excitations, that turn out to be
described by kets like $\left\vert M_{\mathbf{1}},M_{\mathbf{2}},M_{\mathbf{3%
}},M_{\mathbf{4}}\right\rangle $ with $M_{\mathbf{A}}$ being positive
integers, is highly technical; it is\ reported in the Appendix. However%
\textrm{,} to fix ideas, we give here below their typical structure%
\begin{equation}
\left\vert M_{\mathbf{1}},M_{\mathbf{2}},M_{\mathbf{3}},M_{\mathbf{4}%
}\right\rangle =\prod\limits_{A=1}^{4}\mathcal{T}_{-\mathbf{\beta }%
_{A}}\left\vert \Psi _{A}^{l}\right\rangle
\end{equation}%
where we have set $\mathbf{\beta }_{A}=M_{A}\mathbf{\alpha }_{A}$ and where $%
\mathcal{T}_{-\mathbf{\beta }_{A}}$ are generators of $\mathcal{Y}\left(
sl_{3|2}\right) $. For further details, see the Appendix and also \textrm{%
\cite{1CC}}.

\ \ \ \ 

\textbf{B) the super chain in type IIA}\newline
Regarding the embedding of the superchain in the type IIA string, the brane
configuration realising the $\left\vert \Psi _{A}^{l}\right\rangle $ states
is given by two stacks of branes; a first stack $\{A\}$ labeled by the
subscript $A$, and second stack $\{l\}$ labeled by the upperscript $%
l=1,...,N_{f}.$ Below, we give a type IIA realisation where these stacks are
given by $\{NS5_{\mathbf{A}}\}$ and $\{D6^{l}\}$; and due to Hanany-Witten
transition, this brane system can be promoted to NS5$_{\mathbf{A}}$, D2$_{%
\mathbf{AA+1}},$ D6$^{l}$ and D4$_{\mathbf{A}}^{l}$ as well as strings. In
this regard, notice\textrm{\ that t}he implementation of magnons requires
also F1 strings stretching between $n_{A}$ stacks of D2$_{\mathbf{AA+1}}$
denoted like;%
\begin{equation}
\text{(D2}_{\mathbf{AA+1}}\text{)}^{n_{A}}
\end{equation}%
t\textrm{hey will be hidden below}. \textrm{So by restricting to }NS5$_{%
\mathbf{A}}$, D2$_{\mathbf{AA+1}},$ D6$^{l}$ \textrm{and} D4$_{\mathbf{A}%
}^{l}$, we have the following intersections of branes within the 10D
spacetime directions of the type IIA string 
\begin{equation}
\begin{tabular}{|c||c|c|c|c|c|c|}
\hline
{\small 10D} & $\mathbb{R}_{{\small 0}}$ & $\mathbb{R}_{{\small 1}}$ & $%
\mathbb{C}_{{\small 23}}$ & $\mathbb{R}_{{\small 4}}$ & $\mathbb{R}_{{\small %
5}}$ & $\mathbb{\tilde{R}}^{4}$ \\ \hline\hline
${\small NS5}_{\mathbf{A}}$ & x & x & x & $\{{\small \xi _{A}^{4}\}}$ & $%
\left\{ {\small \xi _{A}^{5}}\right\} $ & $\mathbb{\tilde{R}}_{\left\vert 
\mathbf{A}\right\vert }^{2}$ \\ \hline
${\small D2}_{\mathbf{AA+1}}$ & x & x & $\{z_{\mathbf{A}}{\small \}}$ & $%
{\small [\xi _{A}^{4},\xi }_{{\small A+1}}^{{\small 4}}{\small ]}$ & $%
\left\{ {\small \xi _{A}^{5}}\right\} $ & $0^{4}$ \\ \hline
${\small D4}_{\mathbf{A}}^{l}$ & x & x & $\{w_{\mathbf{A}}^{l}{\small \}}$ & 
$\left\{ {\small \xi _{A}^{4}}\right\} $ & ${\small [\xi _{A}^{5},\varrho }_{%
{\small l}}^{{\small 5}}{\small ]}$ & $\mathbb{\tilde{R}}_{\left\vert 
\mathbf{A}\right\vert }^{2}$ \\ \hline
$D6^{l}$ & x & x & $\{u^{l}{\small \}}$ & \textrm{x} & $\left\{ {\small %
\varrho }_{{\small l}}^{{\small 5}}\right\} $ & $\mathbb{\tilde{R}}^{4}$ \\ 
\hline\hline
\end{tabular}
\label{r4}
\end{equation}%
\begin{equation*}
\end{equation*}%
where the $D6^{l}$ branes will play a secondary role because the $D4_{%
\mathbf{A}}^{l}$'s should be semi-infinite due to the flavor symmetry U(N$%
_{f}$). Notice that the four dimension euclidian $\mathbb{\tilde{R}}^{4}$
with coordinates $(X^{6},X^{7},X^{8},X^{9})$ in (\ref{r4}) factorises like $%
\mathbb{\tilde{R}}_{\mathbf{\bar{0}}}^{2}\times \mathbb{\tilde{R}}_{\mathbf{%
\bar{1}}}^{2}$ with $\mathbf{\bar{0}}$ and $\mathbf{\bar{1}}$ referring to
the $\mathbb{Z}_{2}$-grading degree of $\left\vert A\right\vert $. In
addition to the orderings $\xi _{{\small A}}^{{\small 4}}<\xi _{{\small A+1}%
}^{{\small 4}}$ and $\xi _{{\small A}}^{{\small 5}}<\varrho _{{\small l}}^{%
{\small 5}}$ as well as ${\small \varrho _{l}^{5}<\varrho }_{{\small l+1}}^{%
{\small 5}}$, we have used the following notations: \newline
$\left( \mathbf{i}\right) $ The cross (x) means that the dimension of type
IIA is filled while the other boxes generate the transverse spaces of the
branes; they give precisely the position degrees of freedom of the branes
interpreted as scalar fields in super QFT at low energies. \newline
$\left( \mathbf{ii}\right) $ The singleton $\left\{ {\small \xi }_{{\small A}%
}^{{\small 4}}\right\} $\ (resp. $\{\varrho _{{\small l}}^{{\small 5}}\}$)
means that the $NS5_{\mathbf{A}}$\ brane (resp. D6$^{l}$) is located on the
fourth-axis at the point $X^{{\small 4}}=\xi _{{\small A}}^{{\small 4}}$\
(resp. the fifth-axis $\{X^{{\small 5}}=\varrho _{{\small l}}^{{\small 5}}\}$%
). \newline
$\left( \mathbf{iii}\right) $ The interval $[\xi _{{\small A}}^{{\small 4}%
},\xi _{{\small A+1}}^{{\small 4}}]$\ belongs to the fourth-axis, it is
filled by $D2_{\mathbf{AA+1}}$ stretching between NS5$_{A}$ and NS5$_{A+1}$;
the lengths $|\xi _{{\small A}}^{{\small 4}}-\xi _{{\small A\prime }}^{%
{\small 4}}|$ give the masses of the propagating quantum states between the
NS5 branes. Similarly, the interval $[\xi _{A}^{5},\varrho _{{\small l}}^{%
{\small 5}}]$\ belongs to the fifth-axis and is filled by $D4_{A}^{l}$
stretching between NS5$_{A}$ and D6$^{l}.$\ However, because the D6$^{l}$s
are flavor branes with symmetry $U(N_{f})$ containing the diagonal $U\left(
1\right) ^{N_{f}},$\ the positions of the $\varrho _{{\small l}}^{{\small 5}%
} $'s must be pushed far away (say to infinity) from the $\xi _{{\small A}}^{%
{\small 4}}$s. As such, the intervals $[\xi _{A}^{5},\varrho _{{\small l}}^{%
{\small 5}}]$\ of the $D4_{\mathbf{A}}^{l}$\ branes should be thought of as $%
{\small [\xi _{A}^{5},+\infty \lbrack }$ in agreement with the flavor
symmetry requirement and in accord with the realisation given in \textrm{%
\cite{nafiz}}. Below, we hide the $D6^{l}$ branes in (\ref{r4}), thus
reducing the brane system to the following 
\begin{equation}
\begin{tabular}{|c||c|c|c|c|c|c|}
\hline
{\small 10D} & $\mathbb{R}_{{\small 0}}$ & $\mathbb{R}_{{\small 1}}$ & $%
\mathbb{C}_{{\small 23}}$ & $\mathbb{R}_{{\small 4}}$ & $\mathbb{R}_{{\small %
5}}$ & $\mathbb{\tilde{R}}^{4}$ \\ \hline\hline
${\small NS5}_{\mathbf{A}}$ & x & x & x & $\{{\small \xi _{\mathbf{A}}^{4}\}}
$ & $\{{\small \xi _{\mathbf{A}}^{5}\}}$ & $\mathbb{\tilde{R}}_{\left\vert 
\mathbf{A}\right\vert }^{2}$ \\ \hline
${\small D2}_{\mathbf{AA+1}}$ & x & x & $\{z_{\mathbf{A}}{\small \}}$ & $%
{\small [\xi _{\mathbf{A}}^{4},\xi }_{\mathbf{A}{\small +1}}^{{\small 4}}%
{\small ]}$ & $\{{\small \xi _{\mathbf{A}}^{5}\}}$ & $0^{4}$ \\ \hline
${\small D4}_{\mathbf{A}}^{l}$ & x & x & $\{w_{\mathbf{A}}^{l}{\small \}}$ & 
$\left\{ {\small \xi _{\mathbf{A}}^{4}}\right\} $ & ${\small [\xi
_{A}^{5},\infty \lbrack }$ & $\mathbb{\tilde{R}}_{\left\vert \mathbf{A}%
\right\vert }^{2}$ \\ \hline\hline
\end{tabular}
\label{ND}
\end{equation}%
\textrm{\ }Furthermore, because of the $\mathbb{Z}_{2}$ grading of the $%
SL(3|2)$ gauge symmetry \textrm{underlying} the 4D Chern-Simons theory (\ref%
{ac}), the 10D space configurations of the branes can be dispatched
according to the charges\footnote{%
The $SL(1)$ symmetry group is the complexification of the usual unimodular
phase group $U(1)$ with element $e^{i\theta \hat{Q}}$. The $\hat{Q}$ is the
generator of abelien $u(1)$, the Lie algebra of $U(1)$, acting on complex
wave functions as $\left[ \hat{Q},\psi \right] =q\psi .$ Similarly, elements
of $SL(1,\mathbb{C})$ are given by $\lambda ^{\hat{Q}}$ with complex
parameter $\lambda =e^{\rho +i\theta }\in \mathbb{C}^{\ast }.$ Here, the $%
\hat{Q}$ is the generator of $sl(1,\mathbb{C}),$ the Lie algebra of the
abelian group $SL(1,\mathbb{C})$. For example, it acts on homogeneous
coordinates $Z_{i}$ of complex projective $CP^{n}$ as $\left[ \hat{Q},Z_{i}%
\right] =qZ_{i}.$} of the even part $sl(3|2)_{\bar{0}}=sl(3)\oplus sl\left(
1\right) \oplus sl(2).$ In this view, the five states $\left\vert \mathbf{%
\epsilon }_{A}\right\rangle $ split like 
\begin{equation}
\left\vert \mathbf{\epsilon }_{A}\right\rangle =\left\vert \mathbf{%
\varepsilon }_{a}\right\rangle _{\frac{2}{5}}\oplus \left\vert \mathbf{%
\delta }_{i}\right\rangle _{\frac{3}{5}}\qquad ,\qquad \left\vert \mathbf{%
\varepsilon }_{a}\right\vert =0,\qquad \left\vert \mathbf{\delta }%
_{i}\right\vert =1  \label{spl}
\end{equation}%
with $\left\vert \mathbf{\varepsilon }_{a}\right\rangle $ \textrm{being} a
triplet of $sl(3)$ and $\left\vert \mathbf{\delta }_{i}\right\rangle $ a
doublet of $sl(2)$. The labels $\frac{2}{5}$ and $\frac{3}{5}$ designate the
charges of the states under $sl\left( 1\right) $. \newline
Using the dictionary of the Table \textbf{\ref{NS}}, \textrm{we learn that
the p-branes} NS5$_{\mathbf{A}}$, D2$_{\mathbf{AA+1}},$ D6$^{l}$ and D4$_{%
\mathbf{A}}^{l}$ carry, in addition to quantum charges under $sl(3|2)_{\bar{0%
}}$, extra $\mathbb{Z}_{2}$ grading charges given by $\left\vert \mathbf{%
\varepsilon }_{a}\right\vert =\bar{0}$ and $\left\vert \mathbf{\delta }%
_{i}\right\vert =\bar{1}$ as in eq(\ref{spl}). Below, we describe the brane
configuration in terms of these charges.

$\bullet $ \emph{the} $sl(3)$ \emph{sector}\newline
Here, we have three bosonic weight vectors $\mathbf{\varepsilon }_{a}$ and
two bosonic simple roots $\mathbf{\alpha }_{1},$ $\mathbf{\alpha }_{2}.$ The
type IIA brane system for this sector is given by%
\begin{equation}
\begin{tabular}{|c||c|c|c|c|c|c|c|c|c|}
\hline
{\small 10D} & $\mathbb{R}_{{\small 0}}$ & $\mathbb{R}_{{\small 1}}$ & $%
\mathbb{C}_{{\small 23}}$ & $\mathbb{R}_{{\small 4}}$ & $\mathbb{R}_{{\small %
5}}$ & $\mathbb{R}_{{\small 6}}$ & $\mathbb{R}_{7}$ & $\mathbb{R}_{8}$ & $%
\mathbb{R}_{9}$ \\ \hline\hline
${\small NS5}_{a}$ & x & x & x & $\{{\small \xi _{a}^{4}\}}$ & $\{{\small %
\xi _{a}^{5}\}}$ & {\small x} & {\small x} & {\small 0} & {\small 0} \\ 
\hline
${\small D2}_{a,a\mathbf{+1}}$ & x & x & $\{z_{a}{\small \}}$ & ${\small %
[\xi _{a}^{4},\xi }_{a{\small +1}}^{{\small 4}}{\small ]}$ & $\left\{ 
{\small \xi _{a}^{5}}\right\} $ & {\small 0} & {\small 0} & {\small 0} & 
{\small 0} \\ \hline
${\small D4}_{a}^{l}$ & x & x & $\{w_{a}^{l}{\small \}}$ & $\left\{ {\small %
\xi _{a}^{4}}\right\} $ & ${\small [\xi _{a}^{5},\infty \lbrack }$ & {\small %
x} & {\small x} & {\small 0} & {\small 0} \\ \hline\hline
\end{tabular}%
\end{equation}%
The p-brane \textrm{worldvolumes} of this system are as follows: First, the
three NS5$_{a}$ are bosonic like; they expand in 5 space directions, in
particular in $\mathbb{R}_{0}\times \mathbb{R}_{1}$ and the complex $\mathbb{%
C=R}_{2}+i\mathbb{R}_{3}$ as well as $\mathbb{R}_{{\small 6}}\times \mathbb{R%
}_{{\small 7}}.$ The two D2 branes are given by D2$_{a,a+1}$ with $a=1,2;$
they are bosonic-like and expand in $\mathbb{R}_{0}\times \mathbb{R}_{1}$
and in $[\xi _{a}^{{\small 4}},\xi _{a{\small +1}}^{{\small 4}}]$ \textrm{%
representing} the 1D space between the NS5$_{a}$ and NS5$_{a+1}$ branes with
positions in the 4-th directions $\xi _{a}^{{\small 4}}$ and $\xi _{a{\small %
+1}}^{{\small 4}}.$ For the $3N_{f}$ bosonic-like D4 branes labeled as D4$%
_{a}^{l}$, they expand in $\mathbb{R}_{0}\times \mathbb{R}_{1}\times {\small %
[\xi _{a}^{5},\infty \lbrack }\times \mathbb{R}_{{\small 6}}\times \mathbb{R}%
_{{\small 7}}$.

$\bullet $ \emph{the} \emph{sl}$\left( 2\right) $\emph{\ sector}\newline
\textrm{The} p-brane \textrm{worldvolumes} in this sector correspond to the
odd sector of $\mathbb{Z}_{2}$. It is described by two graded weight vectors 
$\mathbf{\delta }_{1}$ and $\mathbf{\delta }_{2}$ with a bosonic- like
simple root $\mathbf{\alpha }_{4}.$ The p-brane system in this sector is
given by%
\begin{equation}
\begin{tabular}{|c||c|c|c|c|c|c|c|c|c|}
\hline
{\small 10D} & $\mathbb{R}_{{\small 0}}$ & $\mathbb{R}_{{\small 1}}$ & $%
\mathbb{C}_{{\small 23}}$ & $\mathbb{R}_{{\small 4}}$ & $\mathbb{R}_{{\small %
5}}$ & $\mathbb{R}_{{\small 6}}$ & $\mathbb{R}_{7}$ & $\mathbb{R}_{8}$ & $%
\mathbb{R}_{9}$ \\ \hline\hline
${\small NS5}_{{\small 3+i}}$ & x & x & x & $\{{\small \xi _{3+i}^{4}\}}$ & $%
\{{\small \xi _{3+i}^{5}\}}$ & {\small 0} & {\small 0} & {\small x} & 
{\small x} \\ \hline
${\small D2}_{45}$ & x & x & $\{z_{4}{\small \}}$ & ${\small [\xi
_{4}^{4},\xi }_{6}^{{\small 4}}{\small ]}$ & $\left\{ {\small \xi _{4}^{5}}%
\right\} $ & {\small 0} & {\small 0} & {\small 0} & {\small 0} \\ \hline
${\small D4}_{{\small 3+i}}^{l}$ & x & x & $\{w_{{\small 3+i}}^{l}{\small \}}
$ & $\left\{ {\small \xi _{3+i}^{4}}\right\} $ & ${\small [\xi
_{3+i}^{5},\infty \lbrack }$ & {\small 0} & {\small 0} & {\small x} & 
{\small x} \\ \hline\hline
\end{tabular}%
\end{equation}%
The D2$_{45}$ brane interpolates between the odd NS5$_{4}$ and NS5$_{5}$ and
sits at the points $\{z_{4}{\small \}}$ and $\left\{ {\small \xi _{4}^{5}}%
\right\} $.

$\bullet $ \emph{the} $sl(1)$ \emph{sector}\newline
This sector gives the \textrm{link} between the $sl(3)$ and the $sl(2)$
sectors. This bridging is given by the D2$_{34}$ brane which is a
fermionic-like brane; it interpolates between the bosonic- like NS5$_{3}$
and the fermionic NS5$_{4}$.%
\begin{equation}
\begin{tabular}{|c||c|c|c|c|c|c|c|c|c|}
\hline
{\small 10D} & $\mathbb{R}_{{\small 0}}$ & $\mathbb{R}_{{\small 1}}$ & $%
\mathbb{C}_{{\small 23}}$ & $\mathbb{R}_{{\small 4}}$ & $\mathbb{R}_{{\small %
5}}$ & $\mathbb{R}_{{\small 6}}$ & $\mathbb{R}_{7}$ & $\mathbb{R}_{8}$ & $%
\mathbb{R}_{9}$ \\ \hline\hline
{\small D2}$_{34}$ & x & x & $\{z_{3}{\small \}}$ & ${\small [\xi
_{3}^{4},\xi }_{4}^{{\small 4}}{\small ]}$ & $\{{\small \xi _{3}^{5}\}}$ & 
{\small 0} & {\small 0} & {\small 0} & {\small 0} \\ \hline
\end{tabular}%
\end{equation}

\subsection{Uplift to M- theory}

The embedding of the super chain in M-theory is given by the uplift of type
IIA string description. To fix the ideas, we parameterise the directions of
the 11D space time dimensions like 
\begin{equation}
\mathcal{M}_{1,10}=\mathcal{M}_{1,9}\times \mathbb{S}_{M}^{1}  \label{21}
\end{equation}%
where $\mathcal{M}_{1,9}$ is as in type IIA string and where $\mathbb{S}%
_{M}^{1}\sim R_{M}e^{i\vartheta }$ is the M-theory circle describing the
eleventh direction.

\subsubsection{Uplifting Table(\protect\ref{NS}) to M-branes}

As the NS5- and the D4-\ branes of type IIA string merge into M5-branes of
M-theory and the D2 is mapped into the M2, the M-brane system realising the
super chain can be obtained by promoting \textrm{Table(\ref{NS}) and eq.(\ref%
{r4})} to 11D. Before giving this uplift, it is interesting to recall two
useful features of type IIA string with regard to 11D M-theory. $\left( 
\mathbf{1}\right) $ Under the compactification of $\mathcal{M}_{1,10}$ on
the usual circle $\mathbb{S}_{M}^{1}$ down to $\mathcal{M}_{1,9}$, we have
the\ following M-branes reductions%
\begin{equation}
\begin{tabular}{cc|c}
& $\mathcal{M}_{1,10}$ & $\mathcal{M}_{1,9}$ \\ \cline{2-3}
$\mathbf{i)}$ & $M5$ & $NS5$ \\ \cline{2-3}
$\mathbf{ii)}$ & $M5$ & $M5/\mathbb{S}_{M}^{1}\sim D4$%
\end{tabular}%
\qquad ,\qquad 
\begin{tabular}{cc|c}
& $\mathcal{M}_{1,10}$ & $\mathcal{M}_{1,9}$ \\ \cline{2-3}
$\mathbf{i)}$ & $M2$ & $D2$ \\ \cline{2-3}
$\mathbf{ii)}$ & $M2$ & $M2/\mathbb{S}_{M}^{1}\sim F1$%
\end{tabular}%
\end{equation}%
showing that M5 and M2 lead to various kinds of branes in $\mathcal{M}_{1,9}$%
. The worldvolumes of these branes with applications to the NS5, D4 and D2
branes in the table (\ref{ND}) are as follows%
\begin{equation}
\begin{tabular}{|c|c|c|c|}
\hline
11D & : & 10D & $\mathbb{S}_{M}^{1}$ \\ \hline\hline
M5$_{A}$ & : & NS5$_{A}$ &  \\ \hline
M5$_{A}^{l}$ & : & D4$_{A}^{l}$ & x \\ \hline\hline
\end{tabular}%
\qquad ,\qquad 
\begin{tabular}{|c|c|c|c|}
\hline
{\small 11D} & : & 10D & $\mathbb{S}_{M}^{1}$ \\ \hline\hline
M2$_{AA+1}$ & : & D2$_{AA+1}$ &  \\ \hline
M2 & : & F1 & x \\ \hline\hline
\end{tabular}%
\end{equation}%
where the cross (x) designates a filled direction. $\left( \mathbf{2}\right) 
$ Type IIA string theory with $N_{f}$ coincident D6- branes and $U(N_{f})$
symmetry have a local description in eleven dimensions in terms of M-theory
compactified on the $A_{N_{f}-1}$ ALE singularity. This real 4D geometry can
be described in different but equivalent ways; for example in terms of the
orbifold $\mathbb{C}^{2}/\mathbb{Z}_{N_{f}}$ with discrete group $\mathbb{Z}%
_{N_{f}}$ acting on the two complex coordinates $z$ and $z^{\prime }$ like 
\begin{equation}
z\rightarrow e^{\frac{2i\pi }{N}}z\qquad ,\qquad z^{\prime }\rightarrow e^{-%
\frac{2i\pi }{N}}z^{\prime }
\end{equation}%
We will think of these complex variables $z$ and $z^{\prime }$ in terms of
the spacetime variables $X^{2}+iX^{3}$ and $\exp (-X^{5}+iR_{M}\vartheta )$
respectively. Another interesting realisation of this singular complex
surface is \textrm{given} in terms of its embedding in $\mathbb{C}%
^{3}[u,v,w] $ where it is defined by the algebraic equation $uv=w^{N}$. More
interestingly, the deformation of the $A_{N-1}$ singularity \textrm{is}
described by $uv=\prod\nolimits_{l=1}^{N}\left( w-\zeta _{l}\right) $ which
is generated by the $N_{f}$ complex parameters $\zeta _{l}$. In the language
of 2-cycle homology, the topology of $A_{N-1}$ may be imagined in terms of
the fibration $\mathcal{S}_{N}=\Sigma _{N}\times \mathbb{C}$ with $\Sigma
_{N}$ being a complex curve (2-cycle) given by the intersection of
generating 2-cycles $C_{l}$ with complexified Kahler moduli $t_{l}=\zeta
_{l}-\zeta _{l+1}$. In terms of these objects, the M-brane description of
the type IIA\ realisation (\ref{ND}) is given\ by%
\begin{equation}
\begin{tabular}{|c||c|c|c|c|c|}
\hline
{\small 11D} & $\mathbb{R}_{{\small 0}}$ & $\mathbb{R}_{{\small 1}}$ & $%
\mathcal{S}_{N}$ & $\mathbb{R}_{{\small 4}}$ & $\mathbb{\tilde{R}}^{4}$ \\ 
\hline\hline
${\small M5}_{\mathbf{A}}$ & x & x & $\Sigma _{A}$ & $\{{\small \xi
_{A}^{4}\}}$ & $\mathbb{\tilde{R}}_{\left\vert \mathbf{A}\right\vert }^{2}$
\\ \hline
${\small M2}_{\mathbf{AA+1}}$ & x & x & $(z_{A},z_{A}^{\prime }{\small )}$ & 
${\small [\xi _{A}^{4},\xi }_{{\small A+1}}^{{\small 4}}{\small ]}$ & $0^{4}$
\\ \hline
${\small M5}_{\mathbf{A}}^{l}$ & x & x & $\Sigma _{A}^{l}$ & $\left\{ 
{\small \xi _{A}^{4}}\right\} $ & $\mathbb{\tilde{R}}_{\left\vert \mathbf{A}%
\right\vert }^{2}$ \\ \hline\hline
\end{tabular}
\label{MD}
\end{equation}%
\begin{equation*}
\end{equation*}%
where $\left( i\right) $ the complex line $\Sigma _{A}$ is given by a
fibration $\mathbb{C}\times \{z_{{\small A}}\}$ with $z_{{\small A}}=X_{%
{\small A}}^{2}+iX_{{\small A}}^{3}$ describing the loci of the ${\small M5}%
_{\mathbf{A}}$ in the complex surface $\mathcal{S}_{N}$. $\left( ii\right) $
the complex $z_{A}^{\prime }$ is given by $e^{X_{A}^{5}+i\vartheta _{A}}$
where $X_{A}^{5}=\xi _{A}^{{\small 5}}$ is as in \textrm{Table} (\ref{ND}).
Moreover, the complex curve $\Sigma _{A}^{l}$ is a complex half line
parameterised by a complex variable $\zeta $ with "left end" value given by $%
\zeta =z_{A}^{\prime }$ and "right end" $\zeta _{l}=e^{X_{l}^{5}+i\vartheta
_{l}}$. Here, the $X_{l}^{5}=\varrho _{l}^{5}$ \textrm{is} as in (\ref{ND}),
but due to the flavor \textrm{nature} of U(N$_{f}$), its absolute value $%
\left\vert \zeta _{l}\right\vert =e^{\varrho _{l}^{5}}$ is pushed to
infinity; that $\left\vert \zeta _{l}\right\vert \rightarrow \infty $.%
\newline
From this description, it follows that the correspondence given by Table 
\textbf{\ref{NS}}\textrm{\ with regards} to type IIA generalises in M-theory
as in the Table \textbf{\ref{M52}} 
\begin{table}[h]
\centering\renewcommand{\arraystretch}{1.2} $%
\begin{tabular}{c|c|ccc}
super chain & type IIA & \multicolumn{3}{|c}{M-branes} \\ \hline\hline
$\left\vert \epsilon _{A}\right\rangle $ & NS5$_{\epsilon _{A}}$ & M5$%
_{\epsilon _{A}}$ & $\equiv $ & M5$_{A}$ \\ 
\underline{$\alpha $}$_{A}$ & D2$_{\alpha _{A}}$ & M2$_{\alpha _{A}}$ & $%
\equiv $ & M2$_{AA+1}$ \\ 
$\left\vert e^{l}\right\rangle $ & D6$_{e^{l}}$ & $C^{2}/Z_{N_{f}}$ &  &  \\ 
$\left\vert \epsilon _{A},e^{l}\right\rangle $ & D4$_{\epsilon _{A},e^{l}}$
& M5$_{\epsilon _{A},e^{l}}$ & $\equiv $ & M5$_{A}^{l}$ \\ \hline\hline
\end{tabular}%
$%
\caption{Lie superalgebra/M-brane correspondence. Simple roots are of $%
sl(m|n)$ associated with graded M2-branes stretching between graded M5
pairs. }
\label{M52}
\end{table}
giving the algorithm for the embedding of the $sl\left( 3|2\right) $ chain
in M- theory. The brane intersections of the M- branes in the Table \textbf{%
\ref{M52}} follow from (\ref{int}) for the $sl\left( 3|2\right) $ superspin
sector and for flavor $U\left( N_{f}\right) $. \textrm{We} have%
\begin{equation}
\begin{tabular}{lll}
$\mathbf{\tilde{\alpha}}_{l}\mathbf{.e}_{l}$ & $=$ & $+1$ \\ 
$\mathbf{\tilde{\alpha}}_{l}\mathbf{.e}_{l+1}$ & $=$ & $-1$ \\ 
$\mathbf{\tilde{\alpha}}_{l}\mathbf{.\tilde{\alpha}}_{l^{\prime }}$ & $=$ & $%
\mathcal{A}_{ll^{\prime }}^{sl_{N_{f}}}$%
\end{tabular}
\label{524}
\end{equation}%
where $\mathcal{A}_{ll^{\prime }}^{sl_{N_{f}}}$ is the Cartan matrix of $%
sl\left( N_{f}\right) .$ From the correspondence in Table \textbf{\ref{M52}}
and the transitions between the quantum states in the pseudo vacuum, we
obtain the following \textrm{M-brane} candidates to build the embedding of
the super chain in M theory:

\begin{itemize}
\item Five $sl_{3|2}$ super M5 branes denoted like M5$_{\mathbf{A}}.$

\item Four $sl_{3|2}$ super M2$_{\mathbf{AA+1}}$ branes stretching between M5%
$_{\mathbf{A}}$ and M5$_{\mathbf{A+1}}.$

\item 5N$_{f}$ super M5 branes denoted like M5$_{\mathbf{A}}^{l}.$

\item M-theory on the singularity geometry A$_{N-1}$ requiring by the
promotion of D6.
\end{itemize}

\begin{figure}[tbph]
\begin{center}
\includegraphics[width=16cm]{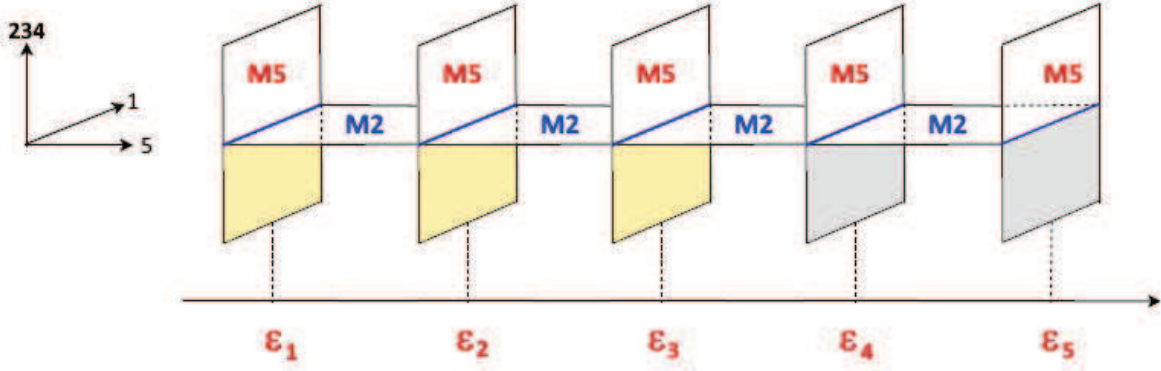}
\end{center}
\par
\vspace{-0.5cm}
\caption{Horizontal M2 branes stretching between pairs of vertical M5s
located at $X_{A}$ and $X_{A+1}$. The M2 and the M5 intersect along an
\textquotedblleft M-string\textquotedblright\ (the 1-direction in blue). The
positions $X_{A}$ can be put in correspondence with the unit weights $%
\protect\epsilon _{A}$ and the $X_{A+1}-X_{A}$ with the simple roots.}
\label{M25}
\end{figure}

A\emph{\ }graphical illustration of this interacting M5-M2 brane system is
depicted by the Figure \textbf{\ref{M25}} where the five basic M5$_{\mathbf{A%
}}$ stacks are represented by 5 vertical sheets and the M2$_{\mathbf{AA+1}}$
brane messengers by 4 horizontal stacks. 

\subsubsection{Graded M-branes}

Using the dictionary of Table \textbf{\ref{NS}}, we learn that the M-branes
M5$_{\mathbf{A}}$, M2$_{\mathbf{AA+1}}$ and M5$_{\mathbf{A}}^{l}$ carry, in
addition to quantum numbers under $sl(3|2)_{\bar{0}}$, extra $\mathbb{Z}_{2}$
charges given by $\left \vert \mathbf{\varepsilon }_{a}\right \vert =\bar{0}$
and $\left \vert \mathbf{\delta }_{i}\right \vert =\bar{1}$. Below, we give
the brane system in terms of these charges.

$\bullet $\emph{\ the} $sl(3)$ \emph{sector}\newline
The M-brane system for this bosonic-like sector is given by%
\begin{equation}
\begin{tabular}{|c|c|c|c|c|c|c|c|c|}
\hline
{\small 11D} & $\mathbb{R}_{0}$ & $\mathbb{R}_{1}$ & $\mathcal{S}_{N}$ & $%
\mathbb{R}_{{\small 4}}$ & $\mathbb{R}_{{\small 6}}$ & $\mathbb{R}_{7}$ & $%
\mathbb{R}_{8}$ & $\mathbb{R}_{9}$ \\ \hline\hline
{\small M5}$_{a}$ & x & x & $\Sigma _{a}$ & $\{{\small \xi _{a}^{4}\}}$ & 
{\small x} & {\small x} & {\small 0} & {\small 0} \\ \hline
{\small M2}$_{aa\mathbf{+1}}$ & x & x & $(z_{a},z_{a}^{\prime }{\small )}$ & 
${\small [\xi _{a}^{4},\xi }_{a{\small +1}}^{{\small 4}}{\small ]}$ & 
{\small x} & {\small x} & {\small 0} & {\small 0} \\ \hline
{\small M5}$_{a}^{l}$ & x & x & $\Sigma _{a}^{l}$ & $\left\{ {\small \xi
_{a}^{4}}\right\} $ & {\small 0} & {\small 0} & {\small 0} & {\small 0} \\ 
\hline\hline
\end{tabular}%
\end{equation}%
where the complex line $\Sigma _{a}$ is given by $\mathbb{C}\times \{z_{%
{\small a}}\}$ with $z_{{\small a}}=X_{{\small a}}^{2}+iX_{{\small a}}^{3}$
describing the loci of the ${\small M5}_{a}$ in the orbifold singularity and
where $z_{A}^{\prime }$ is given by $e^{\xi _{a}^{{\small 5}}+i\vartheta
_{a}}.$

$\bullet $\emph{\ the} \emph{sl}$\left( 2\right) $\emph{\ sector}\newline
The p-brane system in this sector is given by%
\begin{equation}
\begin{tabular}{|c|c|c|c|c|c|c|c|c|}
\hline
{\small 11D} & $\mathbb{R}_{0}$ & $\mathbb{R}_{1}$ & $\mathcal{S}_{N}$ & $%
\mathbb{R}_{{\small 4}}$ & $\mathbb{R}_{{\small 6}}$ & $\mathbb{R}_{7}$ & $%
\mathbb{R}_{8}$ & $\mathbb{R}_{9}$ \\ \hline\hline
{\small M5}$_{3+i}$ & x & x & $\Sigma _{3+i}$ & $\{{\small \xi _{3+i}^{4}\}}$
& {\small 0} & {\small 0} & {\small x} & {\small x} \\ \hline
{\small M2}$_{\mathbf{45}}$ & x & x & $(z_{34},z_{34}^{\prime }{\small )}$ & 
${\small [\xi _{3}^{4},\xi }_{{\small 4}}^{{\small 4}}{\small ]}$ & {\small 0%
} & {\small 0} & {\small x} & {\small x} \\ \hline
{\small M5}$_{3+i}^{l}$ & x & x & $\Sigma _{3+i}^{l}$ & $\left\{ {\small \xi
_{3+i}^{4}}\right\} $ & {\small 0} & {\small 0} & {\small 0} & {\small 0} \\ 
\hline
\end{tabular}%
\end{equation}

$\bullet $\emph{\ the} $sl(1)$ \emph{sector}\newline
This sector gives the bridge between the $sl(3)$ and the $sl(2)$ sectors.
This bridging is given by the M2$_{\mathbf{34}}$ which is a fermionic-like
brane; it interpolates between the bosonic- like M5$_{\mathbf{3}}$ and the
fermionic M5$_{\mathbf{4}}$.%
\begin{equation}
\begin{tabular}{|c|c|c|c|c|c|c|c|c|}
\hline
{\small 11D} & $\mathbb{R}_{0}$ & $\mathbb{R}_{1}$ & $\mathcal{S}_{N}$ & $%
\mathbb{R}_{4}$ & $\mathbb{R}_{{\small 6}}$ & $\mathbb{R}_{7}$ & $\mathbb{R}%
_{8}$ & $\mathbb{R}_{9}$ \\ \hline\hline
{\small M2}$_{\mathbf{34}}$ & x & x & $z_{\mathbf{34}}$ & ${\small [\xi
_{3}^{4},\xi }_{{\small 4}}^{{\small 4}}{\small ]}$ & {\small 0} & {\small 0}
& {\small 0} & {\small 0} \\ \hline
\end{tabular}%
\end{equation}

\section{Geometric interpretation of M5$_{\mathbf{A}}^{l}$}

\begin{figure}[tbph]
\begin{center}
\includegraphics[width=10cm]{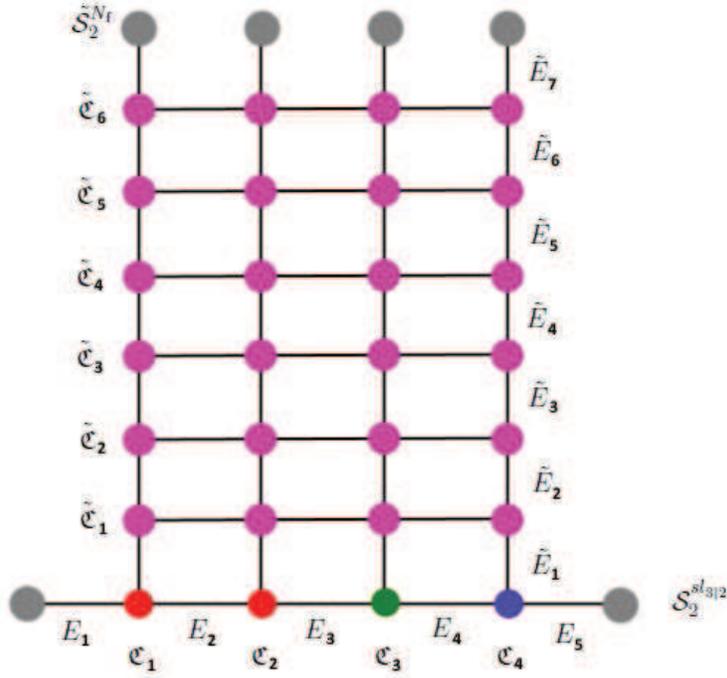}
\end{center}
\par
\vspace{-0.5cm}
\caption{Cycle homology of the 4D complex manifold {\protect\large Y}$_{4}=%
\mathcal{\tilde{S}}_{2}^{N_{\mathrm{f}}}\times \mathcal{S}_{2}^{sl_{3|2}}$.
The horizontal line is given by the gauge surface $\mathcal{S}%
_{2}^{sl_{3|2}} $ and the vertical\ by the flavor surface $\mathcal{\tilde{S}%
}_{2}^{N_{\mathrm{f}}}.$ The colored dots designate 2-cycles $\mathfrak{C}%
_{A}$ and $\mathfrak{\tilde{C}}_{l}$ while the segments represent the
divisors $E_{A}$ and $\tilde{E}_{l}$ }
\label{Lat}
\end{figure}

Here, we use results of\ the sub-section 4.2 regarding super singular
geometry to engineer the M theory manifold $\mathcal{M}_{1,10}$ (\ref{21})
that hosts the M-branes modeling of the $SL(3|2)\times U(N_{f})$ superspin
chain. We show that the brane system is induced by M5/M2 sitting in the
singularity of a complex 4D manifold {\large Y}$_{4}$ that we want to
construct. Under resolution of the singularity, the {\large Y}$_{4}$ has a
real 4-cycle ${\large C}_{4}$ given by the fibration 
\begin{equation}
\begin{tabular}{lll}
${\large \tilde{C}}_{2}^{N_{f}}$ & $\rightarrow $ & ${\large C}_{4}$ \\ 
&  & $\downarrow $ \\ 
&  & ${\large C}_{2}^{sl_{3|2}}$%
\end{tabular}
\label{sc2}
\end{equation}%
that is a 2-cycle fibred over another 2-cycle. A graphical representation of 
${\large C}_{4}\sim {\large \tilde{C}}_{2}^{N_{f}}\times {\large C}%
_{2}^{sl_{3|2}}$ was sketched for $N_{f}=4$ by the Figure \textbf{\ref%
{typeII}}; other equivalent representations will be given below. In this
regard, notice that the 2-cycle base ${\large C}_{2}^{sl_{3|2}}$ in (\ref%
{sc2}) has an $SL(3|2)$ cycle homology generated by four irreducible
2-cycles $\mathfrak{C}_{A}$ according to the distinguished $sl(3|2)$ Dynkin
diagram%
\begin{equation*}
DD_{sl(3|2)}
\end{equation*}%
given by the Figure \textbf{\ref{DSD}}. Similarly, the 2-cycle fiber $%
{\large \tilde{C}}_{2}^{N_{f}}$ in (\ref{sc2}) is a Du Val surface with an $%
SU(N_{f})$ geometry as in the Table eq(\ref{418}). It is generated by $%
N_{f}-1$ irreducible 2-cycles $\mathfrak{\tilde{C}}_{l}$ according to the $%
SU(N_{f})$ Dynkin diagram. 
\begin{equation*}
DD_{su(N_{f})}
\end{equation*}%
To fix the ideas, we give the cycle homology of the 4D manifold {\large Y}$%
_{4}$ in the Figure \textbf{\ref{Lat}} 
as the cross product of the two Dynkin diagrams.%
\begin{equation*}
{\large C}_{4}\sim DD_{sl(3|2)}\times DD_{su(N_{f})}
\end{equation*}

\subsection{From $su(N_{f})$ towards the Du Val geometry}

In this subsection, we construct the complex surface $\mathcal{\tilde{S}}%
^{N_{f}}$ with $su(N_{f})$ singularity and 2-cycle given by ${\large \tilde{C%
}}_{2}^{N_{f}}$. This geometry describes the lifting of the $N_{f}$ flavor D6%
$^{l}$ to M-theory.

\subsubsection{Homology of the 2-cycle ${\protect\large \tilde{C}}_{2}^{N_{%
\mathrm{f}}}$}

\begin{figure}[tbph]
\begin{center}
\includegraphics[width=12cm]{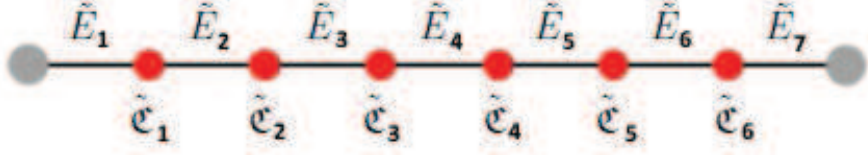}
\end{center}
\par
\vspace{-0.5cm}
\caption{The structure of the 2-cycle ${\protect\large \tilde{C}}_{2}^{N_{%
\mathrm{f}}}$ of the flavor surface with an $su\left( N_{f}\right) $
geometry ($N_{f}=7$). The red dot gives the irreducible 2-cycles. These
homology cycles look like the Dynkin diagram of $su\left( N_{f}\right) $.}
\label{N7}
\end{figure}

Here, we use results of the Algebra/Homology correspondence between $\left(
i\right) $ roots $\mathbf{\tilde{\alpha}}_{l}$/ weights $\mathbf{\tilde{%
\varepsilon}}_{l}$ of the flavor symmetry $su(N_{f})$ on one side; and $%
\left( ii\right) $ 2-cycles $\mathfrak{\tilde{C}}_{l}$/ divisors $\tilde{E}%
_{l}$ of the Du Val complex surface $\mathcal{\tilde{S}}^{N_{f}}$ on the
other side. The surface $\mathcal{\tilde{S}}^{N_{f}}$ is given by the
fibration%
\begin{equation}
\begin{tabular}{lll}
$\mathbb{C}$ & $\rightarrow $ & $\mathcal{\tilde{S}}^{N_{f}}$ \\ 
&  & $\downarrow $ \\ 
&  & ${\large \tilde{C}}_{2}^{N_{f}}$%
\end{tabular}
\label{st}
\end{equation}%
with the base given by the 2-cycle ${\large \tilde{C}}_{2}^{N_{f}}$ that
must have a large volume. Using the A/H correspondence, the compact ${\large 
\tilde{C}}_{2}^{N_{f}}$ is given by $N_{f}-1$ intersecting irreducible
2-cycles $\mathfrak{\tilde{C}}_{l}$ as follows 
\begin{equation}
{\large \tilde{C}}_{2}^{N_{f}}=\sum_{l=1}^{N_{f}-1}\mathfrak{\tilde{C}}_{l}
\label{cpc}
\end{equation}%
with $\mathfrak{\tilde{C}}_{l}$ stretching between two divisors $\tilde{E}%
_{l}$ and $\tilde{E}_{l+1}$ as%
\begin{equation}
\mathfrak{\tilde{C}}_{l}=\tilde{E}_{l}-\tilde{E}_{l+1}
\end{equation}%
As an illustration, we give in the Figure \textbf{\ref{N7}} 
a graphical description of the cycle ${\large \tilde{C}}_{2}^{N_{\mathrm{f}%
}}.$ Using the homology metric 
\begin{equation}
\tilde{E}_{l}.\tilde{E}_{l^{\prime }}=-\delta _{ll^{\prime }}
\end{equation}%
we obtain the intersection matrix between the flavor divisors%
\begin{equation}
\mathfrak{\tilde{C}}_{l}\mathbf{.}\mathfrak{\tilde{C}}_{l^{\prime }}=%
\mathcal{I}_{ll^{\prime }}^{sl_{N_{f}}}
\end{equation}%
with $\mathcal{I}_{ll^{\prime }}^{sl_{N_{f}}}$ given by minus the Cartan
matrix $\mathbf{\tilde{\alpha}}_{l}\mathbf{.\tilde{\alpha}}_{l^{\prime }}=%
\mathcal{A}_{ll^{\prime }}^{sl_{N_{f}}}$ of $sl(N_{f})$; thus leading to%
\begin{equation}
\mathcal{I}_{ll^{\prime }}^{sl_{N_{f}}}=\delta _{ll^{\prime }-1}-2\delta
_{ll^{\prime }}+\delta _{ll^{\prime }+1}  \label{ccp}
\end{equation}

\subsubsection{Defining equation of the surface $\mathcal{\tilde{S}}^{N_{%
\mathrm{f}}}$}

\begin{figure}[tbph]
\begin{center}
\includegraphics[width=16cm]{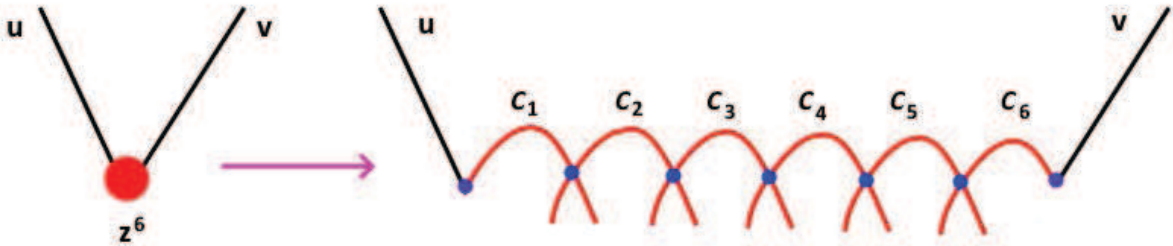}
\end{center}
\par
\vspace{-0.5cm}
\caption{On the left, an ALE surface $uv=z^{6}$ having an $SU\left( 7\right) 
$ singularity at the origin of $\mathbb{C}^{3}$. On the right, its complete
resolution using 6 transversally intersecting complex projective lines. Here
we have set $N_{f}$=7.}
\label{su6}
\end{figure}

The complex surface $\mathcal{\tilde{S}}^{N_{f}}$ is a complex ALE surface
with a resolved $SU(N_{f})$ singularity; its resolution is as illustrated by
Figure \textbf{\ref{su6}. }This graph, and the Figure \textbf{\ref{N7} }are
related by the 2D space duality. 
The defining equation of $\mathcal{\tilde{S}}^{N_{f}}$ embedded in $\mathbb{C%
}^{3}$ with complex local coordinates $\left( u,v,z\right) \ $is given by
the following holomorphic 
\begin{equation}
uv=c\prod\limits_{l=1}^{N_{f}}\left( z-\mu _{l}\right)  \label{sf}
\end{equation}%
where $c$ is a complex number. The $N_{f}$ moduli $\mu _{l}$ are the zeros
of $\mathcal{\tilde{S}}^{N_{f}}$; they can be put in correspondence with the
weight vectors $\mathbf{e}_{l}$ of the $SU\left( N_{f}\right) $ flavor
symmetry (\ref{53}) and the Table \textbf{\ref{NS}}; and by the A/H with the 
$\tilde{E}_{l}$ divisors. So, we have 
\begin{equation}
\mathbf{e}_{l}\leftrightarrow \mu _{l}\leftrightarrow \tilde{E}_{l}
\end{equation}%
By using the Table \textbf{\ref{M5}} linking $SU\left( N_{f}\right) $
weights with M-branes, we deduce that the $\mathrm{\mu }_{l}$'s can be
interpreted in type IIA string as the loci where sit the D6$^{l}$ branes in $%
\mathcal{\tilde{S}}^{N_{f}}$. Similarly, the difference 
\begin{equation}
\upsilon _{ll^{\prime }}=\mu _{l}-\mu _{l^{\prime }}
\end{equation}%
is put in correspondence with the roots $\mathbf{\tilde{\alpha}}_{ll^{\prime
}}=\mathbf{e}_{l}-\mathbf{e}_{l^{\prime }}.$ By the A/H, the $\mathbf{\tilde{%
\alpha}}_{ll^{\prime }}$ describe the 2-cycles $\mathfrak{\tilde{C}}%
_{ll^{\prime }}$ in $\mathcal{\tilde{S}}^{N_{f}}$; and by Table \textbf{\ref%
{M5}}, the roots $\mathbf{\tilde{\alpha}}_{ll^{\prime }}$ and cycles $%
\mathfrak{\tilde{C}}_{ll^{\prime }}$ characterise the branes D2$^{ll^{\prime
}}$ stretching between D6$^{l}$ and D6$^{l^{\prime }}$.\newline
\textrm{At} the end of this construction, notice the following: $\left( 
\mathbf{i}\right) $ for the particular case where all the $\mu _{l}$'s are
equal to $\mu _{0}$, the above complex surface equation becomes 
\begin{equation}
uv=c\left( z-\mu _{0}\right) ^{N_{f}}  \label{nf}
\end{equation}%
having an $SU(N_{f})$ singularity at $z=\mu _{0}$. $\left( \mathbf{ii}%
\right) $ For the case where $N_{f}=2,$ the surface \textrm{is} $uv=c\left(
z-\mu _{1}\right) \left( z-\mu _{2}\right) .$ The condition $\mu _{1}+\mu
_{2}=0$ gives the traceless condition of $sl\left( 2\right) $. $\left( 
\mathbf{iii}\right) $ For the particular case $N_{f}=1,$ the surface in (\ref%
{nf}) reads as $uv=c\left( z-\mu _{0}\right) $ and is regular.

\subsubsection{Flavor symmetry constraints on (\protect\ref{cpc})}

The real 2-cycle $\mathfrak{\tilde{C}}_{2}^{N_{f}}$ living inside the
complex surface $\mathcal{\tilde{S}}^{N_{f}}$ is given by the homological
expansion (\ref{cpc}) with generators $\mathfrak{\tilde{C}}_{l}.$ Because
the $U\left( N_{f}\right) $ is a flavor symmetry, the associated gauge
fields are non dynamical. This property is implemented by demanding that
their gauge field masses are too heavy such that the gauge field dynamics
decouple. As the masses of the gauge particles are related to volumes $%
\upsilon _{l}$ of the $\mathfrak{\tilde{C}}_{l}$ cycles, the flavor symmetry
requires $\upsilon _{l}>>1.$ Therefore, the moduli 
\begin{equation}
\upsilon _{l}=\int_{\mathfrak{\tilde{C}}_{l}}\mathcal{J}
\end{equation}%
has to take big values. The $\mathcal{J}$\ in this relation is the Kahler
2-form $dz\wedge d\bar{z}/2i$. To write down the algebraic equation of the $%
\mathfrak{\tilde{C}}_{l}$'s and derive the flavor symmetry constraints that
we have to impose on the moduli $\mu _{l}$ in (\ref{sf}), we proceed as
follows. \newline
We start from eq(\ref{sf}) defining the non compact surface $\mathcal{\tilde{%
S}}^{N_{f}}$ in terms of the $\mu _{l}$'s, then we sit on the $\mathbb{C}%
^{3} $ patch 
\begin{equation}
u=1\qquad ,\qquad v=\tilde{W}\left( z\right)
\end{equation}%
with two free variables $z$ and $\tilde{W}$. Substituting these values into (%
\ref{sf}), we obtain the complex curve 
\begin{equation}
\tilde{W}^{N_{f}}=\prod\limits_{l=1}^{N_{f}}p\left( z-\mu _{l}\right)
\label{WT}
\end{equation}%
where we have set $p=c^{1/N_{f}}.$ Notice that for the particular situation $%
N_{f}=1,$ the above $\tilde{W}^{N_{f}}$ takes its simplest \textrm{form} $%
\tilde{w}_{1}=p\left( z-\mu _{1}\right) $. Using this property, we can
rewrite (\ref{WT}) as follows%
\begin{equation}
\tilde{W}^{N_{f}}=\prod\limits_{l=1}^{N_{f}}\tilde{w}_{l}
\end{equation}%
with the holomorphic curves $\mathfrak{\tilde{C}}_{l}$ given by $\tilde{w}%
_{l}-\tilde{w}_{l+1}.$ By replacing $\tilde{w}_{l}=p\left( z-\mu _{l}\right) 
$, we end up with $p\left( \mu _{l}-\mu _{l+1}\right) $ characterising the
irreducible $\mathfrak{\tilde{C}}_{l}$ that generate the compact $\mathfrak{%
\tilde{C}}_{2}^{N_{f}}.$ So, the condition for U$\left( N_{f}\right) $ to be
a flavor symmetry is given by 
\begin{equation}
\left\vert \mu _{l}-\mu _{l+1}\right\vert >>1
\end{equation}

\subsection{Geometry of the 11D space time}

In this subsection, we use (\ref{st}) to construct the \emph{11D} space time 
$\mathcal{M}_{1,10}$ (\ref{21}) needed for embedding the $sl(3|2)$ superspin
chain in M theory. As there is a lack in the literature with \textrm{regards}
to the properties of singularities based on supergroups (super
singularities), we develop in the following an attempt to tackle this
problem and give partial results \textrm{in} this matter. \textrm{For} that
purpose, we \textrm{will} think about the space time $\mathcal{M}_{1,10}$ as
follows 
\begin{equation}
\begin{tabular}{lll}
$\mathcal{M}_{1,10}$ & $\sim $ & $\mathcal{M}_{1,9}\times \mathbb{S}_{M}^{1}$
\\ 
$\mathcal{M}_{1,9}$ & $\sim $ & $\mathbb{R}_{t}\times \mathbb{S}_{\theta
}^{1}\times \mathcal{M}_{8}$%
\end{tabular}
\label{M11D}
\end{equation}%
where $\mathcal{M}_{1,9}$ is the space time of type IIA string and $\mathbb{S%
}_{M}^{1}$\ is the M-theory circle.

\subsubsection{The real 8-manifold $\mathcal{M}_{8}$}

In the fibration (\ref{M11D}), the $\mathbb{R}_{t}\times \mathbb{S}_{\theta
}^{1}$ is just the 2D space time of the superspin chain\textrm{\footnote{%
\ \ For closed super chain, the real $\mathbb{R}_{1}$\ is replaced by the
circle $\mathbb{S}_{\theta }^{1}$.}} while the real 8D space $\mathcal{M}%
_{8} $ is thought of as a 4D complex manifold ${\large Y}_{4}$ given by the
following fibration 
\begin{equation}
\begin{tabular}{lll}
$\mathcal{\tilde{S}}_{2}^{N_{f}}$ & $\rightarrow $ & ${\large Y}_{4}$ \\ 
&  & $\downarrow $ \\ 
&  & $\mathcal{S}_{2}^{sl_{3|2}}$%
\end{tabular}
\label{Y4}
\end{equation}%
This factorisation is motivated by the two symmetry factors of the superspin
chain, namely $U\left( N_{f}\right) \times SL(3|2).$ The fiber $\mathcal{%
\tilde{S}}_{2}^{N_{f}}$ is given by (\ref{st}) with large volume cycles as
in (\ref{cpc}-\ref{ccp}). The base surface $\mathcal{S}_{2}^{sl_{3|2}}$ is a
graded complex 2-manifold given by%
\begin{equation}
\begin{tabular}{lll}
$\mathbb{C}$ & $\rightarrow $ & $\mathcal{S}_{2}^{sl_{3|2}}$ \\ 
&  & $\downarrow $ \\ 
&  & ${\large C}_{2}^{sl_{3|2}}$%
\end{tabular}
\label{ts}
\end{equation}%
Using the Super A/H (\ref{pcr}), the compact 2-cycle ${\large C}%
_{2}^{sl_{3|2}}$ is given by the sum of four intersecting graded 2-cycles 
\begin{equation}
{\large C}_{2}^{sl_{3|2}}=\mathfrak{C}_{1}+\mathfrak{C}_{2}+\mathfrak{C}_{3}+%
\mathfrak{C}_{4}  \label{csl2}
\end{equation}%
with the irreducible graded $\mathfrak{C}_{A}$ stretching between two graded
divisors $E_{A}$ and $E_{A+1}$ as follows%
\begin{equation}
\mathfrak{C}_{A}=E_{A}-E_{A+1}
\end{equation}%
Eq(\ref{csl2}) describes the blowing up of the graded singularity of $%
{\large C}_{2}^{sl_{3|2}}$ corresponding to the vanishing of the graded $%
\mathfrak{C}_{A}$ cycles. From the homology view, the blowing of this super
singularity is expressed as follows%
\begin{equation}
\begin{tabular}{lll}
$E_{1}-E_{5}$ & $=$ & $\left( E_{1}-E_{2}\right) +\left( E_{2}-E_{3}\right)
+\left( E_{3}-E_{4}\right) +\left( E_{4}-E_{5}\right) $ \\ 
& $=$ & $\mathfrak{C}_{1}+\mathfrak{C}_{2}+\mathfrak{C}_{3}+\mathfrak{C}_{4}$%
\end{tabular}%
\end{equation}

\subsubsection{Graded 2-cycles}

\begin{figure}[tbph]
\begin{center}
\includegraphics[width=10cm]{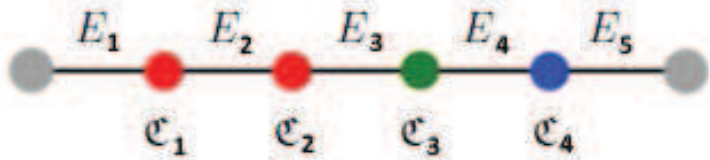}
\end{center}
\par
\vspace{-0.5cm}
\caption{The structure of the 2-cycle ${\protect\large C}_{2}^{sl_{3|2}}$ of
the $SL\left( 3|2\right) $ superspin geometry. The red dots refer to
2-spheres, the green dot to the 2-torus and the blue dot to the double
2-torus.}
\label{C32}
\end{figure}

\begin{figure}[tbph]
\begin{center}
\includegraphics[width=16cm]{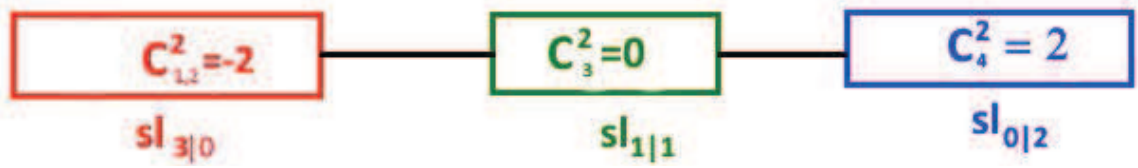}
\end{center}
\par
\vspace{-0.5cm}
\caption{Signs of the square of the simple roots $\mathbf{\protect\alpha }%
^{2}$ for the three gauge factors $sl\left( 3\right) \oplus sl\left(
1\right) \oplus sl\left( 2\right) $ in the super $sl(3|2).$ The five $M5_{%
\mathbf{A}}$ are \textrm{splited} as $M5_{\mathbf{a}}$ in the $sl\left(
3\right) $ sector and $M5_{3+i}$ in the $sl\left( 2\right) $.}
\label{SR}
\end{figure}

Due to the $\mathbb{Z}_{2}$-grading of the $SL\left( 3|2\right) $ gauge
symmetry, the graphical representation of the 2-cycle ${\large C}%
_{2}^{sl_{3|2}}$ is given by a colored Dynkin super diagram as depicted by
Figure \textbf{\ref{C32}}. 
This Figure should be compared with the \emph{DDSD} given by the Figure 
\textbf{\ref{DSD} }where the two red nodes are given by two intersecting
projective lines (2-spheres), the green node \textrm{by }an elliptic curve
(2-torus) and the blue node \textrm{by }a genus $g=2$ Rieman surface. Using
the indefinite divisor metric with (3,2) signature 
\begin{equation}
E_{A}.E_{B}=-g_{AB}
\end{equation}%
as in (\ref{4g}), we get the graded matrix intersection%
\begin{equation}
\mathfrak{C}_{A}\mathbf{.}\mathfrak{C}_{B}=\mathcal{I}_{AB}^{sl_{3|2}}
\end{equation}%
with%
\begin{equation}
\mathcal{I}_{AB}^{sl_{3|2}}=-\mathcal{K}_{AB}^{sl_{3|2}}
\end{equation}%
Due to the superspin algebra $sl(3|2)$ which has two sectors $sl(3|2)_{\bar{0%
}}\oplus sl(3|2)_{\bar{1}}$, the graded simple roots $\mathbf{\alpha }_{A}$
of $sl(3|2)$ as well as the associated graded cycles $\mathfrak{C}_{A}$ are
linked as sketched by Figure \textbf{\ref{SR}}. 
\ \ \ 

\subsection{M5 wrapping 4-cycle ${\protect\large C}_{4}$}

In this subsection, we study the geometry describing the wrapping of M5 over
the compact 4-cycle (\ref{sc2}) of the complex fourfold {\large Y}$_{4}.$
First, we construct the branes M5$_{\mathbf{A}},$ M5$_{\mathbf{A}}^{l}$ and
M2$_{\mathbf{AA+1}}$ given by the table \textbf{\ref{M52}}. Then, we give
the algebraic equation describing {\large Y}$_{4}.$

\subsubsection{Deriving the branes M5$_{\mathbf{A}}^{l}$ and M2$_{\mathbf{%
AA+1}}$}

Because of the fibration (\ref{Y4}), the complex 4D manifold ${\large Y}_{4}$
is imagined as 
\begin{equation}
{\large Y}_{4}\sim \mathcal{\tilde{S}}_{2}^{N_{f}}\times \mathcal{S}%
_{2}^{sl_{3|2}}
\end{equation}%
By substituting $\mathcal{\tilde{S}}_{2}^{N_{f}}$ and $\mathcal{S}%
_{2}^{sl_{3|2}}$ by their expressions (\ref{st}) and (\ref{ts}), we end up
with the following structure 
\begin{equation}
{\large Y}_{4}\sim \mathbb{C}^{2}\times {\large C}_{4}
\end{equation}%
where the 4-cycle \textrm{is fibred} as follows 
\begin{equation}
{\large C}_{4}\sim {\large \tilde{C}}_{2}^{N_{f}}\times {\large C}%
_{2}^{sl_{3|2}}
\end{equation}%
Replacing ${\large \tilde{C}}_{2}^{N_{f}}=\sum_{l}\mathfrak{\tilde{C}}^{l}$
and ${\large C}_{2}^{sl_{3|2}}=\sum_{A}\mathfrak{C}_{A}$, we have 
\begin{equation}
{\large C}_{4}=\sum_{l=1}^{N_{f}-1}\sum_{A=1}^{4}[{\large C}%
_{4}]_{A}^{l}\qquad ,\qquad \lbrack {\large C}_{4}]_{A}^{l}=\mathfrak{\tilde{%
C}}^{l}\times \mathfrak{C}_{A}  \label{c4}
\end{equation}%
By wrapping M5-brane over the 4-cycle (\ref{c4}), we obtain the $5$N$_{f}$
branes M5$_{\mathbf{A}}^{l}$ representing the super chain given by the
wrapped branes%
\begin{equation}
\text{M5}_{A}^{l}=\text{M5}/[{\large C}_{4}]_{A}^{l}
\end{equation}%
The wrapping of M5-brane over the 3-cycles ${\large C}_{A}^{3}=\mathbb{S}%
_{M}^{1}\times \mathfrak{C}_{A}$ gives the four M2$_{AA+1}$ branes
stretching between\textrm{\ the }M5$_{A}$ and M5$_{A+1}$ because $\mathfrak{C%
}_{A}=E_{A}-E_{A+1}$; i.e%
\begin{equation}
\text{M2}_{AA+1}=\text{M5}/{\large C}_{A}^{3}\qquad ,\qquad {\large C}%
_{A}^{3}=\mathbb{S}_{M}^{1}\times \mathfrak{C}_{A}
\end{equation}

\subsubsection{The defining equation of {\protect\large Y}$_{4}$}

We realise the complex fourfold {\large Y}$_{4}$ as a 4D sub-manifold living
in the ambient complex $\mathbb{C}^{5}$ with coordinates $\left(
x,y,z,u,v\right) .$ Because of the fibration ${\large Y}_{4}=\mathcal{\tilde{%
S}}_{2}^{N_{f}}\times \mathcal{S}_{2}^{sl_{3|2}},$ we can think about the
equation defining {\large Y}$_{4}$ inside $\mathbb{C}^{5}$ as follows\textrm{%
\footnote{%
\ \ Notice that if instead of the super $sl(3|2)$ we have the bosonic-like $%
sl(5)$, a simple way to realise the homologue of $Y_{4}$ that we denote as $%
{\large \tilde{Y}}_{4}=\mathcal{\tilde{S}}_{2}^{N_{\mathrm{f}}}\times 
\mathcal{S}_{2}^{sl_{5}}$ is to $\left( i\right) $ embed it in $\mathbb{C}%
^{6}$ with coordinates $\left( u_{1},v_{1},z_{1},u_{2},v_{2},z_{2}\right) $
and $\left( ii\right) $ use two equations as follows%
\begin{equation*}
u_{1}v_{1}=\prod\limits_{l=1}^{N_{\mathrm{f}}}\left( z_{1}-\mathrm{a}%
_{l}\right) ,\qquad u_{2}v_{2}=\prod\limits_{A=1}^{5}\left( z_{2}-\mathrm{%
\gamma }_{\mathbf{A}}\right)
\end{equation*}%
to reduce the number of variables down to 4.}}%
\begin{equation}
uv+h=\prod\limits_{A=1}^{5}c_{\mathbf{A}}\prod\limits_{l=1}^{N_{f}}\left(
z-\mu _{\mathbf{A}}^{l}\right)  \label{s22}
\end{equation}%
with $h$ and $c_{\mathbf{A}}$ two functions as%
\begin{equation}
h=h\left( x,y,z,u,v\right) \qquad ,\qquad c_{\mathbf{A}}=c_{\mathbf{A}%
}\left( x,y\right)
\end{equation}%
We give below four interesting features of the {\large Y}$_{4}$ hypersurface
of $\mathbb{C}^{5}$. \newline
$\left( \mathbf{1}\right) $ If we set the function $h=0$ and hide the $%
\left( x,y\right) $ coordinate variables of the base surface $\mathcal{S}%
_{2}^{sl_{3|2}}$ by thinking of the $c_{\mathbf{A}}$'s as constants, the eq(%
\ref{s22}) reduces to the particular space $\mathbb{C}^{2}\times \mathcal{%
\tilde{S}}_{2}^{(5N_{f})}$ with complex surface $\mathcal{\tilde{S}}%
_{2}^{(5N_{f})}$ 
\begin{equation}
uv=\prod\limits_{A=1}^{5}c_{\mathbf{A}}\prod\limits_{l=1}^{N_{f}}\left(
z-\mu _{\mathbf{A}}^{l}\right)
\end{equation}%
given by the blowing $\mathcal{\tilde{S}}_{2}^{N_{f}}+\mathcal{\tilde{S}}%
_{2}^{N_{f}}+\mathcal{\tilde{S}}_{2}^{N_{f}}+\mathcal{\tilde{S}}_{2}^{N_{f}}+%
\mathcal{\tilde{S}}_{2}^{N_{f}};$ thus having the following Du Val
singularity 
\begin{equation}
SU\left( N_{f}\right) ^{5}
\end{equation}%
$\left( \mathbf{2}\right) $ For a function $h$ vanishing at $\left(
x,y,z,u,v\right) =\left( x,y,\mu _{\mathbf{A}}^{l},0,0\right) ,$ and
functions $c_{\mathbf{A}}\left( x,y\right) $ taking constant values
respecting $SL\left( 3|2\right) $ invariance; say for instance $c_{\mathbf{A}%
}=\left( -\right) ^{\left\vert A\right\vert }c$; we recover the flavor $%
SU\left( N_{f}\right) $ singularity at the points $\left( x,y\right) =\left(
x_{A},y_{A}\right) $ solving 
\begin{equation}
h\left( x,y,\mu _{\mathbf{A}}^{l},0,0\right) =0\qquad ,\qquad c_{\mathbf{A}%
}\left( x,y\right) =\left( -\right) ^{\left\vert A\right\vert }c
\end{equation}%
At the loci $\left( x_{A},y_{A}\right) ,$ the eq(\ref{s22}) has zeros at $%
\left( u,v\right) =\left( 0,0\right) $ and along the z-direction at $z=\mu _{%
\mathbf{A}}^{l}.$ The zeros at $\mu _{\mathbf{A}}^{l}$ give the positions of
the M5$_{A}^{l}$ branes in the surface fiber $\mathcal{\tilde{S}}_{2}^{5N_{%
\mathrm{f}}}$. In terms of these 5N$_{f}$ moduli $\mu _{\mathbf{A}}^{l}$, we
can calculate the volume of 2-cycles and 4-cycles of {\large Y}$_{4}$. In
particular, we are interested in the volumes $\upsilon _{AA+1}^{l}$ and $%
\upsilon _{A}^{ll+1}$ of the 2-cycles associated with the $SL(3|2)$ gauge
symmetry and the $U(N_{f})$ flavor invariance:\newline
$\left( i\right) $ The holomorphic volumes $\upsilon _{AA+1}^{l}$ of the M2$%
_{\mathbf{AA+1}}$ branes stretching between M5$_{\mathbf{A}}$ and M5$_{%
\mathbf{A+1}}$ read as%
\begin{equation}
\upsilon _{AA+1}^{l}=\mu _{\mathbf{A}}^{l}-\mu _{\mathbf{A+1}}^{l}
\end{equation}%
Because $SL(3|2)$ is a gauge symmetry, these holomorphic volumes are
required to take small values; thus constraining the $\mu _{\mathbf{A}}^{l}$%
's like 
\begin{equation}
\left\vert \mu _{\mathbf{A}}^{l}-\mu _{\mathbf{A+1}}^{l}\right\vert <<1
\end{equation}%
$\left( ii\right) $ the holomorphic volumes $\upsilon _{A}^{ll+1}$ of the
flavor brane stretching between M5$_{\mathbf{A}}^{l}$ and M5$_{\mathbf{A}%
}^{l+1}$ are given by 
\begin{equation}
\upsilon _{A}^{ll+1}=\mu _{\mathbf{A}}^{l}-\mu _{\mathbf{A}}^{l+1}
\end{equation}%
Because the $SU\left( N_{f}\right) $ is flavor symmetry, these moduli are
constrained like%
\begin{equation}
\left\vert \mu _{\mathbf{A}}^{l}-\mu _{\mathbf{A}}^{l+1}\right\vert >>1
\end{equation}%
$\left( \mathbf{3}\right) $ The five functions $c_{\mathbf{A}}$ form a super
vector of $sl(3|2)$. Because of the $\mathbb{Z}_{2}$- grading of the $%
sl(3|2) $ superspin, one can put constraints on these $c_{\mathbf{A}}$'s. As
the bosonic $sl(3|2)_{\bar{0}}$ sector of $sl(3|2)$ splits into two
orthogonal symmetries namely $sl(3)$ and $sl(2);$ 
\begin{equation}
\left[ sl(3),sl(2)\right] =0
\end{equation}%
we can split the five $c_{\mathbf{A}}$'s in terms of a triplet $c_{a}$ of $%
sl(3)$ and a doublet $c_{3+i}$ of $sl(2).$ These $sl(3)$ and $sl(2)$
sub-symmetries allow \textrm{us} to \textrm{write} these five functions as
follows 
\begin{equation}
c_{1}=c_{2}=c_{3}=c\qquad ,\qquad c_{4}=c_{5}=c^{\prime }
\end{equation}%
in agreement with the Bose and the Fermi statistics. The first three $c_{a}$%
's are invariant under the permutation group $\mathbb{S}_{3};$ and the $%
c_{4}=c_{5}$ are invariant under their transposition while respecting the
Fermi statistics.\ An interesting choice of these functions is given by $%
c^{\prime }=-c$ such that $c_{\mathbf{A}}c_{\mathbf{B}}=\left( -\right)
^{\left\vert A\right\vert +\left\vert B\right\vert }$ and then%
\begin{equation}
\frac{c_{\mathbf{A}}}{c_{\mathbf{B}}}=\left( -\right) ^{\left\vert
A\right\vert +\left\vert B\right\vert }  \label{cab}
\end{equation}%
where $\left\vert A\right\vert $ refers to the degree of $\left\vert c_{%
\mathbf{A}}\right\vert $. \newline
$\left( \mathbf{4}\right) $ By setting $u=1$ and $v=W_{\mathbf{A}}\left(
z\right) $ in (\ref{s22}); we get 
\begin{equation}
W_{\mathbf{A}}\left( z\right) =c_{\mathbf{A}}\prod\limits_{l=1}^{N_{f}}%
\left( z-\mu _{\mathbf{A}}^{l}\right)
\end{equation}%
Following \cite{nafiz}, a configuration of $n_{A}$ parallel (horizontal) M2$%
_{\mathbf{AA+1}}^{n_{A}}$ branes suspended between the vertical M5$_{\mathbf{%
A}}$ and M5$_{\mathbf{A+1}},$ and located at $z=\sigma _{A}^{n_{A}}$
preserves supersymmetry if and only if the following constraint holds 
\begin{equation}
W_{\mathbf{A}}\left( z\right) =W_{\mathbf{A+1}}\left( z\right)
\end{equation}%
with $z=\sigma _{A}^{n_{A}}.$ Solving this condition, we end up with%
\begin{equation}
\prod\limits_{l=1}^{N_{f}}\frac{\sigma _{A}^{n_{A}}-\mu _{\mathbf{A+1}}^{l}}{%
\sigma _{A}^{n_{A}}-\mu _{\mathbf{A}}^{l}}=\frac{c_{\mathbf{A}}}{c_{\mathbf{%
A+1}}}
\end{equation}%
that \textrm{has} an interpretation in the Bethe equation description 
\textrm{\cite{nafiz}}.

\section{Conclusion and comments}

\begin{figure}[tbph]
\begin{center}
\includegraphics[width=12cm]{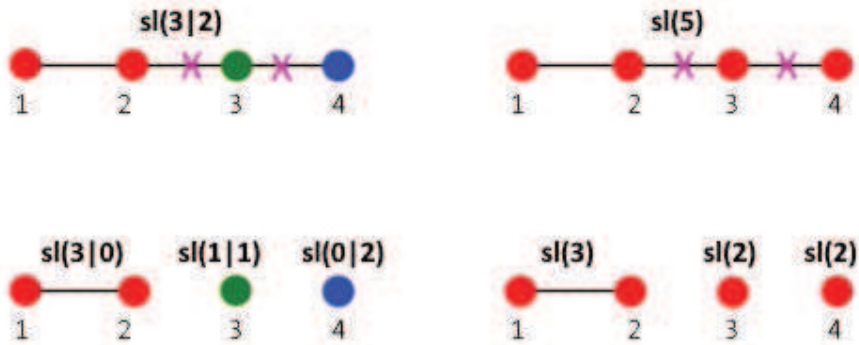}
\end{center}
\par
\vspace{-0.5cm}
\caption{Decompositions of the Dynkin super diagram of the distinguished $%
sl(3|2)$ and of the Dynkin diagram of $sl(5)$. These diagrams are thought of
in terms of intersecting 2-cycles. The colored nodes refer to graded cycles $%
\mathfrak{C}_{A}$ of $sl(3|2)$.}
\label{dec}
\end{figure}

In this paper, we \textrm{investigated the embedding} of integrable $sl(m|n)$
superspin chains and super magnons in M-theory and type II strings with two
main objectives.\newline
First, revisiting the algebraic set up of the $sl(m|n)$ super chain and
employing its degrees of freedom and its algebraic properties to:

\begin{description}
\item[$\left( \mathbf{a}\right) $] Link the quantum chain states to
topological line defects interpreted in this paper in terms of the Wilson
and the 't Hooft lines of 4D Chern-Simons theory. These topological lines,
represented in the Figure \textbf{\ref{TH}}, \textrm{offer} another way to
think about the integrability of the real super chain given in Figure 
\textbf{\ref{HC}} in agreement with the description of \cite{1A}.

\item[$\left( \mathbf{b}\right) $] Motivate brane realisations of the super
chain in type II strings and M-theory as collected by the Tables \textbf{\ref%
{IIA}} and \textbf{\ref{M52}}. This is achieved by using relationships
between the following triad: $\left( \mathbf{i}\right) $ root and weight
systems of Lie algebras. $\left( \mathbf{ii}\right) $ Homology cycles of
complex manifolds with singularities. $\left( \mathbf{iii}\right) $ Branes
in type II strings and M-theory.

\item[$\left( \mathbf{c}\right) $] Relate the exotic features of the
superspin states \textrm{to the} exotic properties of colored Dynkin super
diagrams of Lie superalgebras like those given by Figure \textbf{\ref{10}}.
\end{description}

\ \ \ \newline
\textrm{Secondly,} extending the Algebra/Homology (A/H) correspondence in
the so-called ADE geometries of Du Val to super groups (Super A/H). In this
regard, we used the Super A/H relationship to construct brane realisations
for the family of $sl(m|n)$ with $m>n$ while focussing on \textrm{the}
representative symmetry $sl(3|2)$. We took advantage of the A/H in the Du
Val geometries (\ref{418}) to draw a path towards the $SL(m|n)$ super
singularity. As examples, we studied the cycle-homologies of singular
geometries with the $SL(3|2)$ and $SL(5)$ gauge invariances. The
intersecting 2-cycles of $SL(3|2)$ and $SL(5)$ are illustrated by Figure 
\textbf{\ref{dec}}. 
On the left hand side of this Figure, we have $SL(3|2)$ colored 2-cycles
with the three self intersection values $\mathfrak{C}_{A}^{2}=0,\pm 2$. On
the right hand side, we have $SL(5)$ cycles (in red) with $\mathfrak{\tilde{C%
}}_{l}^{2}=-2$. To our knowledge, super geometries based on $SL(m|n)$
symmetry have not been studied \textrm{enough} in the stringy literature and
the associated super singularities are still an open problem in algebraic
geometry \cite{1E}. The investigation given in this paper constitutes a
contribution to this matter.

\ \ \ \ \ \newline
To undertake this study, we started by investigating the degrees of freedom
of the super chain by using representation group language and revisited the
algebraic set up of properties \textrm{of} the super atoms making the chain.
This analysis showed that there are $\left( m+n\right) !/m!n!$ varieties of
super chains classified by the $\left( m+n\right) !/m!n!$ possible Dynkin
super diagrams of $sl(m|n).$ Recall that contrary to the bosonic Lie
algebras, finite dimensional Lie superalgebras like $sl(m|n)$ have several
Dynkin-like diagrams termed in this paper as \emph{colored Dynkin diagrams}.
These colored diagrams have graded roots that enrich the study of superspin
chains thanks to the different possible varieties\ and the exotic properties
resulting from the three possible \textquotedblleft
lengths\textquotedblright\ $\mathbf{\alpha }_{A}^{2}=0,\pm 2.$ To exhibit
these interesting features, we studied in details the 10 super $sl(3|2)$
chains by using algebra and homology tools. Then, we focused on the
distinguished $sl(3|2)$ chain as representative of $sl(m|n)$ and studied the
geometrisation of properties of the ground state of the super chain in
connection with exotic singularities due to $SL(3|2)$ symmetry. After that,
we extended the Algebra/Homology correspondence regarding Du Val surfaces
with $SU\left( N_{f}\right) $ singularity to the case of super groups and%
\textrm{\ used} this super A/H to approach geometries with $SL(m|n)$ super
singularity.\ \ \newline
Using the obtained results concerning the properties of quantum states of
the super chain as well as the super A/H, we investigated the embedding of
the super chain in type II strings and M-theory. In the type IIA string, the
brane realisation involves varieties of NS5-, D2- and flavored D4-branes as
well as F1 strings. Because of the graded symmetry $SL(m|n)\times SU\left(
N_{f}\right) $ of the super chain, we distinguished the brane sets NS5$_{A}$%
, D6$^{l},$ D4$_{A}^{l}$ and D2$_{AA+1}$ as listed in the Table \textbf{\ref%
{IIA}}. To complete this brane system, notice that there are moreover F1
strings stretching between stacks of $N_{A}$ gauge-like D2$_{AA+1}$ branes
denoted as (D2$_{AA+1}$)$^{N_{A}}$; they describe the super magnons. \textrm{%
Although} not explicitly elaborated, a quite similar construction can be 
\textrm{performed} for type IIB strings.\ \ \ \newline
In M-theory, the brane realisation of the super chain uses various kinds of
M5- and M2-branes as given by the Table \textbf{\ref{M52}}. There, we
distinguished the brane sets M5$_{A}$, M5$_{A}^{l}$ and the gauge-like M2$%
_{AA+1}$ and their wrappings on compact cycles. The super magnons are
described by strings M2/$\mathbb{S}_{M}^{1}$ stretching between stacks of $%
N_{A}$ gauge-like M2$_{AA+1}$ branes given by (M2$_{AA+1}$)$^{N_{A}}$.\ \ \ 
\newline
After that, we constructed the 11D space time $\mathcal{M}_{1,10}=\mathbb{S}%
_{M}^{1}\times \mathcal{M}_{1,9}$ where the super chain is embedded. This
space has the structure $\mathbb{S}_{M}^{1}\times (\mathbb{R}_{t}\times 
\mathbb{S}_{\theta }^{1}\times {\large Y}_{4})$ \textrm{where the} 4D
complex manifold ${\large Y}_{4}$ \textrm{is} given by the fibration $%
\mathcal{\tilde{S}}_{2}^{N_{\mathrm{f}}}\times \mathcal{S}_{2}^{sl_{3|2}}$
with complex flavor surface $\mathcal{\tilde{S}}_{2}^{N_{\mathrm{f}}}$ of Du
Val type and gauge surface $\mathcal{S}_{2}^{sl_{3|2}}$. This gauge surface
has a super geometry with compact 2-cycles given by the colored Dynkin
diagram \textrm{of} Figure \textbf{\ref{dec}}.

\ \ \ \newline
We end these comments by\ mentioning \textrm{that the} brane realisation of
super magnons is to be extended for the families of superspin chains; in
particulat the $sl(m|n)$ and the orthosymplectic $osp(m|2n)$ chain. The
first step towards the construction of these follows from the super Yangian
representation\textrm{\ as} reported in the \textrm{A}ppendix. The next step
regards their brane realisations and demands some technical details
concerning the brane engineering and orientifolds. Progress in this
direction will be reported in a future occasion.

\section*{Appendix: Magnons in Yangian formalism}

In this appendix, we use the Yangian superalgebra $\mathcal{Y}_{sl\left(
m|n\right) }$ to construct the magnons in the $sl\left( m|n\right) $
superspin chain while focussing on the distinguished $sl\left( 3|2\right) .$
To that purpose, we introduce the $\mathcal{Y}_{sl\left( 3|2\right) }$, the
Bethe vectors and the Bethe roots. Then, we build the super magnons for the
closed $sl\left( 3|2\right) $ chain and investigate their algebraic
properties. Their brane realisation will be elsewhere.

\subsection*{Algebraic set up of the $sl\left( 3|2\right) $ magnons}

\begin{figure}[tbph]
\begin{center}
\includegraphics[width=10cm]{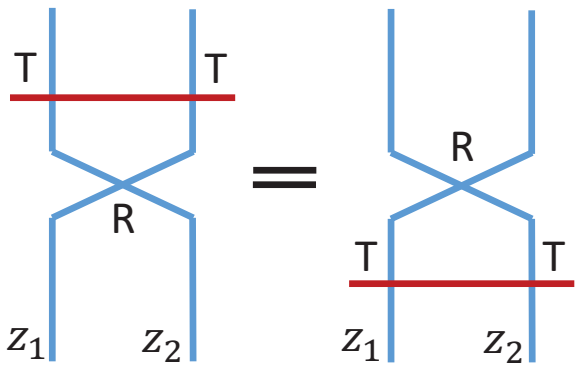}
\end{center}
\par
\vspace{-0.5cm}
\caption{The diagrammatic representation of the RLL relations encoding the
commutation relations between two T-operators at $z_{1}$ and $z_{2}$.}
\label{RTT}
\end{figure}

Magnons in $sl\left( 3|2\right) $ spin chain are quasi-\textrm{particles}
given by excitations of the ground state of \textrm{the} highest weight
representations of the Yangian $\mathcal{Y}_{sl(3|2)}$. This graded algebra
is an infinite dimensional Lie superalgebra given by a particular fibration
of $sl\left( 3|2\right) $ on the complex line $\mathbb{C}$ with coordinate $%
z $ (rapidity/spectral parameter). The $\mathcal{Y}_{sl(3|2)}$ is generated
by the monodromy matrix $T\left( z\right) $ of the $sl\left( 3|2\right) $
superspin chain. The $T\left( z\right) $ operator obeys the RTT
integrability equation defining precisely $\mathcal{Y}_{sl(3|2)}$ \textrm{as}%
\begin{equation}
R\left( z_{1}-z_{2}\right) T\left( z_{1},\zeta _{1}\right) T\left(
z_{2},\zeta _{2}\right) =T\left( z_{2},\zeta _{2}\right) T\left( z_{1},\zeta
_{1}\right) R\left( z_{1}-z_{2}\right)  \label{T}
\end{equation}%
where we have exhibited\ the inhomogeneity\textrm{\ }parameters $\zeta
_{1},\zeta _{2}$; see also Figure \textbf{\ref{RTT}} for a diagrammatic
representation in terms of the intersection of two vertical Wilson lines and
a horizontal 't Hooft line. 
For the explicit expression of 
\begin{equation}
T\left( z,\zeta \right) =\sum E_{AB}\oplus T_{AB}\left( z,\zeta \right)
\end{equation}%
in terms of $z$, $\zeta ,$ and the generators of $sl\left( 3|2\right) ,$ see
for instance \cite{nafiz}. $R\left( z\right) $ in (\ref{T})\ is the usual
R-matrix describing the coupling between two electrically charged line
defects.

\begin{figure}[tbph]
\begin{center}
\includegraphics[width=8cm]{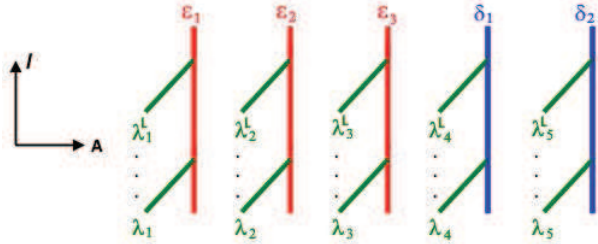}
\end{center}
\par
\vspace{-0.5cm}
\caption{Graphical representation of the pseudo-vacuum of the Yangian
superalgebra. In the horizontal axis, we have the unit $\protect\epsilon %
_{A} $ weight vectors. In the vertical axis, we have the direction of the $%
x_{l}$-chain indexed by the label $1\leq l\leq L.$}
\label{pv1}
\end{figure}

$\bullet $ \emph{Pseudo-vacuum}\newline
Verma modules of $\mathcal{Y}_{sl(3|2)}$ (highest weight representations)
are characterised by a HW state $\left\vert \Omega _{\mathbf{\Lambda }%
}\right\rangle $ often termed as a pseudo-vacuum of the integrable chain.
This ground state is constrained as 
\begin{equation}
\begin{tabular}{lll}
$T_{AA}\left( z,\zeta \right) \left\vert \Omega _{\mathbf{\Lambda }%
}\right\rangle $ & $=$ & $b_{\lambda _{A}}\left( z,\zeta \right) \left\vert
\Omega _{\mathbf{\Lambda }}\right\rangle $ \\ 
$T_{AB}\left( z,\zeta \right) \left\vert \Omega _{\mathbf{\Lambda }%
}\right\rangle $ & $=$ & $0,\qquad A<B$%
\end{tabular}
\label{BE}
\end{equation}%
with the graded labels $A,B=1,...5.$ In these relations, we have $\mathbf{%
\Lambda =}\sum_{l}\mathbf{\lambda }^{l};$ that is the total superspin of the
chain assumed to have $L$ sites with positions $x_{l}$ ($l=1,...,L$). 
\begin{equation}
\mathbf{\Lambda =\lambda }^{1}+\mathbf{\lambda }^{2}+...+\mathbf{\lambda }%
^{L}  \label{lam}
\end{equation}%
Using the weight vector basis $\left( \varepsilon _{1},\varepsilon
_{2},\varepsilon _{3},\delta _{1},\delta _{2}\right) $, we can express (\ref%
{lam}) like $\sum \Lambda _{A}\epsilon _{A}$ with%
\begin{equation}
\Lambda _{A}=\lambda _{A}^{1}+...+\lambda _{A}^{L}  \label{lL}
\end{equation}%
For convenience, we use two useful notations: $\left( \mathbf{i}\right) $ we
set below $\left\vert \Omega _{\mathbf{\Lambda }}\right\rangle \equiv
\left\vert \mathbf{\Lambda }\right\rangle $ or equivalently 
\begin{equation}
\left\vert \Omega _{\mathbf{\Lambda }}\right\rangle \equiv \left\vert
\left\{ \lambda _{A}^{l}\right\} \right\rangle  \label{gs}
\end{equation}%
showing that the ground state $\left\vert \Omega _{\mathbf{\Lambda }%
}\right\rangle $ is characterised by $L\times 5$ quantum numbers $\left\{
\lambda _{A}^{l}\right\} $. $\left( \mathbf{ii}\right) $ We represent
graphically the ground state (\ref{gs}) as shown in Figure \textbf{\ref{pv1}}%
. 

\subsection*{Elementary and composite magnons}

Excitations of the ground state $\left\vert \Omega _{\mathbf{\Lambda }%
}\right\rangle $ of the superspin chain (basis vectors of the $\mathcal{Y}%
_{sl(3|2)}$ Verma module) are obtained by acting with the $T_{AB}^{-}\left(
z\right) $ creators as follows 
\begin{equation}
\prod\limits_{i=1}^{\nu }T_{A_{i}B_{i}}\left( z_{i},\zeta _{i}\right)
\left\vert \Omega _{\mathbf{\Lambda }}\right\rangle \qquad ,\qquad
A_{i}>B_{i}  \label{EB}
\end{equation}%
In addition to the total weight $\mathbf{\Lambda }$ and the simple root
strings $\beta =\sum_{A}N_{A}\mathbf{\alpha }_{A}$ (not necessary roots of $%
sl(3|2)$), the excitation states are characterised by the complex $z_{i}$s
and $\zeta _{i}$s; \textrm{they} can be either elementary magnons or
composites.

$\bullet $ \emph{Elementary super-magnons}\newline
Particular states of such excitations are given by the four following 
\begin{equation}
T_{12}\left( z_{1},\zeta _{1}\right) \left\vert \mathbf{\Lambda }%
\right\rangle ,\quad T_{23}\left( z_{2},\zeta _{2}\right) \left\vert \mathbf{%
\Lambda }\right\rangle ,\quad T_{34}\left( z_{3},\zeta _{3}\right)
\left\vert \mathbf{\Lambda }\right\rangle ,\quad T_{45}\left( z_{4},\zeta
_{4}\right) \left\vert \mathbf{\Lambda }\right\rangle
\end{equation}%
Using the four simple roots $\alpha _{A}=\epsilon _{A}-\epsilon _{A+1}$ of
the superalgebra $sl\left( 2|3\right) $, we can present these four states as%
\begin{equation}
T_{-\alpha _{A}}\left( z_{A},\zeta _{A}\right) \left\vert \mathbf{\Lambda }%
\right\rangle =\left\vert \mathbf{\Lambda }-\alpha _{A};z_{A},\zeta
_{A}\right\rangle
\end{equation}%
These are the four elementary excitations of the $sl\left( 2|3\right) $
superspin chain that we identify with the elementary super-magnons; three of
them are bosonic and one is fermionic. We denote them as 
\begin{equation}
\begin{tabular}{lll}
bosonic & : & $\left[ 1;0;0;0\right] ,\quad \left[ 0;1;0;0\right] ,\quad %
\left[ 0;0;0;1\right] $ \\ 
fermionic & : & $\left[ 0;0;1;0\right] $%
\end{tabular}
\label{BR}
\end{equation}%
These super-magnons are graphically represented as in Figure \textbf{\ref%
{pv2}} that regards the elementary magnon $\left[ 1;0;0;0\right] .$ This
elementary magnon is moreover characterised by the spectral parameter $z_{1}$
and the inhomogeneous $\zeta _{1}$. In the language of \cite{nafiz}, the
Bethe roots associated with eq(\ref{BR}) are given by $\left\{
z_{A}^{1}\right\} .$ 
\begin{figure}[tbph]
\begin{center}
\includegraphics[width=6cm]{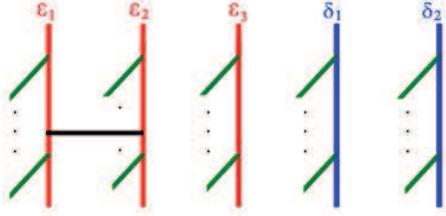}
\end{center}
\par
\vspace{-0.5cm}
\caption{Graphical representation of the\ basic magnon $\left[ 1;0;0;0\right]
$. Similar graphs can be drawn for the three other elementary magnons. The
Bethe root associated with this picture is given by $\left\{
z_{1}^{1}\right\} .$}
\label{pv2}
\end{figure}

\begin{figure}[tbph]
\begin{center}
\includegraphics[width=8cm]{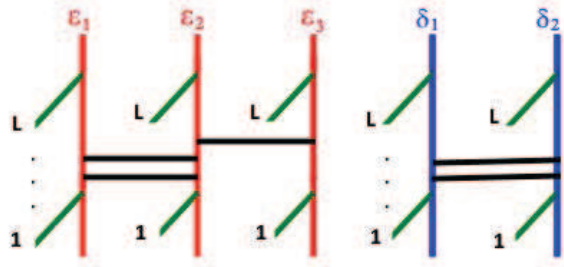}
\end{center}
\par
\vspace{-0.5cm}
\caption{Magnon diagram: A Graphical representation of the magnon $\left[
2;1;0;2\right] $. It consists of five elementary magnons, two elementary
pairs and one singlet. }
\label{pv3}
\end{figure}

\begin{figure}[tbph]
\begin{center}
\includegraphics[width=8cm]{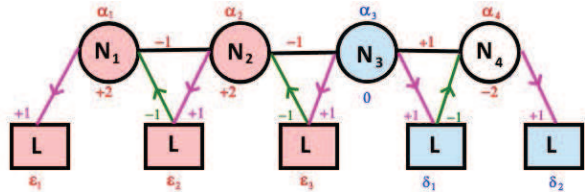}
\end{center}
\par
\vspace{-0.5cm}
\caption{The dual magnon diagram $\left[ N_{1};N_{2};N_{3};N_{4}\right] $.
It is characterised by symmetries of the elementary magnons.}
\label{mag1}
\end{figure}

$\bullet $ \emph{Composite super-magnons}\newline
In the diagrammatic language (Figure \textbf{\ref{pv2})}, composite
super-magnons are condensates of the four elementary super-magnons. They are
characterised by four positive integers $\left[ N_{1};N_{2};N_{3};N_{4}%
\right] $ giving the total numbers of elementary magnons. As such, the
quantum states describing composite super-magnons are denoted like,%
\begin{equation}
\left\vert N_{1},N_{2},N_{3},N_{4}\right\rangle =\left\vert \mathbf{\Lambda }%
-\mathbf{\beta }_{\mathbf{N}}\right\rangle  \label{4N}
\end{equation}%
with $\mathbf{\beta }_{\mathbf{N}}=\sum N_{A}\mathbf{\alpha }_{A}$ and
weight vector $\mathbf{\Lambda }_{\mathbf{N}}=\mathbf{\Lambda }-\mathbf{%
\beta }_{\mathbf{N}}$ reading explicitly as 
\begin{equation}
\begin{tabular}{lll}
$\mathbf{\Lambda }_{\mathbf{N}}$ & $=$ & $\mathbf{\Lambda }%
-\sum\limits_{A}N_{A}\alpha _{A}$ \\ 
& $=$ & $\sum\limits_{A}\left[ \Lambda _{A}\epsilon _{A}-N_{A}\alpha _{A}%
\right] $%
\end{tabular}%
\end{equation}%
\textrm{where }$\Lambda _{A}$ is as in (\ref{lL}). As for the pseudo-vacuum (%
\ref{BE}), the states (\ref{4N}) are also eigenstates of the Bethe
generators $T_{AA}\left( z,\zeta \right) .$ Notice that from the weight
vectors $\mathbf{\Lambda }_{\mathbf{N}}$, we can compute the intersection $%
\mathbf{\Lambda }_{\mathbf{N}}.\mathbf{\alpha }_{B}=Q_{B}$ reading as%
\begin{equation}
Q_{B}=\mathbf{\Lambda }.\mathbf{\alpha }_{B}-\mathbf{\beta }_{\mathbf{N}}.%
\mathbf{\alpha }_{B}
\end{equation}%
and having two contributions $Q_{B}^{0}=\mathbf{\Lambda }.\mathbf{\alpha }%
_{B}$ and $Q_{B}^{\prime }=\mathbf{\beta }_{\mathbf{N}}.\mathbf{\alpha }%
_{B}. $ The $Q_{B}^{0}$\textbf{\ }has an interpretation in quiver gauge
theories as describing fundamental matter while $Q_{B}^{\prime }$ is
interpreted in terms of adjoint and bi-fundamental matter. By substituting $%
\mathbf{\Lambda =}\sum \Lambda _{A}\epsilon _{A}$, we obtain%
\begin{equation}
\begin{tabular}{lll}
$Q_{B}^{0}$ & $=$ & $\sum\limits_{A}\Lambda _{A}\mathcal{G}_{AB}$ \\ 
$Q_{B}^{\prime }$ & $=$ & $-\sum\limits_{A}N_{A}\mathcal{K}_{AB}$%
\end{tabular}%
\end{equation}%
with $\mathcal{K}_{AB}$ as in (\ref{car}) and the generalised matrices $%
\mathcal{G}_{AB}=\left( \mathbf{\epsilon }_{A}.\mathbf{\alpha }_{B}\right) $%
\ given by%
\begin{equation}
\mathcal{G}_{AB}=\left( -\right) ^{B}\delta _{A,B}-\left( -\right)
^{B+1}\delta _{A,B+1}
\end{equation}%
\textrm{which explicitly reads as} 
\begin{equation}
\mathcal{G}_{AB}=\left( 
\begin{array}{ccccc}
1 & 0 & 0 & 0 & 0 \\ 
-1 & 1 & 0 & 0 & 0 \\ 
0 & -1 & 1 & 0 & 0 \\ 
0 & 0 & 1 & -1 & 0 \\ 
0 & 0 & 0 & 1 & -1%
\end{array}%
\right)
\end{equation}%
Another example of composites made of five elementary magnons is described
by the following state%
\begin{equation}
\left\vert 2,1,0,2\right\rangle =\left\vert \mathbf{\Lambda }-2\alpha
_{1}-\alpha _{2}-2\alpha _{4}\right\rangle
\end{equation}%
It consists \textrm{of a p}air $[1,0,0,0]{\small ,}$ a pair $[0,0,0,1]$ and
a singlet $[0,1,0,0].$ Its graphical representation is given by the Figure 
\textbf{\ref{pv3}}. 
From this description, we learn two interesting features: \newline
$\left( \mathbf{1}\right) $ The magnon diagram has intrinsic symmetries due
to the indistinguishable property of elementary magnons of\textrm{\ }the%
\textrm{\ }same nature. This symmetry factorizes like $G_{g}\times G_{f}$
with $\left( \mathbf{a}\right) $ $G_{g}$ standing for gauge symmetry due to
internal magnons (stretching between two neighboring vertical lines). $%
\left( \mathbf{b}\right) $ $G_{f}$ referring to a flavor symmetry concerning
the $L\times 5$ external lines (in Green color). For the magnon diagram of
Figure \textbf{\ref{pv3}}, we have the following symmetries 
\begin{equation}
\begin{tabular}{lll}
$G_{g}$ & $=$ & $U\left( 2\right) \times U\left( 1\right) \times U\left(
2\right) $ \\ 
$G_{f}$ & $=$ & $U\left( L\right) ^{5}$%
\end{tabular}%
\end{equation}%
$\left( \mathbf{2}\right) $ The magnon diagram has a dual representation
which looks like the well known diagrams of quiver gauge theories. Using 1D
duality mapping lines to points and points to lines, it is easy to see that
the dual of the magnon diagram \textbf{\ref{pv3}} is given by the quiver
graph of the Figure \ref{mag1}. This description done for $\mathcal{Y}%
_{sl_{3|2}}$ extends straightforwardly to $\mathcal{Y}_{sl_{m|n}}.$%
\begin{equation*}
\end{equation*}


\begin{thebibliography}{99}
\bibitem{1A} E. Witten, Integrable Lattice Models From Gauge Theory,
arXiv:1611.00592 [hep-th], https://doi.org/10.48550/arXiv.1611.00592.

\bibitem{2A} K. Costello, Integrable lattice models from four-dimensional
field theories, in String-Math 2013, vol. 88 of Proc. Sympos. Pure Math.,
pp. 3--23, Amer. Math. Soc., Providence, RI, (2014), 1308.0370, DOI.

\bibitem{2A0} Kevin Costello, Supersymmetric gauge theory and the Yangian,
(2013), arXiv:1303.2632 [hep-th].

\bibitem{2AA} Nathan Haouzi, Christian Schmid, Little String Origin of
Surface Defects, Journal of High Energy Physics volume 2017, Article number:
82 (2017), arXiv:1608.07279 [hep-th]

\bibitem{3A} N. A. Nekrasov and S. L. Shatashvili, Quantum integrability and
supersymmetric vacua, Prog. Theor. Phys. Suppl. 177 (2009) 105 [0901.4748].

\bibitem{3Aa} N. A. Nekrasov and S. L. Shatashvili, Supersymmetric vacua and
Bethe ansatz, Nucl. Phys. B Proc. Suppl. 192/193 (2009) 91 [0901.4744].

\bibitem{3AA} M. Yamazaki, New Integrable Models from the Gauge/YBE
Correspondence, arXiv:1307.1128 [hep-th],
https://doi.org/10.1007/s10955-013-0884-8.

\bibitem{3AB} D. Orlando and S. Reffert, Relating gauge theories via
gauge/Bethe correspondence, JHEP 10 (2010) 071, 30 [1005.4445]. 

\bibitem{3AC} D. Orlando, A stringy perspective on the quantum integrable
model/gauge correspondence, arXiv:1310.0031 [hep-th],
https://doi.org/10.48550/arXiv.1310.0031.

\bibitem{3AD} E. Witten, Fivebranes and knots, Quantum Topol. 3 (2012) 1
[1101.3216].

\bibitem{3AE} A. Hanany and E. Witten, Type IIB superstrings, BPS monopoles,
and three-dimensional gauge dynamics, Nucl. Phys. B 492 (1997) 152
[hep-th/9611230].

\bibitem{3AF} D. Volin, String hypothesis for gl(n|m) spin chains: a
particle/hole democracy, Lett. Math.Phys. 102 (2012) 1 [1012.3454].

\bibitem{3AG} E.H Saidi, Quantum line operators from Lax pairs, Journal of
Mathematical Physics 61, 063501 (2020), arXiv:1812.06701 [hep-th].

\bibitem{3AJ} EH Saidi, Twisted 3D supersymmetric YM on deformed lattice,
Journal of Mathematical Physics 55 (1), 012301.

\bibitem{3AH} T. Okuda, Line operators in supersymmetric gauge theories. In
New dualities of super gauge theories (pp. 195-222). Springer, (2016),
arXiv:1412.7126 [hep-th]

\bibitem{nafiz} N. Ishtiaque, S. F. Moosavian, S. Raghavendran, J. Yagi,
Superspin chains from superstring theory,
https://doi.org/10.48550/arXiv.2110.15112 [hep-th].

\bibitem{wtn1} K. Costello, E. Witten, M. Yamazaki, Gauge Theory and
Integrability, I, arXiv:1709.09993 [hep-th],
https://doi.org/10.4310/ICCM.2018.v6.n1.a6.

\bibitem{8B} V. Mikhaylov, E. Witten, Branes And Supergroups,
arXiv:1410.1175 [hep-th], https://doi.org/10.1007/s00220-015-2449-y.

\bibitem{wtn2} K. Costello, E. Witten, M. Yamazaki, Gauge Theory and
Integrability, II, arXiv:1802.01579 [hep-th],
https://doi.org/10.4310/ICCM.2018.v6.n1.a7.

\bibitem{wtn3} K. Costello, E. Witten, M. Yamazaki, Gauge Theory and
Integrability, III, arXiv:1908.02289 [hep-th],
https://doi.org/10.48550/arXiv.1908.02289.

\bibitem{4A} K. Maruyoshi, T. Ota, J. Yagi, Wilson-'t Hooft lines as
transfer matrices. Journal of High Energy Physics, 2021(1), 1-31, (2021),
arXiv:2009.12391 [hep-th].

\bibitem{5A} E.H Saidi, L.B Drissi,\ 5D N = 1 super QFT: symplectic quivers,
Nucl Phys B 2021.

\bibitem{6A} K. Maruyoshi, Wilson-'t Hooft Line Operators as Transfer
Matrices. Progress of Theoretical and Experimental Physics, (2021).

\bibitem{1B} K. Costello, D. Gaiotto, J. Yagi, Q-operators are 't Hooft
lines, arXiv:2103.01835 [hep-th], https://doi.org/10.48550/arXiv.2103.01835;

\bibitem{2B} \textrm{M. Ashwinkumar, }M. Tan, Q. Zhao, Branes and
Categorifying Integrable Lattice Models, arXiv:1806.02821 [hep-th],
https://doi.org/10.4310/ATMP.2020.v24.n1.a1.\newline
Meer Ashwinkumar, Meng-Chwan Tan, Unifying Lattice Models, Links and Quantum
Geometric Langlands via Branes in String Theory,
Adv.Theor.Math.Phys.24:1681-1721, 2020, arXiv:1910.01134v3 [hep-th].

\bibitem{3B} K. Costello, J. Yagi, Unification of integrability in
supersymmetric gauge theories, arXiv:1810.01970 [hep-th],
https://doi.org/10.4310/ATMP.2020.v24.n8.a1.

\bibitem{4B} Youssra Boujakhrout, El Hassan Saidi, On Exceptional 't Hooft
Lines in 4D-Chern-Simons Theory, Nucl.Phys.B 980 (2022) 115795,
10.1016/j.nuclphysb.2022.115795, arXiv:2204.12424.

\bibitem{4BA} E.H Saidi, M.B Sedra, Hyper-Kaehler Metrics Building and
Integrable Models, Modern Physics Letters A 9 (34), 3163-3173.

\bibitem{4BB} E.H Saidi, M.B Sedra, On the Gelfand-Dickey algebra GD(SLn)
and the Wn-symmetry. I. The bosonic case Journal of Mathematical Physics 35
(6), 3190-3210.

\bibitem{5B} Youssra Boujakhrout, El Hassan Saidi, Minuscule ABCDE Lax
operators from 4D Chern-Simons theory, Nucl.Phys.B 981 (2022) 115859.

\bibitem{5BA} Y. Boujakhrout, E.H Saidi, R. Ahl Laamara, L.B Drissi, 't
Hooft lines of ADE-type and Topological Quivers, LPHE-MS preprint-2022,
under consideration in Physics Review D.

\bibitem{6B} R. J. Baxter, Exactly Solved Models in Statistical Mechanics,
Academic Press, Inc.[Harcourt Brace Jovanovich, Publishers], London, (1989)

\bibitem{7B} N. Nekrasov, Superspin chains and supersymmetric gauge
theories, arXiv:1811.04278 [hep-th],
https://doi.org/10.1007/JHEP03\%282019\%29102.

\bibitem{paper3} Y. Boujakhrout, E.H Saidi, R. Ahl Laamara, L.B Drissi, Lax
operator and superspin chains from 4D CS gauge theory, J. Phys. A: Math.
Theor. 55 415402, arXiv:2209.07117v1 [hep-th].

\bibitem{9B} M. Yamazaki, New T-duality for Chern--Simons theory, JHEP 12
(2019) 090 [1904.04976].

\bibitem{10B} N. Nekrasov and E. Witten, The Omega deformation, branes,
integrability and Liouville theory, JHEP 09 (2010) 092 [1002.0888].

\bibitem{1CA} Jiaju Zhang, M. A. Rajabpour, Entanglement of magnon
excitations in spin chains, JHEP 02 (2022) 072, arXiv:2109.12826v2
[cond-mat.stat-mech].

\bibitem{1CB} Ning Wu, Hosho Katsura, Sheng-Wen Li, Xiaoming Cai, Xi-Wen
Guan, arXiv:2106.14809v4 [cond-mat.stat-mech], Phys. Rev. B 105, 064419
(2022).

\bibitem{1CC} S. Bao, J. Wang, W. Wang, Z. Cai, S. Li, Z. Ma, D. Wang, K.
Ran, Z. Dong, D. L. Abernathy, S. Yu, X. Wan, J. Li, J. Wen, Discovery of
coexisting Dirac and triply degenerate magnons in a three-dimensional
antiferromagnet, Nature communications, 9(1), 1-7.

\bibitem{1C} N. A. Nekrasov and S. L. Shatashvili, Supersymmetric vacua and
Bethe ansatz, Nucl. Phys. B Proc. Suppl. 192/193 (2009) 91 [0901.4744].

\bibitem{2C} C. Kristjansen1, D. Muller and K. Zarembo, Overlaps and
Fermionic Dualities for Integrable Super Spin Chains,
https://doi.org/10.48550/arXiv.2011.12192 [hep-th].

\bibitem{3C} C. Kristjansen1, D. Muller and K. Zarembo, Duality Relations
for Overlaps of Integrable Boundary States in AdS/dCFT, J. High Energ. Phys.
2021, https://doi.org/10.1007/JHEP09(2021)004

\bibitem{1D} S. Katz, P. Mayr, and C. Vafa, Mirror symmetry and exact
solution of 4-D N=2 gauge theories: 1., Adv.Theor.Math.Phys. 1 (1998)
53-114, arXiv:hep-th/9706110.

\bibitem{2D} El Hassan Saidi, Mutation Symmetries in BPS Quiver Theories:
Building the BPS Spectra, arXiv:1204.0395, JHEP, 2012, Volume 2012, Number
8, 18.

\bibitem{3D} R. Ahl Laamara, M. Ait Ben Haddou, A Belhaj, L. B Drissi, E. H
Saidi, RG Cascades in Hyperbolic Quiver Gauge Theories, Nucl.Phys. B702
(2004) 163-188, arXiv:hep-th/0405222.

\bibitem{4D} M. Ait Ben Haddou, A. Belhaj, E. H. Saidi, Geometric
Engineering of N=2 CFT\_ \{4\}s based on Indefinite Singularities:
Hyperbolic Case, Nucl.Phys. B674 (2003) 593-614, arXiv:hep-th/0307244 .

\bibitem{4DA} Rachid Ahl Laamara, Lalla Btissam Drissi, El Hassan Saidi,
D-string fluid in conifold: I. Topological gauge model,
Nucl.Phys.B743:333-353,2006, arXiv:hep-th/0604001v1.

\bibitem{4DB} Rachid Ahl Laamara, Lalla Btissam Drissi, El Hassan Saidi,
D-string fluid in conifold: II. Matrix model for D-droplets on S3 and S2,
Nuclear Physics B 749(1):206-224, arXiv:hep-th/0605209v1.

\bibitem{5D} Malika Ait Benhaddou, El Hassan Saidi, Explicit Analysis of
Kahler Deformations in 4D N=1 Supersymmetric Quiver Theories, Physics
Letters B575(2003)100-110, arXiv:hep-th/0307103.

\bibitem{ABC} \textrm{Hirotaka Hayashi, Takuya Okuda, Yutaka Yoshida}, ABCD
of 't Hooft operators, J. High Energy. Phys. 2021, 241 (2021).

\bibitem{6D} C. N. Yang, Some exact results for the many-body problem in one
dimension with repulsive delta-function interaction, Physical Review Letters
19 (23) (1967) 1312.

\bibitem{7D} R. J. Baxter, Partition function of the eight-vertex lattice
model, Annals of Physics 70 (1) (1972) 193--228.

\bibitem{8D} R. Frassek, Oscillator realisations associated to the D-type
Yangian: Nucl. Phys B, 956, 115063, (2020), arXiv:2001.06825 [math-ph].

\bibitem{1I} T. Okuda, Line operators in supersymmetric gauge theories. In
New dualities of super gauge theories (pp. 195-222). Springer, (2016),
arXiv:1412.7126 [hep-th].

\bibitem{2I} Kapustin, A. Wilson-'t Hooft operators in four-dimensional
gauge theories and S-duality. Physical Review D, 74(2), 025005, (2006),
arXiv:hep-th/0501015.

\bibitem{3I} Anton Kapustin, Natalia Saulina, The algebra of Wilson-'t Hooft
operators, Nucl.Phys.B814:327-365,2009, arXiv:0710.2097 [hep-th].

\bibitem{Volin} D. Volin, String hypothesis for gl(n|m) spin chains: a
particle/hole democracy,\ Lett. Math. Phys. 102 (2012) 1--29,
arXiv:1012.3454 [hep-th].

\bibitem{DV} Fabio Perroni, Orbifold Cohomology of ADE-singularities, PhD
thesis at SISSA, Trieste (Italy), 113 pages, arXiv:math/0510528v1 [math.AG],
https://doi.org/10.48550/arXiv.math/0510528.

\bibitem{HEPA} El Hassan Saidi, Mutation Symmetries in BPS Quiver Theories:
Building the BPS Spectra, Journal of High Energy Physics, 2012, Volume 2012,
Number 8, 18, arXiv:1204.0395 [hep-th].

\bibitem{HEPB} E. H. Saidi, Weak Coupling Chambers in N=2 BPS Quiver Theory,
Nuclear Phys B Volume 864, Issue 1, 2012, Pages 190-202, arXiv:1208.2887
[hep-th].

\bibitem{DW} L. B. Drissi, E. H. Saidi, Domain Walls in Topological
Tri-hinge Matter, European Physical Journal Plus 136, (68) (2021).

\bibitem{DW1} L.B. Drissi, E.H. Saidi, A Signature Index for Third Order
Topological Insulator. J. Conden. Matter Phys. 5; 32(36), 365704 (2020)

\bibitem{SH1} Nafiz Ishtiaque, Seyed Faroogh Moosavian, Yehao Zhou,
Topological Holography: The Example of The D2-D4 Brane System, SciPost Phys.
9, 017 (2020), arXiv:1809.00372.

\bibitem{SH2} Jihwan Oh, Yehao Zhou, Feynman diagrams and deformed M-theory,
SciPost Phys. 10, 029 (2021). arXiv:2002.07343.

\bibitem{SH3} Simeon Hellerman, Domenico Orlando, Susanne Reffert, J. High
Energ. Phys. 2012, 61 (2012)., arXiv:1204.4192 [hep-th]

\bibitem{SH4} Simeon Hellerman, Domenico Orlando, Susanne Reffert, J. High
Energ. Phys. 2012, 148 (2012), arXiv:1106.0279 [hep-th].

\bibitem{1E} Pierre Henry-Labordere, Bernard Julia, Louis Paulot, Borcherds
symmetries in M-theory, JHEP 0204 (2002) 049, arXiv:hep-th/0203070v2.
\end{thebibliography}
\end{document}